\documentclass[12pt,preprint]{aastex}

\def\linebreak{\hfil\break}

\def\
{\hfil\linebreak}

\def\degree{\ifmmode {^\circ}\else {$^\circ$}\fi}
\def\mum{\ifmmode {\rm \mu {\rm m}}\else $\rm \mu {\rm m}$\fi}
\def\arcsec{\ifmmode ^{\prime \prime}\else $^{\prime \prime}$\fi}

\def\inch{\ifmmode ^{\prime \prime}\else $^{\prime \prime}$\fi}
\def\arcmin{\ifmmode ^{\prime}\else $^{\prime}$\fi}

\def\mearth{M$_\oplus$}
\def\msun{M$_\odot$}
\def\lsun{L$_\odot$}
\def\2470{[24]--[70]}

\newbox\grsign \setbox\grsign=\hbox{$>$} \newdimen\grdimen \grdimen=\ht\grsign
\newbox\simlessbox \newbox\simgreatbox
\setbox\simgreatbox=\hbox{\raise.5ex\hbox{$>$}\llap
     {\lower.5ex\hbox{$\sim$}}}\ht1=\grdimen\dp1=0pt
\setbox\simlessbox=\hbox{\raise.5ex\hbox{$<$}\llap
     {\lower.5ex\hbox{$\sim$}}}\ht2=\grdimen\dp2=0pt

\begin{document}

\title{Variations on Debris Disks: Icy Planet Formation 
at 30--150~AU for 1--3 \msun\ Main Sequence Stars}
\vskip 7ex
\author{Scott J. Kenyon}
\affil{Smithsonian Astrophysical Observatory,
60 Garden Street, Cambridge, MA 02138} 
\email{e-mail: skenyon@cfa.harvard.edu}

\author{Benjamin C. Bromley}
\affil{Department of Physics, University of Utah, 
201 JFB, Salt Lake City, UT 84112} 
\email{e-mail: bromley@physics.utah.edu}
%
%

\begin{abstract}

We describe calculations for the formation of icy planets and debris 
disks at 30--150~AU around 1--3~\msun\ stars. Debris disk formation
coincides with the formation of planetary systems. As protoplanets
grow, they stir leftover planetesimals to large velocities. A cascade 
of collisions then grinds the leftovers to dust, forming an observable
debris disk. Stellar lifetimes and the collisional cascade limit the
growth of protoplanets. The maximum radius of icy planets,
$r_{max} \approx$ 1750 km, is remarkably independent of initial disk 
mass, stellar mass, and stellar age. These objects contain $\lesssim$ 
3\%--4\% of the initial mass in solid material. Collisional cascades
produce debris disks with maximum luminosity $\sim 2 \times 10^{-3}$ 
times the stellar luminosity.  The peak 24~$\mu$m excess varies from 
$\sim$ 1\% times the stellar photospheric flux for 1 \msun\ stars to 
$\sim$ 50 times the stellar photospheric flux for 3 \msun\ stars.  
The peak 70--850 $\mu$m excesses are $\sim$ 30--100 times the stellar 
photospheric flux.  For all stars, the 24--160~$\mu$m excesses rise at 
stellar ages of 5--20~Myr, peak at 10--50~Myr, and then decline. The 
decline is roughly a power law, $ f \propto t^{-n}$ with $n \approx$ 
0.6--1.0. This predicted evolution agrees with published observations 
of A-type and solar-type stars. The observed far-IR color evolution of 
A-type stars also matches model predictions.

\end{abstract}


\section{INTRODUCTION}

During the past 25 years, observations from {\it IRAS}, {\it ISO},
and {\it Spitzer} have revealed substantial mid-infrared (mid-IR)
excesses associated with hundreds of normal main sequence stars 
\citep[e.g.,][]{bac93,hab01,rie05,bry06,moor06,rhe07a}. 
Current samples include stars with spectral types A--M and ages 
$\sim$ 5~Myr to $\sim$ 10 Gyr 
\citep[e.g.,][]{che05,kim05,rie05,bei06,su06,hil08}.
Although binary stars and single stars in dense clusters and in
the field are roughly equally likely to have IR excesses 
\citep{stau05,su06,bry06,gor06,cur07a,sie07,tri07}, the frequency 
of excess emission declines from $\sim$ 30\%--40\% for A-type stars 
\citep{su06} to $\sim$ 10\%--20\% for solar-type stars 
\citep{gre03,tri08,mey08}. Thus, this phenomenon is common among 
main sequence stars and may depend on stellar mass.

High quality images demonstrate that dust orbiting the central star 
produces the excesses \citep{smi84,bra04,sta04,kal05,mey07}. In 
$\beta$ Pic and AU Mic, the dust is in a geometrically thin, edge-on 
disk with an outer radius of $a \sim$ 200--1000~AU 
\citep{smi84,tel88,gol93,kal04,liu04a,aug06}. 
In these disks, the small scale height of the dust, $H/a \sim$ 0.1, 
is consistent with material in roughly circular orbits 
\citep[e.g.,][and references therein]{bac93,kal04}.  Although broad 
tori of dust are visible in many other systems 
\citep[e.g.][]{gre98,au99,hol03,kal06,su06,fit07},
narrow rings of dust produce the emission in $\alpha$ PsA and HR 4796A 
\citep{jay98,sch99,gre00,tel00,kal05b}.
For systems with face-on rings and tori, the total emission constrains 
the scale height, $H/a \sim$ 0.1. Thus, the dust in these systems is 
as highly flattened as the structures in $\beta$ Pic and AU Mic.

Broadband spectral energy distributions constrain the luminosity, size, 
temperature, and total mass of the dust \citep{bac93,lag00,den00,wol03}.  
Some stars have excesses from grains plausibly associated with the 
terrestrial zone 
\citep[e.g.,][]{bei05,abs06,cur07b,rhe07b,mey08,lis07c}. 
Optically thin emission from cooler material with temperature $T \sim$ 
20--150~K is more typical \citep[e.g.][]{su06,tri08,hil08}. 
For systems with submm observations, the measured fluxes suggest grains 
with sizes $\sim$ 1~$\mu$m--1~cm and total mass $\sim$ 
0.01~\mearth\ \citep{liu04b,naj05,che06,wil06}. 
The grains have compositions similar to dust in the asteroid belt, 
comets, or the trans-Neptunian region of the Solar System 
\citep{gru95,bro97,lis07a,lis07b}.
Because the dust mass in these systems lies between the initial mass 
of solids in protostellar disks 
\citep[$\sim$ 100--1000~\mearth;][]{nat2000,and05}
and the dust mass in the Solar System
\citep[$\lesssim 10^{-4}$ \mearth;][]{hah02,lan02,nes06}, the 
dusty structures in these systems are often called `debris disks' 
\citep{bac93,lag00}.

In addition to the dust properties, several other observations suggest
plausible links between debris disks and the formation of planetary systems. 
Observations of A-type stars suggest a `rise and fall' of debris disk 
emission \citep{cur08a}, with a clear increase in the typical 24~$\mu$m 
excess at 5--10~Myr, a peak at 10--20~Myr, and a decline for $t \gtrsim$ 
20--30~Myr. The rise in debris disk emission roughly coincides with the 
disappearance of optically thick emission from protostellar disks 
\citep[e.g.][]{hai01,sic06,her06,cur07a,cur07c}. 
The broad plateau occurs at a time when radiometric dating 
\citep{yin02} and theory \citep{cha01,kb06} suggest the Earth 
contained $\sim$ 90\% of its final mass. The decline of dusty debris 
around A-type stars is at roughly the same time as a gradual decrease 
in the cratering rate of objects in the Solar System
\citep{swi93,mel93,wad00,koe03}. These results suggest that the evolution 
of dust in debris disks parallels the evolution of larger solid objects
in the Solar System.

Simple physical arguments also link debris disks with the formation of
planetary systems. Because radiation removes 1--100~$\mu$m grains on 
timescales shorter than the stellar age, some process replenishes the dust. 
To maintain the observed dust masses for long timescales, normal stars 
must have a large reservoir, $\sim$ 10--100 \mearth, of unseen objects
that continuously collide at large velocities and fragment into smaller 
objects. Remnant material from planet formation satisfies both needs.
The growth of 1000 km or larger planets in a disk of small grains
naturally leaves behind an ensemble of `leftover' 1--10~km `planetesimals'
on eccentric orbits \citep[][2004b]{kb04a}. For a mass of $\sim$ 
10--100~\mearth\ in leftovers, high velocity collisions produce enough
dust for most debris disks \citep[e.g.,][]{bac93,hab01,kb04b}. If this 
interpretation is correct, debris disks provide conclusive evidence 
for the formation of Pluto-mass or larger planets around many, if not 
most, main sequence stars.

In addition to these considerations, numerical calculations suggest that 
an evolving swarm of 1--10~km planetesimals explains several observed 
trends in the properties of debris disks.  Starting with an ensemble 
of $\lesssim$ 1 km-sized planetesimals, \citet[][2004a, b, 2005]{kb02b} 
show that 
collisions and mergers form 500--1000~km-sized objects in 1--50 Myr.
These protoplanets stir up leftover planetesimals along their orbits. 
Destructive collisions among the leftovers then produce a collisional 
cascade -- where collisions gradually grind large objects into smaller 
ones -- along with copious amounts of dust 
\citep[see also][]{wil94,dur97,qui07}.  \citet{dom03}, \citet{wya07a}, and 
\citet{loh08} show that collisional evolution in a belt of high velocity 
planetesimals naturally produces a dust luminosity that declines roughly 
inversely with time \citep[see also][2004b, 2005]{kb02a}, explaining 
the observed time evolution -- $L_d \propto t^{-n}$, with $n \approx$ 
0.5--1 -- suggested by recent observations of A-type stars
\citep[e.g.,][]{kal98,hab01,dec03,gre03,rie05,rhe07a}.  To account for 
the large observed range of IR excesses among stars of similar ages, 
\citet{wya07a} propose belts with a range of initial masses and semimajor
axes, as suggested from submm observations of protostellar disks 
\citep[][2007b]{and05}.

Here, we continue to explore the evolution of dusty debris arising from 
planet formation in a disk of icy planetesimals. Our suite of calculations 
for disks at 30--150~AU around 1--3~\msun\ stars yields robust predictions 
for the maximum sizes of icy planets as a function of semimajor axis and 
stellar age. Results for the long-term evolution of IR excesses account
for many fundamental aspects of the data.  These calculations are the 
first to explain the `rise and fall of debris disks' around A-type stars 
\citep{cur08a} and the apparent peak in the 70--160~$\mu$m excesses of G-type 
stars with ages of $\sim$ 100~Myr \citep{hil08}. Comparisons between our 
models and current observations suggest that the minimum stable grain size 
and the slope of the IR emissivity law are critical parameters. 

The models suggest a set of further critical observations.  Spatially
resolved images of debris disks around A-type and solar-type stars can 
improve our understanding of the minimum stable grain size. Larger
samples of debris disks with high quality submm data from ALMA, Herschel,
and SOFIA can place better constraints on the slope of the emissivity law.
Together, these data can test our predictions for the time evolution
of debris disk emission around 1--3 \msun\ stars and provide input for 
more complete calculations that include the formation and dynamical
evolution of giant planets.

We outline our model in \S2. We describe results for the formation of 
icy planets in \S3 and the evolution of debris disks in \S4.  After
discussing several applications of our calculations in \S5, we conclude 
with a brief summary in \S6.

\section{THE MODEL}

\citet[][2002a, 2004a, 2004c]{kb01} and \citet{bk06} describe our hybrid 
multiannulus numerical model for planetesimal growth.  \citet[][1999]{kl98}, 
\citet[][2002a]{kb01}, and \citet{bk06} compare results with analytical 
and numerical calculations.  We adopt the \citet{saf69} statistical 
approach to calculate the collisional evolution of an ensemble of 
planetesimals in orbit around a star of mass $M_{\star}$ 
\citep[see also][]{spa91,wei97,kri06,the07,loh08}. The model grid contains 
$N$ concentric annuli with widths $\delta a_i$ centered at semimajor axes 
$a_i$.  Calculations begin with a differential mass distribution $n(m_{ik}$) 
of objects with horizontal and vertical velocities $h_{ik}(t)$ and 
$v_{ik}(t)$ relative to a circular orbit.  The horizontal velocity is 
related to the orbital eccentricity, 
$e_{ik}^2(t)$ = 1.6 $(h_{ik}(t)/V_{K,i})^2$, where $V_{K,i}$ is the circular 
orbital velocity in annulus $i$.  The orbital inclination depends on the 
vertical velocity, $i_{ik}^2(t)$ = sin$^{-1}(2(v_{ik}(t)/V_{K,i})^2)$.

The mass and velocity distributions evolve in time due to inelastic collisions, 
drag forces, and gravitational forces.  For inelastic collisions, we solve 
the coagulation equations for a particle in mass batch $k$ of annulus $i$ 
colliding with another particle in mass batch $l$ of annulus $j$, 
\begin{equation}
\delta n_{i^{\prime}k^{\prime}} = \delta t \left [ \epsilon_{ijkl} A_{ijkl} n_{ik} n_{jl}
~ - ~ n_{i^{\prime}k^{\prime}} A_{i^{\prime}jk^{\prime}l} n_{jl} \right ] ~ + ~ \delta n_{i^{\prime}k^{\prime},f}~ - ~ \delta n_{i^{\prime}k^{\prime},gd}
\label{eq:deltan}
\end{equation}
\begin{equation}
\delta M_{i^{\prime}k^{\prime}} = \delta t ~ m_{i^{\prime}k^{\prime}} \left [ \epsilon_{ijkl} A_{ijkl} n_{ik} n_{jl} ~ - ~ n_{i^{\prime}k^{\prime}} A_{i^{\prime}jk^{\prime}l} n_{jl} \right ] ~ + ~ \delta M_{i^{\prime}k^{\prime},f} - ~ \delta M_{i^{\prime}k^{\prime},gd}
\label{eq:deltam}
\end{equation}
where $t$ is time,
$M_{i^{\prime}k^{\prime}}$ is the total mass in mass bin 
$k^{\prime}$ in annulus $i^{\prime}$, $A_{ijkl}$ is the cross-section,
$\epsilon_{ijkl} = 1/2$ for $i = j$ and $k = l$, and
$\epsilon_{ijkl} = 1$ for $k \ne l$ and any $i, j$.
The terms in these equations represent (i) mergers of $m_{ik}$ and $m_{jl}$
into a body of mass $m_{i^{\prime}k^{\prime}} = m_{ik} + m_{jl} - m_{e,ijkl}$,
(ii) loss of $m_{i^{\prime}k^{\prime}}$ through mergers with other objects,
(iii) addition of mass from debris produced by the collisions of other 
objects \citep{kl99}, and (iv) loss of mass by gas drag \citep{kl98}.  
In each equation, the second term includes the possibility that a 
collision can produce debris but no merger 
\citep[rebounds; see][and references therein]{dav85,kl99}.

The collision cross-section is
\begin{equation}
A_{ijkl} = \alpha_{coll}~\left ( \frac{1}{4~H_{ijkl}~\langle a_{ij} \rangle~ \langle \Delta a_{ij} \rangle} \right ) ~V_{ijkl}~F_{g,ijkl}~(r_{ik} + r_{jl})^2 ~ ,
\label{eq:Across}
\end{equation}
where
$\alpha_{coll}$ is a constant \citep{wet93, kl98},
$H_{ijkl}$ = $\sqrt{2 ~ (v_{ik}^2 + v_{jl}^2)} / \langle \Omega_{ij} \rangle$ 
is the mutual scale height, $\langle a_{ij} \rangle $ and 
$ \langle \Delta a_{ij} \rangle $ are the average heliocentric 
distance and width for the two interacting annuli,
$\langle Omega_{ij} \rangle$ is the average angular velocity,
$V_{ijkl}$ is the relative particle velocity,
$F_{g,ijkl}$ is the gravitational focusing factor, and
$r_{ik}$ and $r_{jl}$ are the particle radii.
We adopt the piecewise analytic approximation of \citet{spa91}
for the gravitational focusing factor in the dispersion regime 
and the collisional cross-sections of \citet{gre91} in the 
shear-dominated regime \citep[see also][1992]{grz90}. For more 
details of this algorithm, see \citet{kl98}, \citet{kb02a}, 
\citet{kb04a}, and \citet{bk06}.

To choose among possible collision outcomes, we use an energy-scaling 
algorithm.  If $Q_d^*$ is the collision energy needed to eject half 
the mass of a pair of colliding planetesimals and $Q_c$ is the center 
of mass collision energy, the mass of the ejecta is 
\begin{equation}
m_{e,ijkl} = 0.5 ~ (m_{ik} + m_{jl}) \left ( \frac{Q_c}{Q_d^*} \right)^{9/8} ~ , 
\label{eq:mej}
\end{equation}
where $m_{ik}$ and $m_{jl}$ are the masses of the colliding planetesimals. 
This approach allows us to derive ejected masses for catastrophic 
collisions with $Q_c \sim Q_d^*$ and cratering collisions with
$Q_c \ll Q_d^*$ \citep[see also][]{wet93,st97,kl99}. Consistent with 
N-body simulations of collision outcomes \citep[e.g.,][]{ben99}, we set
\begin{equation}
Q_d^* = Q_b r_{ijkl}^{\beta_b} + Q_g \rho_g r_{ijkl}^{\beta_g}
\label{eq:Qd}
\end{equation}
where $r_{ijkl}$ is the radius of a merged object with mass 
$m_{ik}$ + $m_{jl}$, $\rho_g$ is the mass density of a planetesimal,
$Q_b r^{\beta_b}$ is the bulk component of the binding energy, and
$Q_g \rho_g r^{\beta_g}$ is the gravity component of the binding energy.

\citet{kb05} and \citet{kbod08} describe how collisional evolution 
depends on various choices for $Q_d^*$. For icy objects, detailed
numerical collision simulations yield 
$Q_b \lesssim 10^7$ erg cm$^{-\beta_b}$ g$^{-1}$,
$-0.5 \lesssim \beta_b \lesssim$ 0, 
$\rho_g \approx$ 1--2 g cm$^{-3}$,
$Q_g \lesssim$ 1--2 erg cm$^{3-\beta_g}$ g$^{-2}$, and 
$\beta_g$ $\approx$ 1--2 \citep[e.g.,][]{ben99,lein08}.  
Calculations for the breakup of comet Shoemaker-Levy 9 
suggest a smaller component of the bulk strength, 
$Q_b r^{\beta_b} \sim 10^3$~erg~g$^{-1}$ 
\citep[e.g.,][]{asph96}, which yields smaller disruption 
energies for smaller objects.  Because nearly all models 
for collisional disruption yield similar results for objects
with $r \gtrsim$ 1~km \citep[e.g.,][]{kb04c,kbod08}, collisional 
evolution is relatively independent of these uncertainties as 
planetesimals grow into larger objects. Thus, we choose standard 
values -- $Q_g$ = 1.5 erg cm$^{1.75}$ g$^{-2}$, $\rho_g$ = 
1.5~g~cm$^{-3}$, and $\beta_g$ = 1.25 -- for the gravity component 
of $Q_d^*$. To check how the evolution of the small planetesimals 
depends on $Q_d^*$, we consider a broad range in the bulk component 
of the strength, $Q_b$ = 1--$10^5$ erg g$^{-1}$ with $\beta_b$ = 0 
\citep{pan05,kb04c,kb05,kbod08}.

To compute velocity evolution, we include collisional damping from 
inelastic collisions, gas drag, and gravitational interactions. Our
equations for the evolution of the velocity dispersion are
\begin{equation}
\frac{dh_{ik}^2}{dt} = 
\frac{dh_{in,ik}^2}{dt} + \frac{dh_{gd,ik}^2}{dt} + \frac{dh_{lr,ik}^2}{dt} +
\frac{dh_{sr,ik}^2}{dt} 
\label{eq:dhdt}
\end{equation}
for the horizontal component
and
\begin{equation}
\frac{dv_{ik}^2}{dt} = 
\frac{dv_{in,ik}^2}{dt} + \frac{dv_{gd,ik}^2}{dt} + \frac{dv_{sr,ik}^2}{dt}
\label{eq:dvdt}
\end{equation}
for the vertical component, where the subscripts refer to the
contributions from collisional damping (`in'), gas drag (`gd'), 
and long-range (`lr') and short-range (`sr') gravitational 
interactions.

For collisional damping, we adopt  
\begin{equation}
\frac{dh_{in,ik}^2}{dt} = \sum_{j=0}^{j=N} \sum_{l=0}^{l=l_{max}} \frac{C_{in}}{2}~(m_{jl} h_{jl}^2 - m_{ik} h_{ik}^2 - (m_{ik} + m_{jl}) h_{ik}^2)~I_e(\beta_{ijkl})
\label{eq:dhdtin}
\end{equation}
and
\begin{equation}
\frac{dv_{in,ik}^2}{dt} = \sum_{j=0}^{j=N} \sum_{l=0}^{l=l_{max}} \frac{C_{in}}{\beta_{ijkl}^2}~(m_{jl}
v_{jl}^2 - m_{ik} v_{ik}^2 - (m_{ik} + m_{jl}) v_{ik}^2)~I_i(\beta_{ijkl})
\label{eq:dvdtin}
\end{equation}
where $C_{in} = \alpha_{coll} ~ f_{ijkl} ~ \epsilon_{ijkl} ~ \rho_{jl} ~
V_{ijkl} ~ F_{g,ijkl} ~ (r_{ik} + r_{jl})^2$, 
$\beta_{ijkl}^2 = (i_{ik}^2 + i_{jl}^2)/(e_{ik}^2 + e_{jl}^2)$,
and $\rho_{jl}$ is the volume density of planetesimals with mass 
$m_{jl}$ in annulus $j$ \citep{oht92,wet93}.  
In the second summation, $l_{max} = k$ when $i = j$; $l_{max}$ = $M$ 
when $i \neq j$ \citep[see also][1999]{kl98}.  We add a term, 
$f_{ijkl}$, to treat the overlap between adjacent zones; $f_{ijkl}$ 
= 1 when $i = j$ and $f_{ijkl} \leq 1$ when $i \neq$ j \citep{kb01}.  
The integrals $I_e$ and $I_i$ are elliptic integrals described in 
previous publications \citep{wet93,ste00,oht02}.

For velocity damping from gas drag, we follow \citet{wet93} and write
\begin{equation}
\frac{dh_{gd,ik}}{dt} = -\beta_{ik}~ \frac{\pi C_D}{2m_{ik}} \rho_{gas} V_{gas}^2 r_{ik}^2 ,
\label{eq:dhdtgd}
\end{equation}
and
\begin{equation}
\frac{dv_{gd,ik}}{dt} = -(1 - \beta_{ik}) ~ \frac{\pi C_D}{2m_{ik}} \rho_{gas} V_{gas}^2 r_{ik}^2 ,
\label{eq:dvdtgd}
\end{equation}
where $C_D$ = 0.5 is the drag coefficient,
$\beta_{ik} = h_{ik} / (h_{ik}^2 + v_{ik}^2)^{1/2}$, 
$\rho_{gas}$ is the gas density,
$\eta$ is the relative gas velocity, and
$V_{gas} = (V_{ik} (V_{ik} + \eta))^{1/2}$ 
is the mean relative velocity of the gas
\citep[see][]{ada76,wei77b,wet93}.

For gravitational interactions, we compute long-range stirring from
distant oligarchs \citep{wei89} and short-range stirring from the
swarm \citep{oht02}. The long-range stirring only has a horizontal
component,
\begin{equation}
\frac{dh_{lr,ik}^2}{dt} = \sum_{j=1}^{j=N} \sum_{l=1}^{l=M} C_{lr,e} ~ x_{ijkl} ~ \frac{G^2 \rho_{jl} M_{jl}}{\langle \Omega_{ij} \rangle} \left ( \frac{{\rm tan^{-1}}(H_{ijkl}/D_{min})}{D_{min}} - \frac{{\rm tan^{-1}}(H_{ijkl}/D_{max})}{D_{max}} \right )
\label{eq:dhdtlr-cont}
\end{equation}

\noindent
for continuum objects and
\begin{equation}
\frac{dh_{lr,ik}^2}{dt} = \sum_{j=1}^{j=N} \sum_{l=1}^{l=M} \frac{G^2}{\pi \Omega a} \left ( \frac{C_{lr,e}^{\prime} m_{jl}^2}{(\delta a^2 + 0.5 H_{jl}^2)^2} \right)
\label{eq:dhdtlr-oli}
\end{equation}
for individual oligarchs,
where $x_{ijkl}$ is the fraction of objects with mass $m_{ik}$ in annulus $i$ 
that approach no closer than 2.4 $R_H$ of the objects with mass $m_l$ in 
annulus $j$, 
$D_{min} = {\rm max} (2.4 R_H, 1.6(h_{ik}^2 + h_{jl}^2)^{1/2})$,
$D_{max} = 0.5 ~ {\rm max} (w_{ik}, w_{jl})$,
$\delta a$ = $ | a_i - a_j | $, $C_{lr,e}$ = 23.5, and
$C_{lr,e}^{\prime}$ = 5.9 \citep[see also][]{kb01}.

For short-range gravitational interactions, the stirring depends on
the ratio of the relative collision velocity to the mutual Hill
velocity,
\begin{equation}
v_H \approx \langle \Omega_{ij} \rangle ~ \langle a_{ij} \rangle ~ [(m_{ik} + m_{jl})/3 M_{\star}]^{1/3} ~ .
\label{eq:vhill}
\end{equation}
In the high velocity regime, the collision 
velocity exceeds the Hill velocity.  Statistical solutions to the 
Fokker-Planck equation then yield accurate estimates for the stirring 
rates \citep[e.g.,][]{hor85,wet93,ste00,kb01}. At low velocities,
$n$-body calculations provide good estimates. We follow \citet{oht02}
and write the stirring as the sum of rates in the two regimes:
\begin{equation}
\frac{dh_{sr,ik}^2}{dt} = \frac{dh_{high,ik}^2}{dt} ~ + ~ \frac{dh_{low,ik}^2}{dt}
\label{eq:dhdtsr}
\end{equation}
and
\begin{equation}
\frac{dv_{sr,ik}^2}{dt} = \frac{dv_{high,ik}^2}{dt} ~ + ~ \frac{dv_{low,ik}^2}{dt} ~ ,
\label{eq:dvdtsr}
\end{equation}
where the subscripts `high' and `low' indicate the velocity regime
\citep[e.g., Eq. (25) of][]{oht02}. 

In the high velocity regime, the stirring is \citep[e.g.][]{ste00,kb01}
\begin{equation}
\frac{dh_{high,ik}^2}{dt} = \sum_{j=1}^{j=N} \sum_{l=1}^{l=M} f_{ijkl} ~ C_{high} ~ ( (h_{ik}^2 + h_{jl}^2)~m_{jl} ~ J_e(\beta_{ijkl}) +  1.4 ~ (m_{jl}h_{jl}^2 - m_{ik}h_{ik}^2)~H_e(\beta_{ijkl}) )
\label{eq:dhdthsr}
\end{equation}
and
\begin{equation}
\frac{dv_{high,ik}^2}{dt} = \sum_{j=1}^{j=N}  \sum_{l=1}^{l=M} f_{ijkl} ~ \frac{C_{high}}{\beta_{ijkl}^2}~((v_{ik}^2 + v_{jl}^2)~m_{jl}~J_z(\beta_{ijkl}) + 1.4 ~ (m_{jl}v_{jl}^2 - m_{ik}v_{ik}^2)~H_z(\beta_{ijkl})) ~ ,
\label{eq:dvdthsr}
\end{equation}
where
$f_{ijkl}$ the fraction of objects with mass $m_{ik}$ in annulus $i$ that approach
within 2.4 $R_H$ of the objects with mass $m_{jl}$ in annulus $j$ and $C_{high}$
 = 0.28 $A_{\Lambda} ~ G^2 ~ \rho_{jl} / ((h_{ik}^2 + h_{jl}^2)^{3/2})$. In the
expression for $C_{high}$, $A_{\Lambda}$ = ln $( \Lambda^2 + 1)$ and
\begin{equation}
\Lambda = \left ( \frac{0.19 ~ (h_{ik}^2 + h_{jl}^2 ~ + ~ 1.25 ~ (v_{ik}^2 + v_{jl}^2)) ~ (v_{ik}^2 + v_{jl}^2)^{1/2}}{v_H^3} \right )^2 ~ .
\label{eq:lambda}
\end{equation}
The functions $H_e$, $H_z$, $J_e$, and $J_z$ are definite integrals 
defined in Stewart \& Ida (2000).

In the low velocity regime, the evolution equations are \citep{oht02}:
\begin{equation}
\frac{dh_{low,ik}^2}{dt} = \sum_{j=1}^{j=N} \sum_{l=1}^{l=M} f_{ijkl} ~ C_{low} ~ ( m_{jl} ~ \chi_1 ~ + ~ (m_{jl} h_{jl}^2 ~ - ~ m_{ik} h_{ik}^2) ~ \chi_3 )
\label{eq:dhdtlsr}
\end{equation}
and
\begin{equation}
\frac{dv_{low,ik}^2}{dt} = \sum_{j=1}^{j=N}  \sum_{l=1}^{l=M} f_{ijkl} ~ C_{low} ~ ( m_{jl} ~ \chi_2 ~ + ~ (m_{jl} h_{jl}^2 - m_{ik} h_{ik}^2) ~ \chi_4 )
\label{eq:dvdtlsr}
\end{equation}
where the $\chi$'s are simple functions of the Hill radius 
\begin{equation}
r_{H,ijkl}  = a [(2 ~ (m_{ik} + m_{jl}))/3 M_{\star}]^{1/3} ~ .
\label{eq:rhill}
\end{equation}
and the normalized eccentricity and inclination 
\citep[][see also Ida 1990; Ida \& Makino 1992]{oht02}.
For the low velocity limit of the horizontal velocity
\begin{equation}
\chi_1 = 73 ~ C_1 ~ r_{H,ijkl}^4
\label{eq:chi1}
\end{equation}
and
\begin{equation}
\chi_2 = C_2 ~ (4 ~ \tilde{i}_{ij}^2 + 0.2 ~ \tilde{e}_{ij}^2 ~ (\tilde{e}_{ij}^2 \tilde{i}_{ij}^2)^{1/2}) ~ r_{H,ijkl}^4 ~ .
\label{eq:chi2}
\end{equation}
For the low velocity limit of the vertical velocity
\begin{equation}
\chi_3 = 10 ~ C_3 ~ \tilde{e}_{ij}^2 ~ r_{H,ijkl}^4 / (h_{ik}^2 + h_{jl}^2)
\label{eq:chi3}
\end{equation}
and
\begin{equation}
\chi_4 = 10 ~ C_3 ~ \tilde{i}_{ij}^2 ~ r_{H,ijkl}^4 / (h_{ik}^2 + h_{jl}^2) ~ .
\label{eq:chi4}
\end{equation}
Here, $C_{low} = 0.625 ~ \langle a_{ij} \rangle ~ \rho_{jl} ~ H_{ijkl} ~ V_{K,i}^3 / (m_{ik} + m_{jl})^2$,
$\tilde{e}_{ij}^2 = (e_{ik}^2 + e_{jl}^2) / r_{H,ijkl}^2$, and
$\tilde{i}_{ij}^2 = (i_{ik}^2 + i_{jl}^2) / r_{H,ijkl}^2$. 
The constants $C_1$, $C_2$, and $C_3$ are identical to 
those in Eq. (26) of \citet{oht02}.

Several tests indicate that these expressions provide an accurate
treatment of velocity evolution for planetesimals in the high and 
low velocity regimes.
Figs. 5--7 of \citet{oht02} show comparisons with results from $n$-body
simulations \citep[see also][]{ida90,ida92}. Our simulations confirm 
this analysis.  \citet{wei97} and \citet{kb01} compare variants of 
this formalism with other $n$-body calculations. \citet{gol04} 
demonstrate that our numerical results agree with analytic estimates. 

To follow the evolution of the most massive objects more accurately, our 
code includes an $n$-body algorithm. When objects have masses exceeding
a `promotion mass' $m_{pro}$, we promote them into an $n$-body code that 
directly solves their orbits. The $n$-body code incorporates algorithms
to allow collisions among $n$-bodies and interactions between $n$-bodies 
and coagulation particles. \citet{bk06} describe this code in detail.
Because dynamical interactions among large oligarchs are rare and occur
at late stages in the evolution, we set $m_{pro} = 10^{26}$ g. We 
describe several test calculations with smaller $m_{pro}$ in \S3.3.

To treat the time evolution of the gas volume density $\rho_{gas}$, we 
use a simple nebular model with gas surface density 
$\Sigma_{gas}(a,t) = \Sigma_{gas,0} a^{-3/2} e^{-t/t_{gas}}$,
gas-to-solids ratio $\Sigma_{gas,0}(a) / \Sigma (a)$ = 100 -- where
$\Sigma$ is the surface density of solids, and scale height 
$H_{gas}(a) = H_{gas,0} (a/a_0)^{1.125}$ \citep{kh87}.  To approximate gas 
removal on a timescale $t_{gas}$, the gas density declines exponentially 
with time.  We set $t_{gas}$ = 10~Myr. During the early stages of 
calculations at 30--150~AU, velocity damping is important for particles
with $r \lesssim$ 100 m. However, particle losses from gas drag are small, 
$\sim$ 1\% or less of the initial mass. By the time viscous stirring
dominates the velocity evolution of small objects, the gas disk has 
dispersed.  Inward drift and velocity damping are then negligible 
\citep[see also][]{wet93}.

The initial conditions for these calculations are appropriate for a disk 
with an age of $\lesssim$ 1--2~Myr \citep[e.g.][]{dul05,nom06,cie07,gar07}.  
We consider systems of $N$ annuli in disks with $a_i$ = 30--150~AU 
and $\delta a_i/a_i$ = 0.025.  The disk is composed of small 
planetesimals with radii of $\sim$ 1--1000 m and an initial mass 
distribution $n_i(m_{ik})$ in each annulus. The mass ratio between 
adjacent bins is $\delta = m_{ik+1}/m_{ik}$ = 1.4--2. At the start 
of the calculations, each bin has the same total mass, eccentricity 
$e_0 = 1-3~\times~10^{-4}$, and inclination $i_0 = e_0/2$.  We assume 
a power law variation of the initial surface density of solid material 
with semimajor axis, 
\begin{equation}
\Sigma_i = \Sigma_0(M_{\star}) ~ x_m ~ (a_i/a_0)^{-3/2} ~ , 
\label{eq:sigma}
\end{equation}
where $x_m$ is a scaling factor.  For a 1 $M_{\odot}$ central star, 
models with $\Sigma_0 \approx$ 0.1--0.2 g cm$^{-2}$ at $a_0$ = 30~AU 
have a mass in icy solids comparable to the minimum mass solar nebula 
\citep[MMSN hereafter;][]{wei77a,hay81}.  Consistent with observations of 
disks surrounding pre-main sequence stars \citep[e.g.,][]{nat2000,sch2006}, 
we scale the reference surface density with the stellar mass, 
$\Sigma_0 (M_{\star})$ = 0.18 $(M_{\star} / M_{\odot}$) g cm$^{-2}$.

Table 1 lists the ranges in $M_{\star}$ and $x_m$ we consider. The table
also lists the main sequence lifetime, $t_{ms}$, defined as the time to
reach core hydrogen exhaustion in the $X$ = 0.71, $Y$ = 0.27, and $Z$ =
0.02 stellar evolution models of \citet{dem04}, where $X$ is the initial
mass fraction of hydrogen, $Y$ is the mass fraction of helium, and $Z$
is the metallicity.  For most of our calculations, 
the number of annuli in the disk is $N$ = 64. To check these results, 
we also calculated models for disks with $N$ = 32 around 1 $M_{\odot}$ 
stars. Because the growth of planets has large stochastic variations, 
we repeated the calculations 5--12 times for each set of starting 
conditions, $M_{\star}$, $N$, $x_m$, and $Q_b$.  Table 1 lists 
the number of calculations for each ($M_{\star}$, $x_m$) pair.

Our calculations follow the time evolution of the mass and velocity 
distributions of objects with a range of radii, $r_{ik} = r_{min}$
to $r_{ik} = r_{max}$.  The upper limit $r_{max}$ is always larger
than the largest object in each annulus.  To save computer time in
our main calculation, we do not consider small objects which do not 
affect significantly the dynamics and growth of larger objects, 
$r_{min}$ = 100 cm.  
Erosive collisions produce objects with $r_{ik}$ $< r_{min}$ which
are `lost' to the model grid. Lost objects are more likely to be 
ground down into smaller objects than to collide with larger objects 
in the grid \citep[see][2004a]{kb02a}.

To estimate the amount of dusty debris produced by planet formation,
we perform a second calculation. Each main calculation yields
$\dot{M}_i (t)$, the amount of mass lost to the grid per annulus 
per timestep, and $H_{i0}(t)$, the scale height of the smallest
particle ($r = r_{min}$) in each annulus of the coagulation grid.  
Objects with sizes smaller than $r_{min}$ contain a small fraction 
of the mass in each annulus; thus, the scale height for objects with 
$r < r_{min}$ is $H_{i0}(t)$ \citep{gol04}.  The total amount of mass 
lost from the planetesimal grid in each timestep is 
$\dot{M}(t) = \sum_{i=1}^{N} \dot{M}_i (t)$.  The debris has a known size 
distribution, $n^{\prime}_{ij} = n^{\prime}_{i0} r_i^{-\beta}$, where
$\beta$ is a constant \citep[see][and references therein]{st97,kl99}.  
The normalization constant $n_{i0}^{\prime}$
depends only on $\beta$ and $\dot{M} (t)$, which we derive at each 
timestep in the main calculation.  To evolve the dust distribution 
in time, we use a simple collision algorithm that includes 
Poynting-Robertson drag and radiation pressure\footnote{Because the 
collisional cascade begins after the gas disk dissipates, we ignore 
gas drag.}.  The optical depth $\tau$ of the dust follows from 
integrals over the size distribution in each annulus. The optical 
depth and a radiative transfer solution then yield the luminosity 
and radial surface brightness of the dust as a function of time.  
\citet{kb04a} describe this calculation in more detail.

Throughout the text, we use simple scaling relations to show how our
results depend on initial conditions and the properties of the grid. 
For each set of calculations (Table 1), we derive median results for
the size distribution, the size of the largest object as a function of 
$a$ and $t$, and other physical variables. Substituting the inter-quartile
range for the dispersion, we then perform least-squares fits to relate 
these median results to input parameters (e.g., $x_m$) and the properties 
of the grid (e.g., $a$). For parameters where analytic theory predicts 
a relation (e.g., the growth time as a function of $a$), we derive the 
best-fitting coefficient, test whether different fitting functions provide 
better fits to our results, and keep the result that minimizes $\chi^2$
per degree of freedom. When analytic theory provides no guidance, we 
derive fitting functions that yield the sensitivity of our results to 
all important physical variables. Thus, our fits test some aspects of 
analytic theory and guide other aspects.

\section{PLANET FORMATION CALCULATIONS}

\subsection{Icy Planet Formation in Disks around 1 $M_{\odot}$ Stars}

We begin with a discussion of planet formation in disks at 30--150~AU 
around a 1 $M_{\odot}$ star. For most disks around low mass stars, the 
timescale for planet formation is shorter than the main sequence lifetime. 
Thus, the growth of planetesimals into planets and the outcome of the 
collisional cascade depend more on the physics of solid objects than on 
stellar physics.  Here, we review the stages in planet growth and describe 
the outcome of the collisional cascade. For a suite of calculations of 
planet formation in disks of different masses, we derive basic relations 
for the growth time and the maximum planet mass as a function of initial 
disk mass. We also show how the dust production rate and the mass in small 
objects depend on initial disk mass and time.

The next section compares these results with calculations for 
1.5--3~$M_{\odot}$ stars. For disks around more massive stars, the
planet formation timescale is comparable to the main sequence lifetime.
Thus, the central star evolves off the main sequence before planet
formation and the collisional cascade reach a natural end-state. 
During post-main sequence evolution, the star brightens considerably 
\citep[e.g.,][]{dem04} and develops a powerful stellar wind 
\citep[e.g.,][and references therein]{kna85},
melting icy objects in the inner disk and ejecting small grains
throughout the disk. Thus, we halt our calculations when the star
evolves off the main sequence. Here, we show how the physics of
main sequence stars changes the results derived for planet formation 
around 1~\msun\ stars.

\subsubsection{Growth of Large Objects}

The formation of icy planets in the outer regions of a quiescent 
planetesimal disk has three distinct stages \citep{kl99,kb04a}.
Planetesimals with $r \lesssim$ 1~km grow slowly. As they grow, 
dynamical friction damps $e$ for the largest objects; dynamical
friction and viscous stirring raise $e$ for the smallest objects 
\citep[e.g.,][]{gre84,wet93,gol04}. After $\sim$ 0.1--1 Myr, 
gravitational focusing enhances the collision cross-sections by 
factors of 10--100. Slow, orderly growth ends. Runaway growth begins. 
At the inner edge of the disk, the largest objects take $\sim$ 3~Myr 
to grow to $\sim$ 100~km and $\sim$ 30~Myr to grow to $\sim$ 1000~km. 
Throughout runaway growth, the gas disk dissipates. Thus, velocity
damping by gas drag ceases; dynamical friction and viscous stirring 
increase $e$ for the smallest objects. Stirring reduces gravitational 
focusing factors, slowing the growth of the largest objects relative to 
one another and relative to leftover planetesimals \citep{ida93,wet93}.  
Runaway growth ends; oligarchic growth begins \citep{kok98,raf03,cha06,naga07}. 
After 30--100 Myr, the largest objects -- oligarchs -- grow slowly 
and contain an ever increasing fraction of the remaining mass in the 
disk.

During the transition from runaway to oligarchic growth, collisions
start to produce copious amount of dust. Once oligarchs reach sizes 
$\sim$ 500~km, collisions between 1--10~km objects produce debris 
instead of mergers \citep[][and references therein]{kbod08}. Once 
fragmentation begins, continued stirring leads to a collisional 
cascade, where leftover planetesimals are ground to dust. For dust 
grains with sizes $\gtrsim$ 10 $\mu$m,
the collision time is much shorter than the time to remove particles 
by gas drag \citep{ada76} or by Poynting-Robertson drag \citep{bur79}. 
Thus, the cascade proceeds to particle sizes $\lesssim$ 1--10 $\mu$m, 
where radiation pressure removes material on the dynamical time scale 
\citep{bur79}.  Because runaway growth leaves most of the mass in 1--10~km 
objects, the collisional cascade effectively removes a significant 
fraction of the solid material in the disk.

\begin{figure}
\epsscale{1.1}
\plottwo{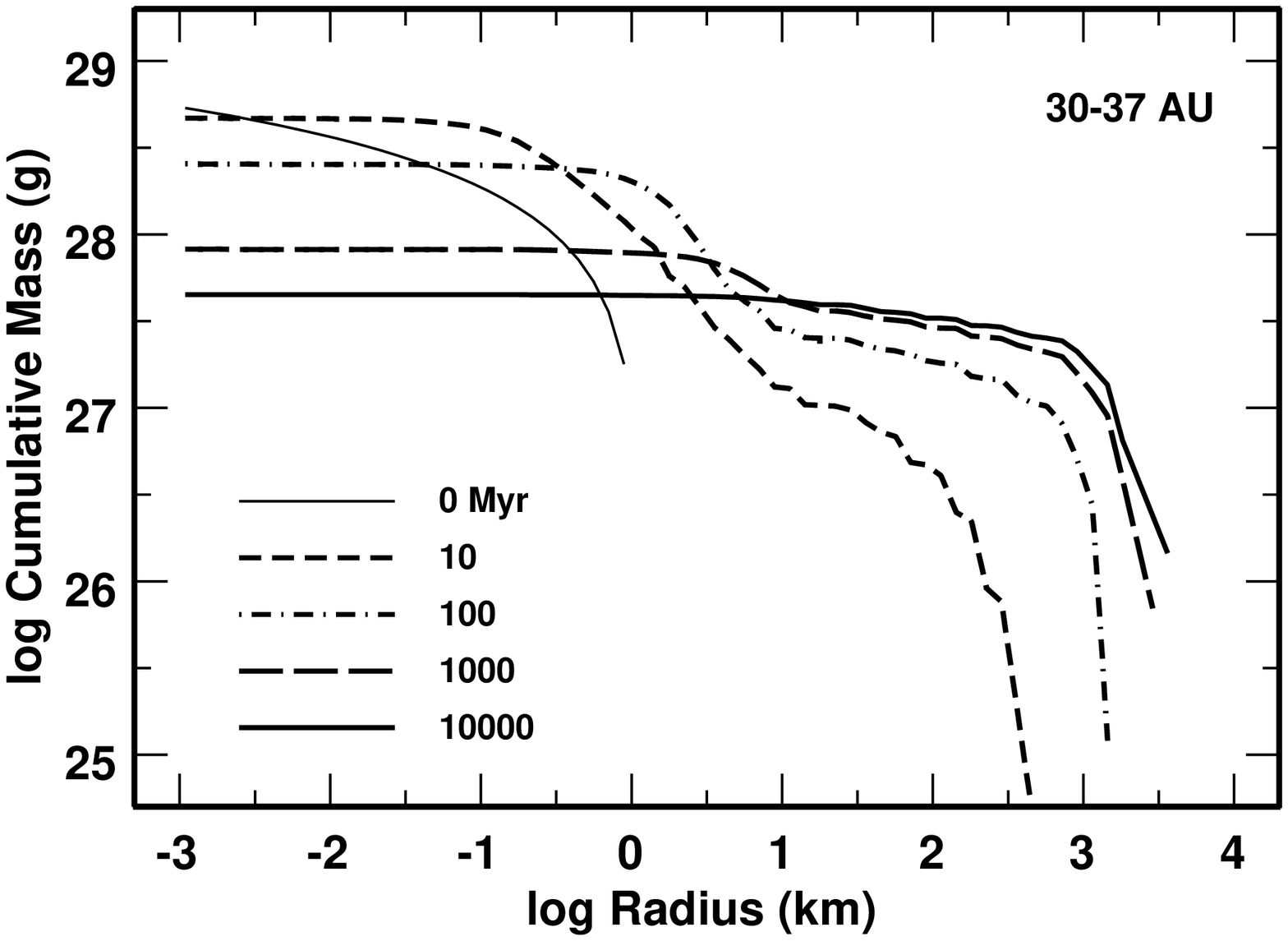}{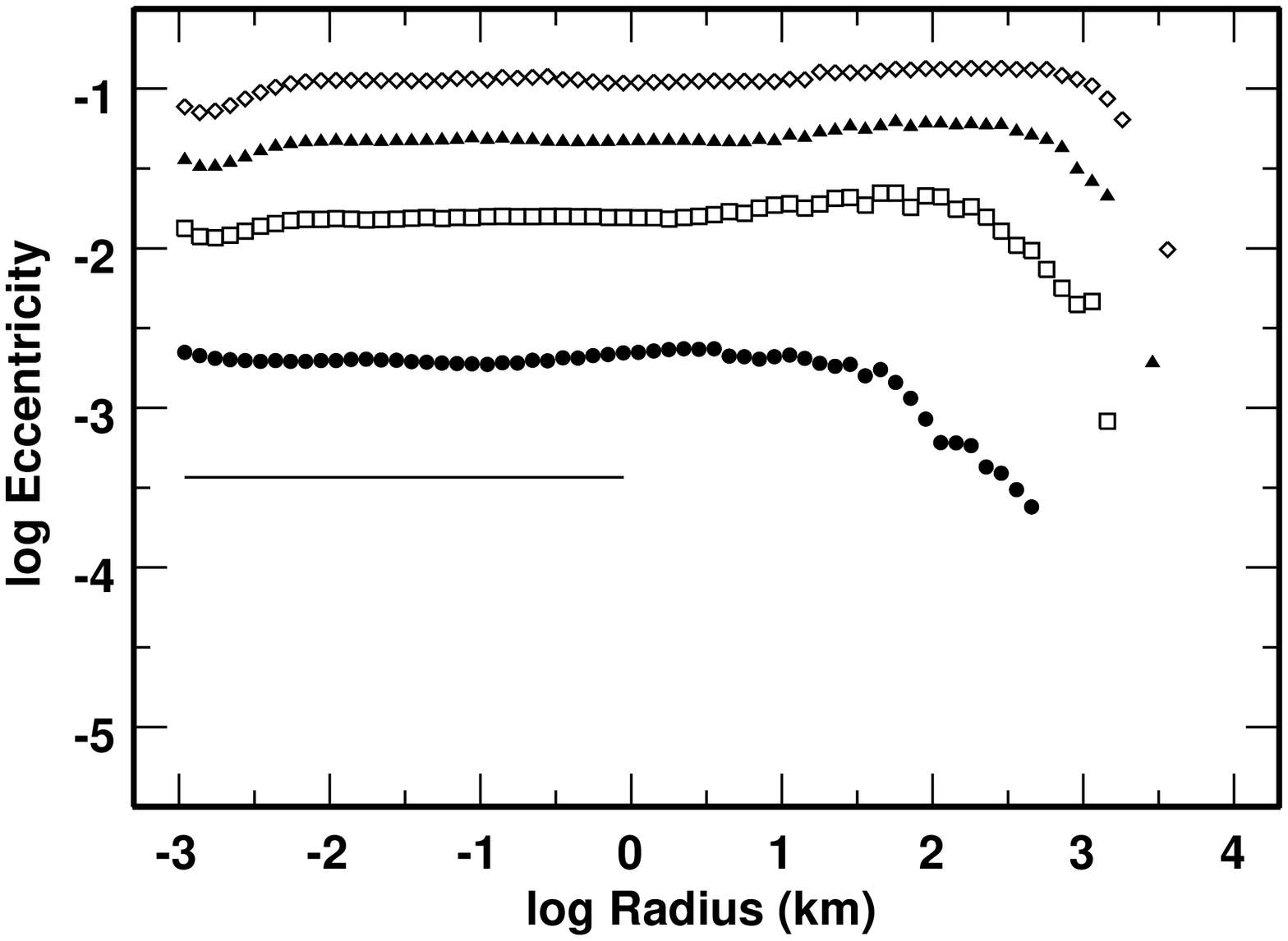}
\figcaption[f1.eps]
{
Evolution of a multiannulus coagulation model with
$\Sigma = 0.18 (a_i/{\rm 30~AU})^{-3/2}$~g~cm$^{-2}$
at 30--37~AU around a 1 \msun\ star.  
{\it Left}: median cumulative mass distribution at times
indicated in the legend.
{\it Right}: median eccentricity distributions at
$t$ = 0 (light solid line),
$t$ = 10~Myr (filled circles),
$t$ = 100~Myr (open boxes),
$t$ = 1~Gyr (filled triangles), and
$t$ = 10~Gyr (open diamonds).
As large objects grow in the disk, they stir up
the leftover planetesimals to $e \sim$ 0.1.
Disruptive collisions then deplete the population
of 0.1--10~km planetesimals, which limits the growth
of the largest objects.
\label{fig:sd1}
}
\end{figure}

Fig. \ref{fig:sd1} shows the time evolution of the eccentricity and the 
mass distributions for objects in the innermost 8 annuli of a disk with 
initial mass distribution similar to the MMSN. To minimize stochastic 
variations, these plots show median results for 15 calculations. 
During slow growth and the early stages of runaway growth, dynamical 
friction damps $e$ for the largest objects and raises $e$ for the
smallest objects (Fig. \ref{fig:sd1}; right panel, 10 Myr). The 
mass distribution develops a pronounced shoulder from 10~km to 
$\sim$ 300--500~km. As the evolution proceeds, growth concentrates
more mass in the largest objects; stirring excites the orbits of the 
smallest objects. After 100 Myr, the collisional cascade removes mass
efficiently from the smallest objects but leaves the mass distribution 
of the largest objects unchanged. By $\sim$ 5--10 Gyr, almost all of
the small objects have been removed.

\begin{figure}
\includegraphics[width=6.5in]{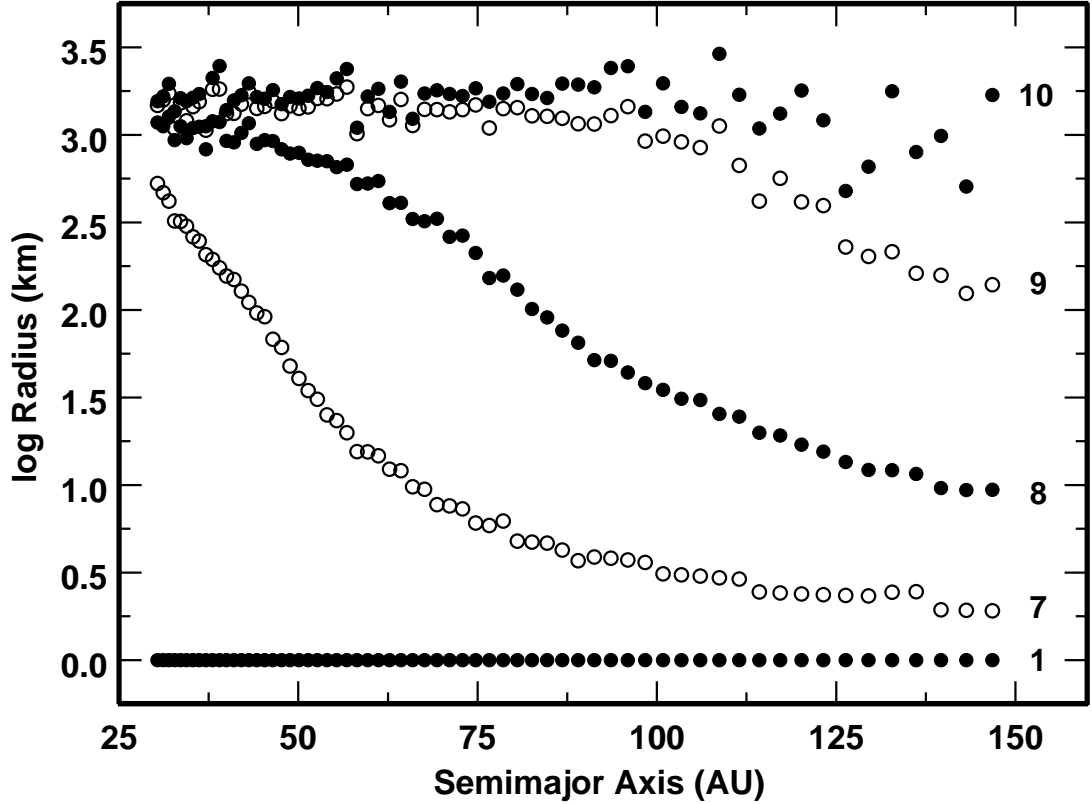}
\figcaption[f2.eps]
{Evolution of the radius of the largest object in each annulus
for a MMSN disk around a 1 $M_{\odot}$ star. The number to the
right of each set of points indicates log $t$ in yr from the
start of the calculation. Large objects with $r \sim$ 1000~km
form at the inner edge of the disk in 10--100~Myr, in the middle
of the disk in 0.3--1~Gyr, and at the edge of the disk in 10~Gyr.
\label{fig:radevol1}}
\end{figure}

In these calculations,
the rate of planet formation is very sensitive to semimajor axis
(Fig. \ref{fig:radevol1}).  For collisional processes, the growth time 
in the disk is $t \propto P / \Sigma$, where $P$ is the orbital period
\citep[see the Appendix; also][]{lis87,kl98,gol04}. For $P \propto a^{3/2}$ and 
$\Sigma \propto x_m a^{-3/2}$ (Eq. (\ref{eq:sigma})), the growth 
time is $t \propto a^3 x_m^{-1}$.  Thus, although 
it takes only $\sim$ 10 Myr for the largest planets to reach radii of 
300--600~km at 30~AU, the largest objects at $a \gtrsim$ 100~AU still 
have $r \sim$ 3--5~km. By 100 Myr, 100~km objects form at 75--80~AU. 
After 1 Gyr, 100~km objects form beyond 125~AU. By the end of the 
calculation at 10 Gyr, 1000~km objects form throughout the disk.
 
\begin{figure}
\includegraphics[width=6.5in]{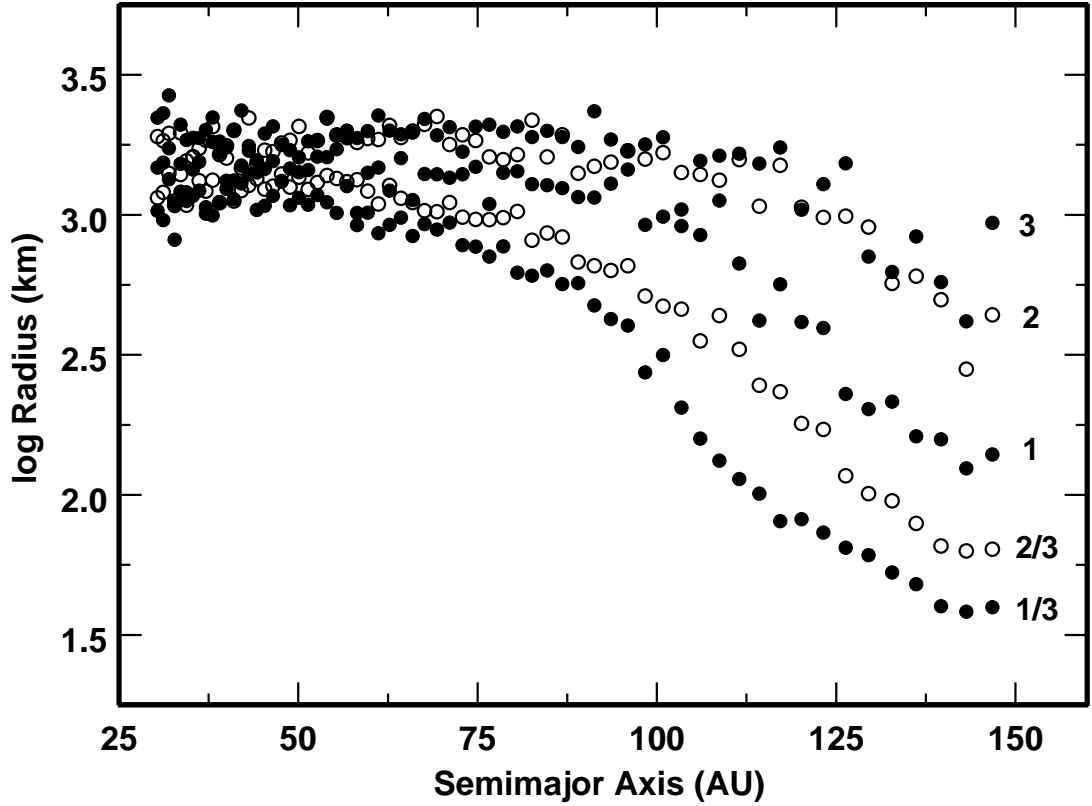}
\figcaption[f3.eps]
{Median radii of the largest objects at 1~Gyr for disks around a 
1~$M_{\odot}$ star. The numbers to the right of each set of points 
indicate $x_m$, the disk mass in units of the MMSN.  Planets form 
earlier in the inner portions of the most massive disks.  Icy planet 
formation saturates at maximum radii $r \sim$ 1500 km.
\label{fig:radevol2}}
\end{figure}

The formation rate also depends on the initial disk mass (Fig. 
\ref{fig:radevol2}). For an expected growth time $t \propto a^3 x_m^{-1}$,
planets grow faster in more massive disks. At 100~AU, planets with
$r \sim$ 2000~km form in a massive disk ($x_m$ = 3) within 1 Gyr.
In a low mass disk with $x_m \sim$ 1/3, the largest planet at 100~AU
grows to $r \sim$ 300~km in 1 Gyr and $r \sim$ 2000~km in 10 Gyr. 
For all simulations of disks around 1 $M_{\odot}$ stars, the median 
timescale for the formation of the first 1000~km object is
\begin{equation}
t_{1000} \approx 475 ~ x_m^{-1.15} ~ \left ( \frac{a}{\rm 80~AU} \right )^3 ~ {\rm Myr} ~ .
\label{eq:t1000}
\end{equation}
This relation fits our results for the median growth time to $\approx$ 
5\% for $a$ = 30--150~AU and for $x_m$ = 1/3 to 3. For each initial disk 
mass, the inter-quartile range for the formation time is $\sim$ 20\%. 
Thus, there is a modest range of outcomes for identical starting conditions.

In equation \ref{eq:t1000}, there is a small difference between the result 
expected from simple theory ($t \propto x_m^{-1}$) and the result derived 
in our calculations ($t \propto x_m^{-1.15}$). We show in the Appendix how 
gas drag during runaway and oligarchic growth can modify the simple theory 
and explain the result of our calculations.

Although the timescale to produce the first 1000~km object is a strong 
function of initial disk mass and semimajor axis, the evolution at late 
times is less sensitive to the starting conditions. To derive a simple 
relation for the median size $r_{max}$ of the largest object as a function 
of initial disk mass and semimajor axis, we adopt a simple function
\begin{equation}
r_{max}(a) = r_0 ~ e^{-(a_i / a_0)^{\alpha_r}}
\label{eq:rmax}
\end{equation}
and use an amoeba algorithm \citep{pre92} to derive the fitting parameters 
$a_0$, $r_0$, and $\alpha_r$ as a function of time.  For stellar ages 
1 Gyr $\lesssim t_{\star} \lesssim$ 10 Gyr, the ensemble of calculations 
yields
\begin{equation}
r_0 \approx 1650 ~ x_m^{0.2} \left ( \frac{t}{\rm 3~Gyr} \right )^{0.06} ~ {\rm~km}
\label{eq:r0}
\end{equation}
for the radius of the largest object,
\begin{equation}
a_0 \approx 190 ~ x_m^{0.1} \left ( \frac{t}{\rm 3~Gyr} \right )^{0.1} ~ {\rm~AU}
\label{eq:a0}
\end{equation}
for the scale length, and $\alpha_r \approx$ 5--6 for the exponent. These
relations match our results to $\pm$5\%. The uncertainties are $\pm$3\% 
in the coefficients and $\pm$5\% in the exponents.

These calculations produce relatively low mass icy planets with radii 
$\sim$ 50\% larger than the radius of Pluto \citep[][2007]{you94,ell03}. 
Although these objects form relatively rapidly in the inner disk, they 
grow very slowly at late times. Between 1--10~Gyr, most large object grow 
by $\sim$ 10--20\% in radius ($\sim$ 50\% in mass). Because the size 
of the largest object depends weakly on the initial disk mass, nearly 
all disks form Pluto-mass planets.  These objects stir leftover 
planetesimals effectively; thus, nearly all disks develop a collisional
cascade.

Despite the general formation of Pluto-mass planets in any disk, the lowest 
mass disks ($x_m \lesssim$ 1/3) form Plutos inefficiently.  The scale length, 
$a_0 \gtrsim$ 150~AU, derived from our calculations exceeds the outer 
radius of the disk. Thus, planet formation does not proceed to completion 
at large $a$ for the lowest mass disks. In these systems, the largest icy 
planets at $a \approx$ 125--150~AU are factors of 3--10 smaller than $r_0$.
The large exponent, $\alpha_r \sim$ 5--6, derived in our fits implies a 
rapid transition -- $\sim$ 10--20~AU -- between disk regions where
the largest objects are planets with $r_{max}$ $\approx r_0$ and where
the largest objects have $r_{max} \lesssim$ 300--500~km.

In our calculations, the collisional cascade limits the size of the largest 
objects. Once a few objects have radii $\gtrsim$ 1000~km, they stir up leftover 
planetesimals to the disruption velocity. When the collisional cascade 
begins, the timescale for 1~km planetesimals to collide and fragment 
into smaller objects is shorter than the timescale for oligarchs to 
accrete leftover planetesimals. Thus, the growth of the largest objects 
stalls at $\sim$ 1000--2000~km ($\sim$ 0.01--0.02 $M_{\oplus}$).
Occasionally, runaway and oligarchic growth produce a very large object 
with $r \sim$ 5000~km ($\sim$ 0.1 $M_{\oplus}$), but these objects form 
in only $\sim$ 5--10\% of the simulations.  These objects form at random
semimajor axes and tend to form in more massive disks. 

\begin{figure}
\epsscale{1.1}
\plottwo{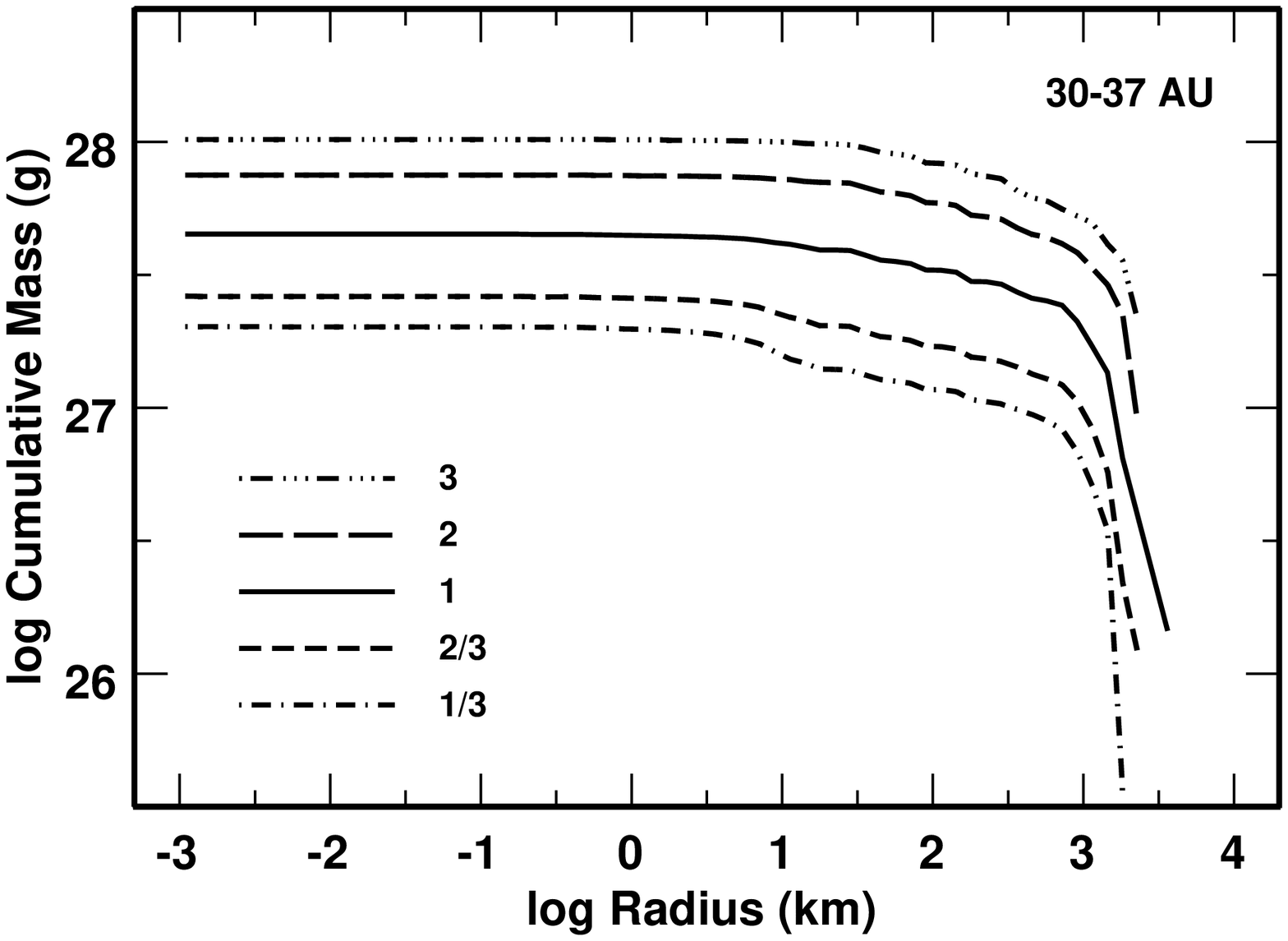}{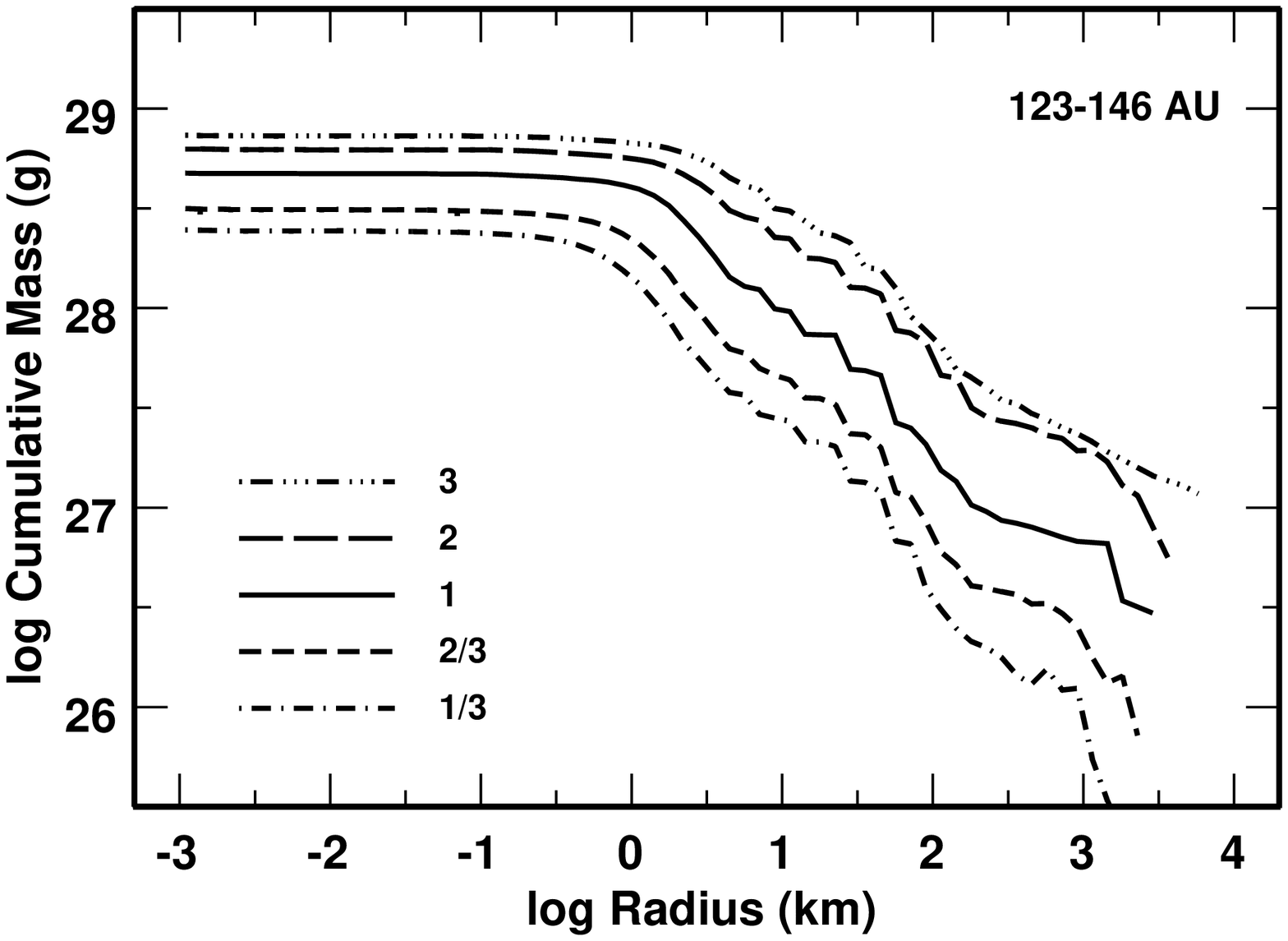}
\figcaption[f4.eps]
{Median cumulative mass distributions at 1 Gyr for annuli at
30--37~AU (left panel) and at 123--146~AU (right panel) around a 
1 \msun\ star. The legend indicates $x_m$, the initial disk mass in 
units of the scaled MMSN. In the inner disk, many large planets
form and the collisional cascade removes nearly all of the 
material in objects with $r \lesssim$ 1--10~km. In the outer disk, 
few large planets form; collisions are inefficient at removing 
material in small objects.
\label{fig:sd2}
}
\end{figure}

The large radial variation in the formation time produces dramatic differences
in the mass distribution as a function of semimajor axis (Fig. \ref{fig:sd2}). 
In the inner disk, rapid growth leads to many objects with $r \gtrsim$ 
1000~km (Table \ref{tab:rad1000.1}). With many large objects stirring 
the leftover planetesimals in the inner disk, the collisional cascade 
removes most of the mass in small objects (Fig. \ref{fig:sd2}, left panel). 
In the outer disk, slow growth results in a handful of Pluto-mass 
objects. A few large objects cannot stir leftover planetesimals 
efficiently. Thus, the collisional cascade is weak and leaves a 
substantial amount of mass in 1--10~km planetesimals 
(Fig. \ref{fig:sd2}, right panel).

\begin{figure}
\includegraphics[width=6.5in]{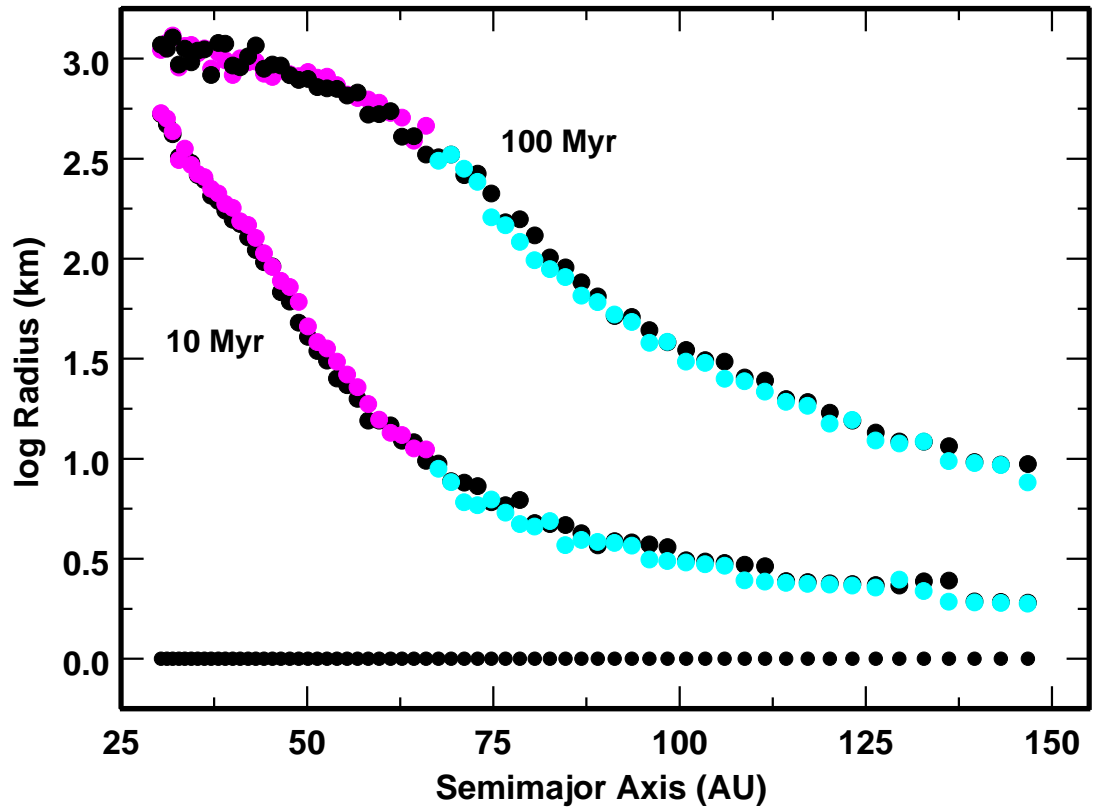}
\figcaption[f5.eps]
{Radius of the largest object in each annulus at 10~Myr and at 100~Myr
for a MMSN disk around a 1 $M_{\odot}$ star. The black points indicate
results for calculations with 64 annuli; the magenta and cyan points show 
results for calculations with 32 annuli. The good agreement between the
32 annulus and 64 annulus calculations shows that planet formation is
not sensitive to the size of the grid.
\label{fig:radevol3}}
\end{figure}

The growth of objects as a function of semimajor axis and time is not 
sensitive to the size of the model grid (Fig. \ref{fig:radevol3}). 
For two sets of calculations with 32 annuli (cyan and magenta points), 
the median radius of the largest object in each annulus is nearly 
identical to results for calculations with 64 annuli (black points).  
The results for Eqs. 
(\ref{eq:t1000}--\ref{eq:frac100}) are also independent of the grid.
In principle, long-range stirring from planets at small $a$ can
influence runaway growth of objects at large $a$. Our results suggest
that icy planet formation at large semimajor axes is not influenced by 
the formation of small icy planets at small semimajor axes.  

To conclude this discussion of the formation of large objects in a 
planetesimal disk, we quote several simple relations for the amount 
of solid material in small and large objects as a function of initial 
disk mass and semimajor axis for the ensemble of calculations around
a 1 \msun\ star. At 3--10 Gyr, the median fraction of solids remaining 
in the disk is
\begin{equation}
f_s \approx 0.3 \left ( \frac{a}{\rm 100~AU} \right ) \left ( \frac{1}{x_m} \right )^{1/4} ~ .
\label{eq:frac10gy}
\end{equation}
For a MMSN with $x_m$ = 1, the amount of mass remaining in the disk
at 3--10~Gyr ranges from 9\% of the initial mass at 30~AU to roughly 50\% 
of the initial mass at 150~AU. Thus, the inner disk is substantially
depleted, while the outer disk contains a significant fraction of its
initial mass.

For each $x_m$, the median fraction of the initial disk mass in 1000~km 
and larger objects is
\begin{equation}
\label{eq:frac1000}
f_{1000} = 0.035 \left ( \frac{\rm 30~AU}{a} \right ) ~.
\end{equation}
The median fraction of the mass in 100~km and larger objects is roughly
50\% larger,
\begin{equation}
\label{eq:frac100}
f_{100} = 0.06 \left ( \frac{\rm 30~AU}{a} \right ) ~.
\end{equation}
For the ensemble of calculations, the typical inter-quartile range is 
$\sim$ 0.1 for $f_s$ and $\sim$ 20\% for $f_{100}$ and $f_{1000}$.
Thus, the mass distributions in our calculations are top-heavy, with
more mass in 1000+~km objects than in 100--1000~km objects. 

These relations demonstrate that planet formation at 30--150~AU is very 
inefficient.  For all disk radii in this range, only $\sim$ 6\% or less 
of the initial population of 1~km objects is incorporated into large 
objects with radii exceeding 100~km. In the inner disk (30--50~AU), the 
collisional cascade is very efficient at removing leftover 1--10~km
objects. Thus, at 3--10~Gyr, the large objects contain most of the mass
in the inner disk.  In the outer disk (100--150~AU), the collisional
cascade does not have enough time to remove leftover planetesimals.
Thus, small objects with radii of 1--10~km contain most of the remaining 
mass at 100--150~AU. 

\subsubsection{Evolution of Dust}

At all semimajor axes, the collisional cascade converts a large fraction 
of the initial mass in solids into small dust grains. Because oligarchs
and leftover planetesimals are unobservable with current techniques,
dust emission provides the sole observational diagnostic of the growth
of icy planets at 30--150 AU around other stars. Here, we describe the 
evolution of these small particles and demonstrate that the collisional 
cascade is observable.

Two physical processes set the visibility of dust grains in a debris 
disk. Once significant fragmentation begins, collisions gradually grind 
the fragments to dust. When dust grains are small enough, radiative 
processes remove them from the disk.  For disks at 30--150~AU, radiation 
pressure dominates mass loss for $t \lesssim$ 1--3 Gyr and removes 65\%
to 70\% of the total mass loss.  Poynting-Robertson drag removes material 
at late times and is responsible for 30\% to 35\% of the total mass loss. 
Because the gas density is negligible once the collisional cascade begins, 
gas drag is unimportant.

To describe our results, we divide dusty debris into 
`large grains' with 1 mm $\lesssim r \lesssim$ 1 m, 
`small grains' with 1 $\mu$m $\lesssim r \lesssim$ 1 mm, 
and `very small grains' with $r \lesssim$ 1 $\mu$m.
Collisions dominate the evolution of large grains at all times.
For $t \lesssim$ 1--3 Gyr, collisions dominate the
evolution of small grains; Poynting-Robertson drag
then removes grains with radii of 1--100 $\mu$m on Gyr
timescales.  Radiation pressure removes very small
grains on the local dynamical timescale. Thus, 
radiation pressure produces a `wind' of very small grains 
in the disk midplane. 

\begin{figure}
\includegraphics[width=7.0in]{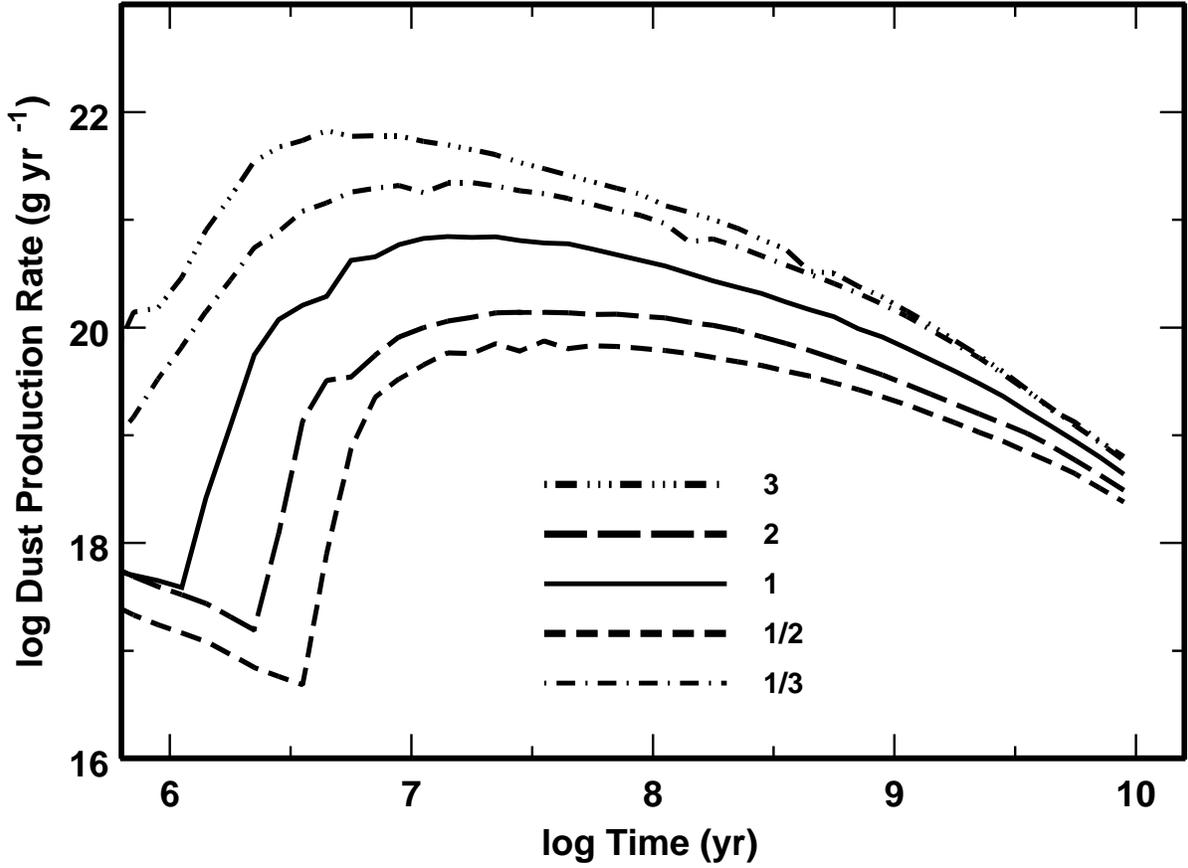}
\figcaption[f6.eps]
{Median production rate of 0.01--1 $\mu$m objects as a function of time 
for 30--150~AU disks around a 1 \msun\ central star.  The legend indicates 
$x_m$, the initial disk mass in units of the scaled MMSN.  As large 
objects grow during the early stages of the evolution, the dust 
production rate declines. Once large objects start to stir up
the leftover planetesimals, debris production rises rapidly. After
dust production peaks, the collision rate and dust production decline
slowly with time. For all stars, more massive disks eject more 
material into a wind of small particles.
\label{fig:dust1}}
\end{figure}

Fig. \ref{fig:dust1} shows the time evolution of the dust production rate 
for very small grains as a function of initial disk mass. At the start
of each calculation, dynamical friction and collisions damp orbital 
eccentricities. Thus, collisions produce less and less debris; the 
dust production rate declines with time. As oligarchs reach radii of 
$\sim$ 500~km, they stir leftover planetesimals along their orbits. 
Dust production increases. Because oligarchs continue to grow, they
stir leftover planetesimals to larger and larger velocities. Collision
energies rapidly exceed the critical disruption energy; the dust production 
rate increases dramatically (Eq. \ref{eq:mej}--\ref{eq:Qd}).  When oligarchs 
start to reach their maximum radii in the inner disk (Eq. (\ref{eq:rmax})), 
the dust production rate peaks. As oligarchs grow at larger and larger 
disk radii, the dust production rate slowly declines. 

Although the outer disk contains more mass, the global dust production rate 
declines with time for two reasons. Large oligarchs form at late times 
in the outer disk (Fig. \ref{fig:radevol1}), but the smaller disk surface 
density and the longer orbital periods lead to smaller collision rates. 
Smaller collision rates yield smaller dust production rates.  Initially, 
collisions dominate Poynting-Robertson drag; thus, radiation pressure 
ejects the smallest grains on the local orbital timescale
\citep[e.g.,][]{kri00,wya05}. The dust production rate then declines 
roughly as $t^{-1}$. At late times, the collision rates decline.  
Poynting-Robertson drag then removes larger grains from the disk, which 
reduces the population of very small grains from erosive collisions.  
The dust production rate then declines with evolution time as $t^{-2}$
\citep[see also][]{dom03,kb04a,kb05,wya05,wya07a,wya07b}.

The time evolution of the collision rate in the disk yields a simple 
relation between the maximum dust production rate $\dot{M}_{max}$ 
and the initial disk mass.  For the complete set of calculations,
\begin{equation}
\dot{M}_{max} \approx 6.5 \times 10^{20} ~ x_m^2 ~ {\rm g~yr^{-1}} ~ .
\label{eq:mdot1}
\end{equation}
For a MMSN with $x_m$ = 1, the maximum rate is $\sim$ 0.1 $M_{\oplus}$ 
every million years. The collision rate scales with the square of the 
number density of objects; thus, the dust production rate grows as the 
square of the initial disk mass (e.g., $\dot{M}_{max} \propto x_m^2$). 

The timescale for the peak in dust production is shorter than the
timescale for the production of 1000~km objects in the disk,
\begin{equation}
t_{\dot{M}_{max}} \approx 14 ~ x_m^{-1} ~ {\rm Myr} ~ .
\label{eq:tmdot1}
\end{equation}
This time corresponds roughly to the time of peak stirring of
leftover planetesimals by oligarchs growing in the inner disk,
starting the collisional cascade. Oligarchs form faster in more
massive disks; thus, the dust production rate peaks earlier
in more massive disks.

\begin{figure}
\includegraphics[width=7.0in]{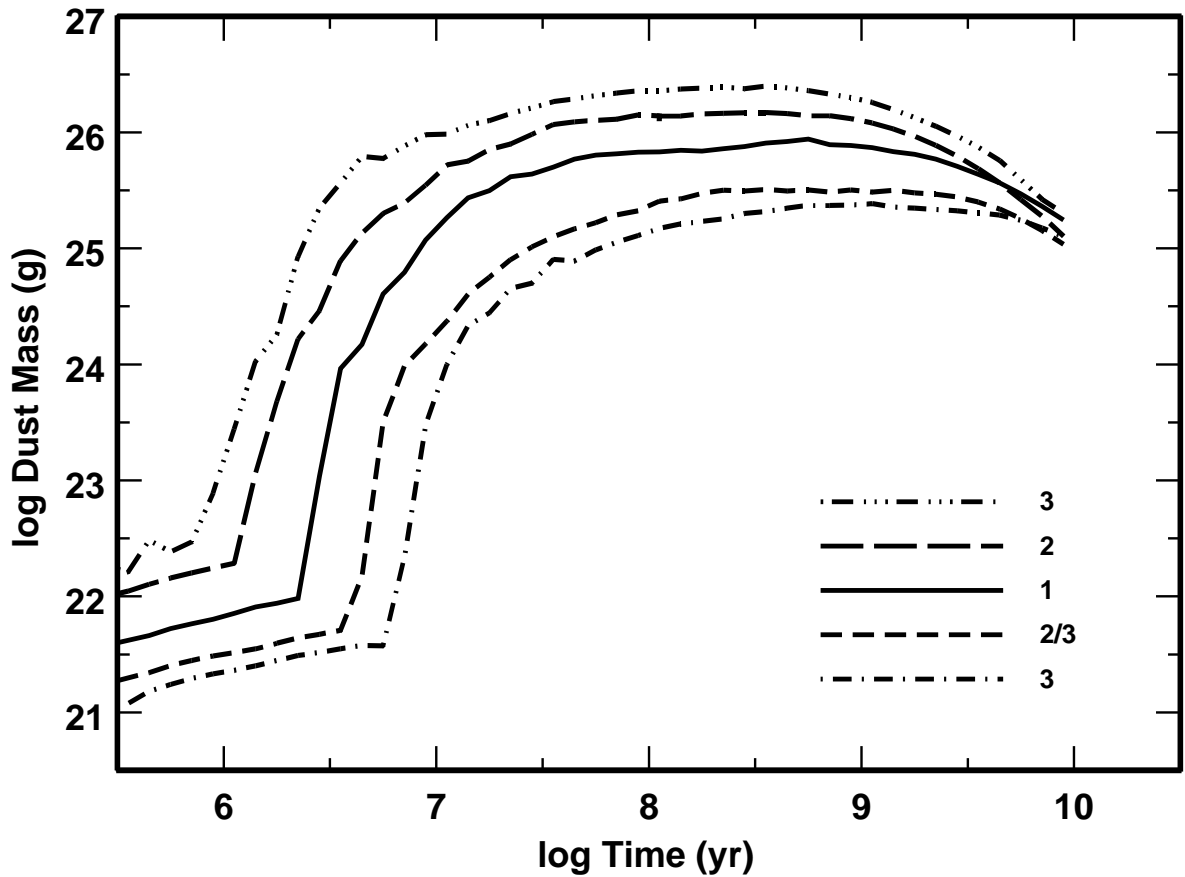}
\figcaption[f7.eps]
{Median mass in 0.001--1~mm objects as a function of time for 
30--150~AU disks around a 1~\msun\ central star. The legend
indicates $x_m$, the initial disk mass in units of the scaled MMSN.
During runaway growth, the median dust mass is small and roughly
constant in time. As planet formation propagates through the disk,
the dust mass grows with time. Once planets form in the outer disk,
collisions and Poynting-Robertson drag removes small grains from
the disk.
\label{fig:dust2}}
\end{figure}

Fig. \ref{fig:dust2} shows the time evolution of the mass in small grains 
for disks with a range of initial masses. Initially, the dust production
rates are small (Fig. \ref{fig:dust1}) and the dust mass increases slowly
with time. Once large oligarchs form in the inner disk, the dust production 
rate -- and thus the mass in small grains -- grows rapidly. For all disks, 
it takes $\sim$ 5--10~Myr for the mass in small grains to grow from $10^{-6}$
$M_{\oplus}$ (which is unobservable with current techology) to $\sim$
$1-10 \times 10^{-4}$ $M_{\oplus}$ (which is observable). After this rapid
rise, oligarchs form at larger and larger disk radii, leading to enhanced
dust production farther and farther out in the disk. The dust mass then
rises slowly with time. Once oligarchs form at the outer edge of the disk, 
the collisional cascade removes more and more solid material throughout
the disk. The dust mass then begins to decline.

The maximum mass in small grains scales with the initial disk mass,
\begin{equation}
M_{max,small} \approx 0.013 ~ x_m ~ M_{\oplus} ~ .
\label{eq:msmall}
\end{equation}
Because the derived size distributions are dominated by collisional
processes, the maximum mass in large grains is roughly 40 times 
larger \citep[e.g.,][]{doh69,wil94,kb04c,kri06},
\begin{equation}
M_{max,large} \approx 0.5 ~ x_m ~ M_{\oplus} ~ .
\label{eq:mlarge}
\end{equation}
In both cases, the larger collision rate in more massive disks 
leads to more dust.  Although these dust masses are significant,
they are small compared to the mass in objects with $r \gtrsim$ 
100~km. The typical mass in large grains is $\lesssim$ 10\% of 
the mass in 100~km and larger objects. The mass in small grains 
is $\lesssim$ 0.25\% of the mass in the largest objects.

The timescale to reach the maximum dust mass is a function of 
the particle size. For the small grains,
\begin{equation}
t_{max,small} \approx 270 ~ x_m^{-1/2} ~ {\rm Myr} ~ .
\label{eq:tdustsm}
\end{equation}
For the large grains,
\begin{equation}
t_{max,large} \approx 600 ~ x_m^{-1/2} ~ {\rm Myr} ~ .
\label{eq:tdustlg}
\end{equation}

Several features of the collisional cascade set these timescales. Early 
in the evolution, the collision timescale for all particle sizes is 
smaller than the timescale for Poynting-Robertson drag. Thus, the cascade 
erodes small particles until radiation pressure ejects them. As planet
formation propagates out through the disk, collisions produce more and
more small grains. Because the mass in grains is set by a balance between 
the collision time, which scales as $x_m^{-1}$ and the local dynamical 
time, which scales as $x_m^{-1/2}$, the timescale to reach the maximum 
grain mass is proportional to $x_m^{-1/2}$. As the collision rate declines,
Poynting-Robertson drag starts to remove mass from the disk. This drag
removes smaller particles from the disk more effectively than it removes 
large particles. Thus, the mass in small grains peaks before the mass in 
larger grains.  

These results suggest that the mass in collisional debris is large, 
roughly a lunar mass in 0.001--1 mm grains, throughout the lifetime 
of a 1 $M_{\odot}$ star. Although the Solar System has much less dust
\citep[e.g.,][and references therein]{lan02,nes06}, these large disk 
masses are comparable to the mass in dust grains detected in many 
debris disks around other stars \citep[e.g.,][]{bei06,tri08,mor08}.
Because our dust production rates are observable, the large range in 
dust masses as a function of initial disk mass and time implies a 
correspondingly large range in the observable properties of debris 
disks, such as the disk luminosity and the IR excess, at fixed
stellar age.  Because the dust production rate declines with time, 
we expect the disk luminosity and IR excesses to decline with
time. We derive detailed predictions for this evolution in \S4
and compare these results with observations in \S5.

\subsection{Icy Planet Formation in 1.5--3 $M_{\odot}$ Stars}

Stellar evolution is an important feature of icy planet formation
at 30--150 AU. Because the main sequence lifetime 
\citep[$t_{ms} \propto M_{\star}^{-n}$, with $n \approx$ 3--3.5; 
e.g.,][]{ibe67,dem04} is more sensitive to stellar mass than the 
timescale to produce large icy planets 
($t \propto M_{\star}^{-3/2}$; see below), massive stars evolve off 
the main sequence before oligarchic growth and the collisional cascade 
remove solid material in the outer disk.  After a 1--3~\msun\ star 
evolves off the main sequence, it becomes more luminous (as a red giant) 
and hotter (as a post-AGB star). During this evolution, icy planetesimals 
and planets melt, decreasing collision rates and changing the outcome of 
the collisional cascade\footnote{We assume that melting is accompanied 
by a loss of volatiles and an increase in the mass density of leftover
planetesimals.}.  Short main sequence lifetimes of massive stars thus 
lead to clear differences in the amount of solid material in large and 
small objects in the outer disk.

The stellar mass also affects the outcome of icy planet formation.  The 
timescale for planet formation scales with orbital period and the surface 
density, $t \propto P/\Sigma$ (see the Appendix). For a disk with 
$\Sigma = \Sigma_0 ~ x_m ~ a^{-3/2}$ (Eq. (\ref{eq:sigma})) and fixed 
$\Sigma_0 ~ x_m$, the formation time is $t \propto a^3 ~ M_{\star}^{-1/2}$. 
Thus, at fixed $a$ in disks with identical surface density distributions, 
planets form faster around more massive stars. However, disk masses 
in the youngest stars scale with stellar mass \citep[e.g., 
$M_d \propto M_{\star}$;][]{nat2000,sch2006}. Thus, $\Sigma_0$ scales with 
stellar mass, $\Sigma_0 \propto M_d \propto M_{\star}$. Combining these 
relations leads to a formation time $t \propto a^3 ~ M_{\star}^{-3/2}$.
Thus, at fixed $a$ in typical disks, icy planets form $\sim$ 5 times 
faster around 3 \msun\ stars than around 1 \msun\ stars. 

\begin{figure}
\includegraphics[width=7.0in]{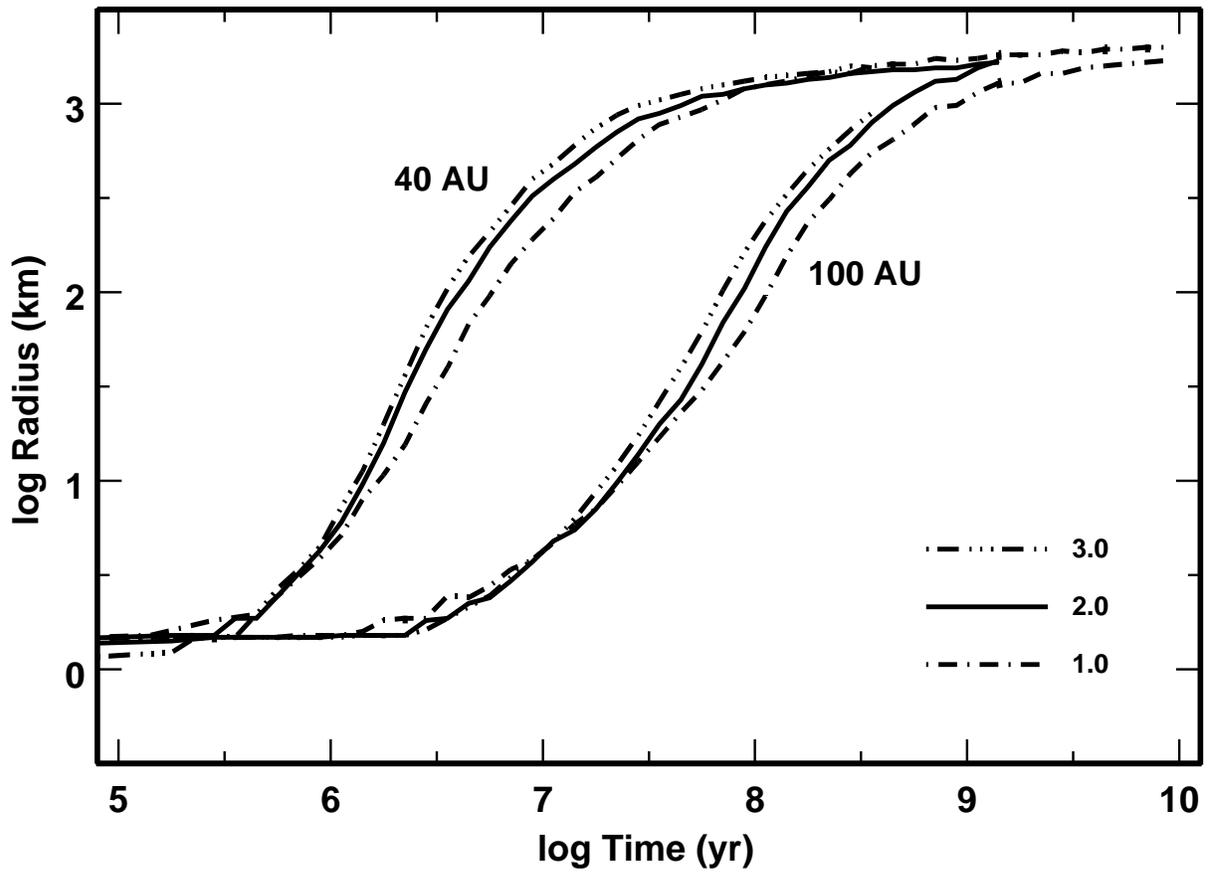}
\figcaption[f8.eps]
{Time evolution of the radius of the largest object at 40~AU and
at 100~AU for identical disks around 1~\msun\ (dot-dashed curves), 
2~\msun\ (solid curves), and 3~\msun\ (triple dot-dashed curves) 
stars. Planets grow faster around more massive stars and in the
inner disks of all stars.
\label{fig:rad}}
\end{figure}

To illustrate how stellar mass and evolution affect planet formation,
we begin with the growth of large objects at 40~AU and at 100~AU (Fig.
\ref{fig:rad}). For disks with identical initial surface density
distributions, planets at the same $a$ in disks around 3 \msun\ stars 
grow $\sim$ 1.7 times faster than planets around 1 \msun\ stars. Fig.
\ref{fig:rad} also shows the clear scaling of growth time with semimajor
axis, $t \propto a^3$ for a disk with $\Sigma \propto a^{-3/2}$. The
simple scaling of the growth time with disk mass and orbital period 
in our calculations leads to a general relation for the median timescale 
for the formation of the first 1000~km object in disks at 30--150~AU,
\begin{equation}
t_{1000} \approx 145 ~ x_m^{-1.15} ~ \left ( \frac{a}{\rm 80~AU} \right )^3 ~ \left ( \frac{2~M_{\odot}}{M_{\star}} \right )^{3/2} {\rm Myr} ~ .
\label{eq:t1000allm}
\end{equation}
For 1--3 \msun\ stars, this relation fits our results for the median growth 
time to $\approx$ 7\% for $a$ = 30--150~AU and for $x_m$ = 1/3 to 3. For 
all initial disk masses, the inter-quartile range for the formation time 
is $\sim$ 20\%. Thus, there is a modest range of outcomes for identical 
starting conditions around 1--3 \msun\ stars.

Aside from the extra factor of $x_m^{-0.15}$, this relation follows the
prediction of $t \propto x_m^{-1} M_{\star}^{-3/2} a^3$ from standard
coagulation theory. As outlined in the Appendix, velocity damping from 
gas drag can speed up planet formation in more massive disks.

\begin{figure}
\includegraphics[width=7.0in]{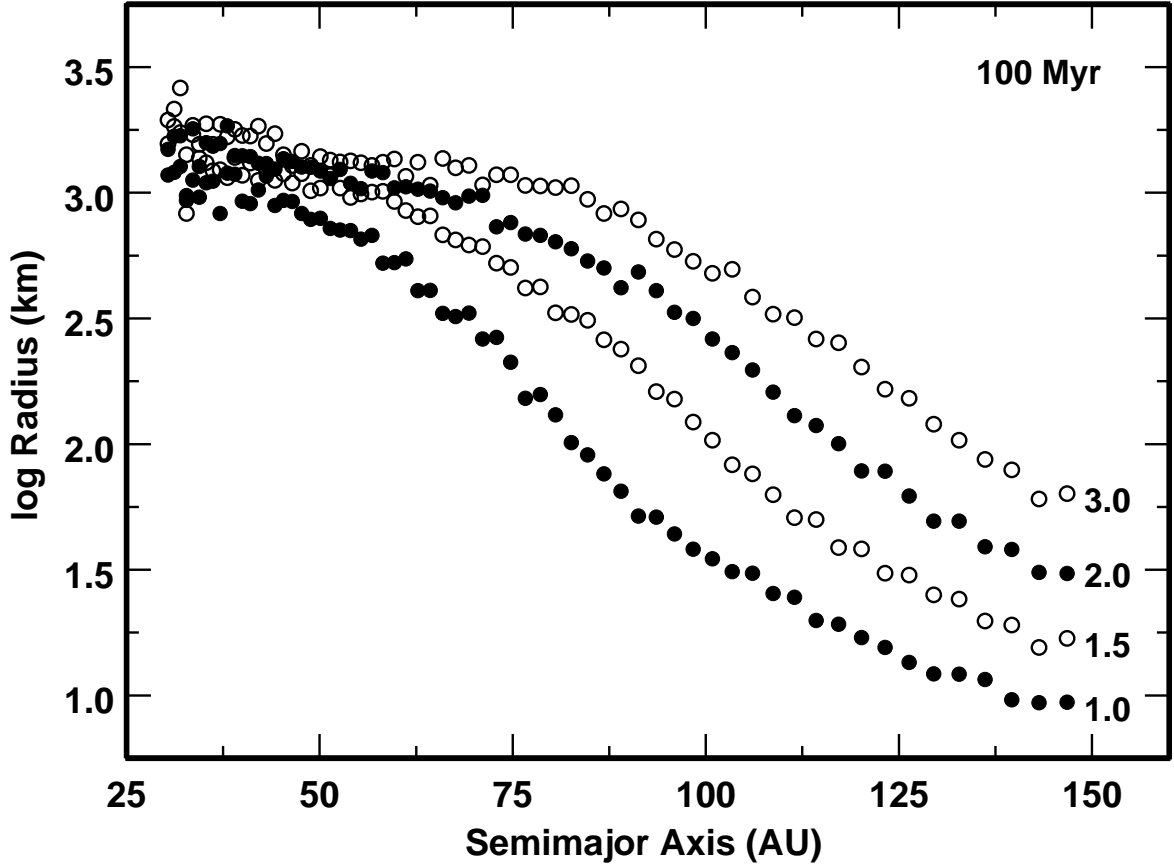}
\figcaption[f9.eps]
{Median radius of the largest object at 100~Myr in each annulus for a 
scaled MMSN disk ($x_m$ = 1) around 1--3~$M_{\odot}$ stars. The number 
to the right of each set of points indicates the stellar mass in 
\msun.  At all disk radii, large objects form faster around more 
massive stars. At 30--60~AU, planet formation saturates at radii 
$r \sim$ 1000--2000~km on relatively short timescales, $t \sim$ 
100~Myr (see also Eq. (\ref{eq:t1000allm})). At larger disk radii,
planets form more slowly and do not reach the maximum radius unless 
the formation time is shorter than the main sequence lifetime.
\label{fig:radevol4}}
\end{figure}

Fig. \ref{fig:radevol4} further shows how the growth time varies with 
$a$ and $M_{\star}$ for constant $x_m$. At 100 Myr, icy planets are 
close to their maximum sizes at 30--50~AU in the inner disk. At large 
disk radii ($a \sim$ 100--150~AU), planet formation is clearly faster 
around more massive stars. For all stars with $M_{\star}$ = 1--3 \msun, 
the $r_{max}(a)$ relations have a similar morphology, consisting of a 
plateau at $r_{max} \approx$ 1000--2000~km and a steep decline of
$r_{max}$ with increasing $a$. As in \S3.1.1, we fit our results to 
a simple relation between $r_{max}$, $a$, and time (Eq. (\ref{eq:rmax})).
For all of our calculations, we derive an exponent $\alpha_r \approx$ 
5--6 and a characteristic maximum radius
\begin{equation}
r_0 \approx 1750 ~ x_m^b \left( \frac{M_{\star}}{2~M_{\odot}} \right)^{0.09} \left ( \frac{3t}{t_{ms}} \right )^{0.06} ~ {\rm~km}
\label{eq:r0allm}
\end{equation}
with the exponent $b = 0.22 + 0.033~M_{\star}/M_{\odot}$. The disk
scale length is 
\begin{equation}
a_0 \approx 190 ~ x_m^{0.1} \left ( \frac{3 t}{t_{ms}} \right )^{0.1} ~ {\rm~AU}
\label{eq:a0allm}
\end{equation}
These results are valid for late times, $t \approx$ 0.1--1 $t_{ms}$.

Eq. (\ref{eq:r0allm}) shows that the maximum sizes of icy planets at 30--150~AU
are relatively insensitive to initial disk mass, stellar mass, or time. For
disks with identical $x_m$ around 1--3 \msun\ stars, the largest icy planets 
around 3 \msun\ stars are only $\sim$ 10\%--20\% larger than the largest icy 
planets around solar-type stars. This range is comparable to the range in 
maximum sizes for planets formed in identical disks around stars of identical 
mass (\S3.1.1). Disks with a factor of 10 range in $x_m$ yield planets with 
a 20\%--30\% range in radii (a factor of $\sim$ 2 in mass). Thus, our 
calculations predict that the largest icy planets at 30--150~AU around 
1--3 \msun\ stars have masses comparable to Pluto and other large Kuiper 
belt objects in the Solar System beyond the orbit of Neptune.

Although the maximum sizes of planets are fairly independent of initial 
conditions, the number of Pluto-mass objects $n_P$ is sensitive to disk mass
and stellar mass (Table \ref{tab:rad1000.2}). In the inner disk (30--60~AU),
$n_P$ scales roughly with initial disk mass and stellar mass.  In the outer
disk (100--150~AU), the formation timescale for icy planets is comparable 
to the main sequence lifetime.  Thus, $n_P$ scales with initial disk mass 
and stellar mass only for the most massive disks.  In lower mass disks, 
stars evolve off the main sequence before disks can produce large numbers 
of Pluto-mass objects. 

\begin{figure}
\epsscale{1.1}
\plottwo{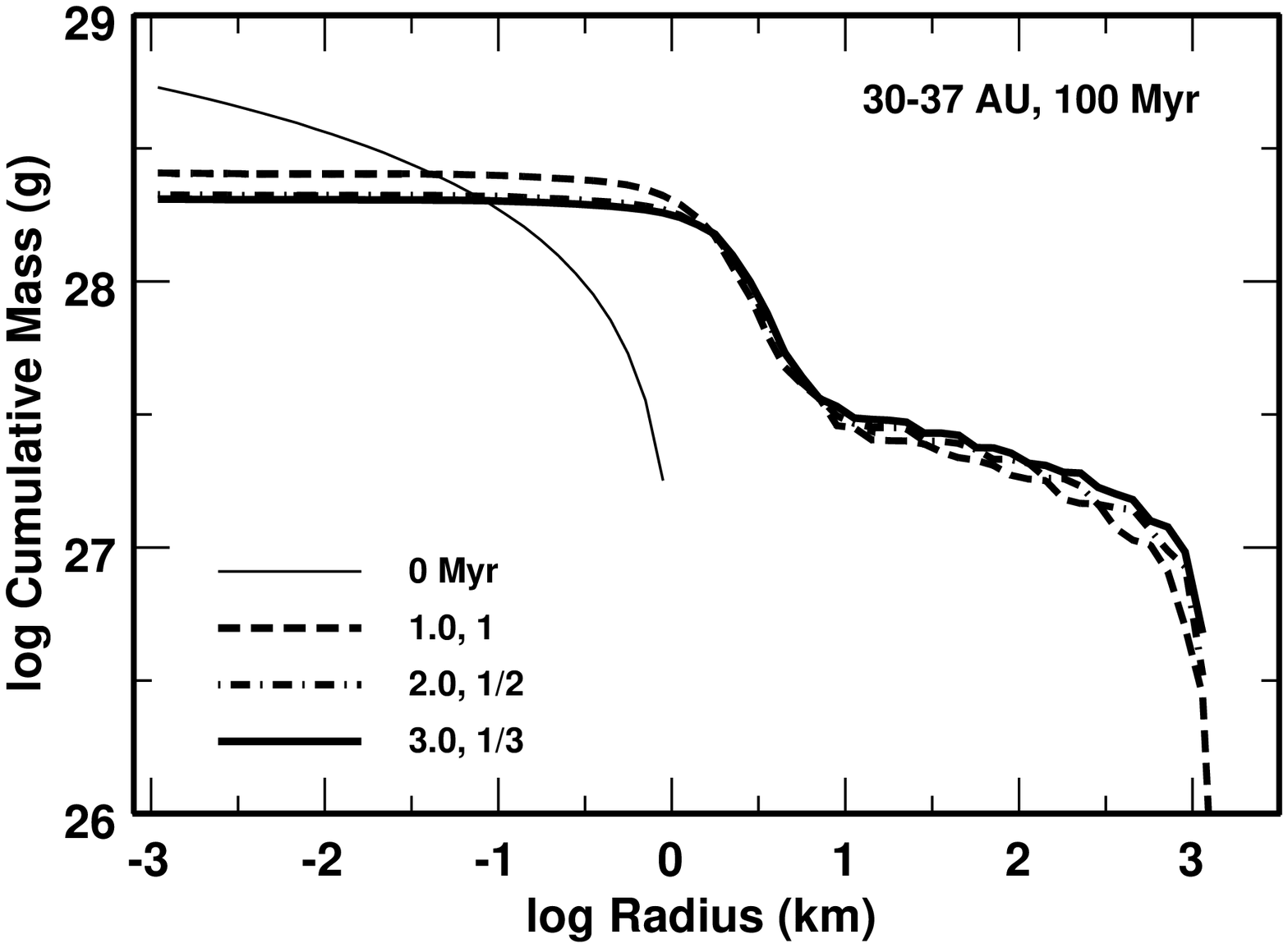}{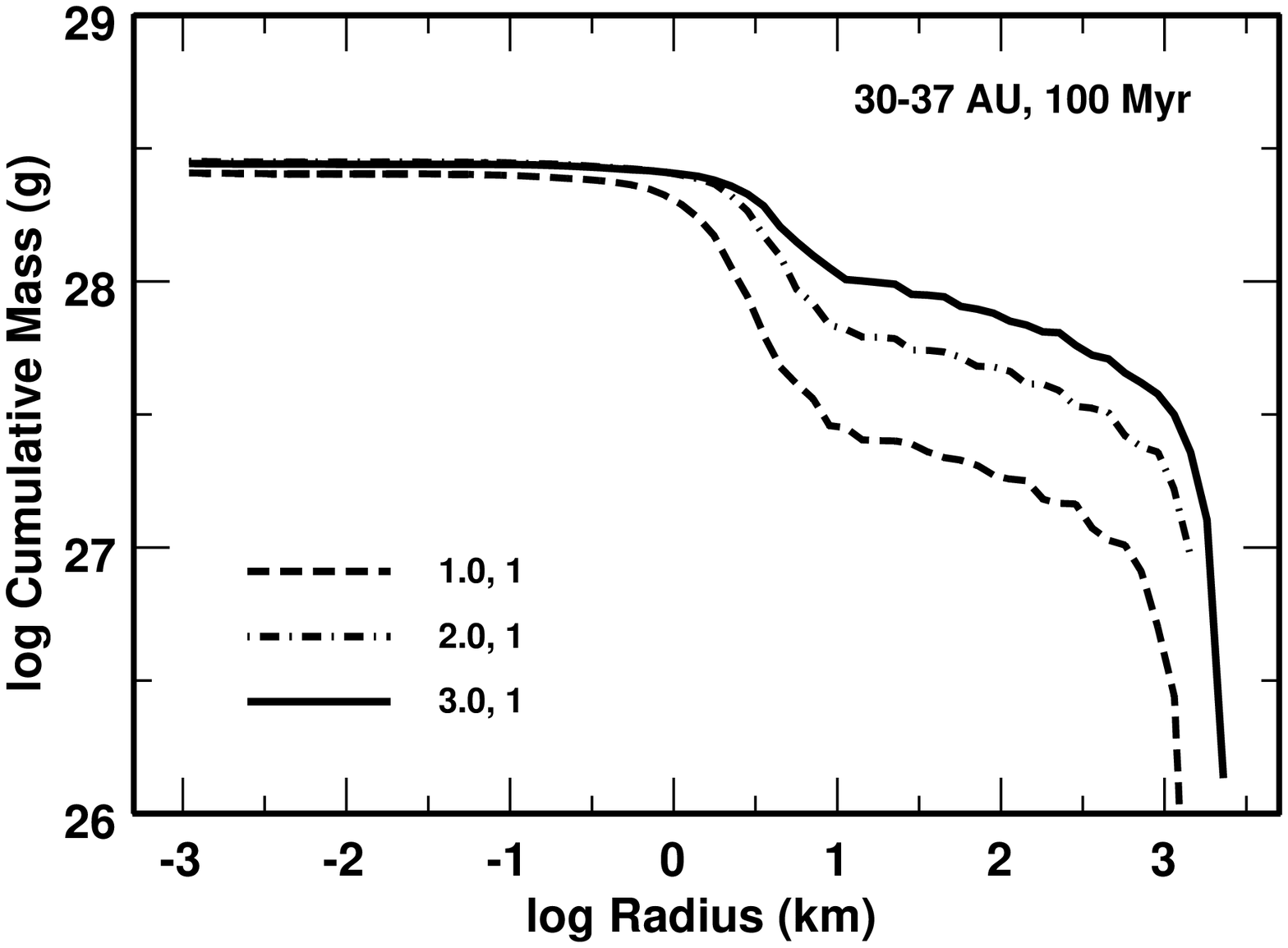}
\figcaption[f10.eps]
{Median cumulative mass distributions at 100~Myr for planet 
formation calculations at 30--37~AU around 1--3 \msun\ stars.
{\it Left}: Results for models with 
$\Sigma = 0.18 ~ (a_i/{\rm 30~AU})^{-3/2}$ g cm$^{-2}$.
The light solid line indicates the initial mass distribution.
The dashed (1 \msun, $x_m$ = 1), dot-dashed (2 \msun, $x_m$ = 1/2), 
and heavy solid (3 \msun, $x_m$ = 1/3) lines show median 
results for the same initial conditions. 
{\it Right}: Results for models with a scaled surface density ($x_m$ = 1),
$\Sigma = 0.18 ~ (a_i/{\rm 30~AU})^{-3/2}$ $(M_{\star} / M_{\odot})$ g cm$^{-2}$,
and different stellar masses (1 \msun: dashed line, 2 \msun: dot-dashed 
line, and 3 \msun: heavy solid line).
Although more massive planets form around more massive stars, 
the collisional cascade leads to a small dispersion in total
disk mass at late times.
\label{fig:sd3}}
\end{figure}

Cumulative mass distributions provide another useful comparison of
icy planet formation as a function of disk mass and stellar mass.
For disks with identical initial surface density distributions at $a$ 
= 30--37~AU, the shape of the mass distribution is fairly independent 
of stellar mass at 100 Myr (Fig. \ref{fig:sd3}; left panel).  Because
large icy planets form first in disks around more massive stars, disks 
of fixed age around 3 \msun\ stars have more mass in larger planets 
and are more collisionally depleted than disks around 1 \msun\ stars. 

For calculations in scaled MMSN (Fig. \ref{fig:sd3}; right panel), 
growth is a stronger function of stellar mass.  As predicted by the 
simple scaling relations, disks around 3 \msun\ stars have $\sim$ 
5 times more mass in large objects than 1 \msun\ stars. More mass 
in large objects produces more stirring, enhancing mass loss in the 
collisional cascade. Despite large difference in initial disk mass, 
the mass distributions of disks around 1--3 \msun\ stars are very 
similar for $r <$ 1~km at similar times.

\begin{figure}
\epsscale{1.1}
\plottwo{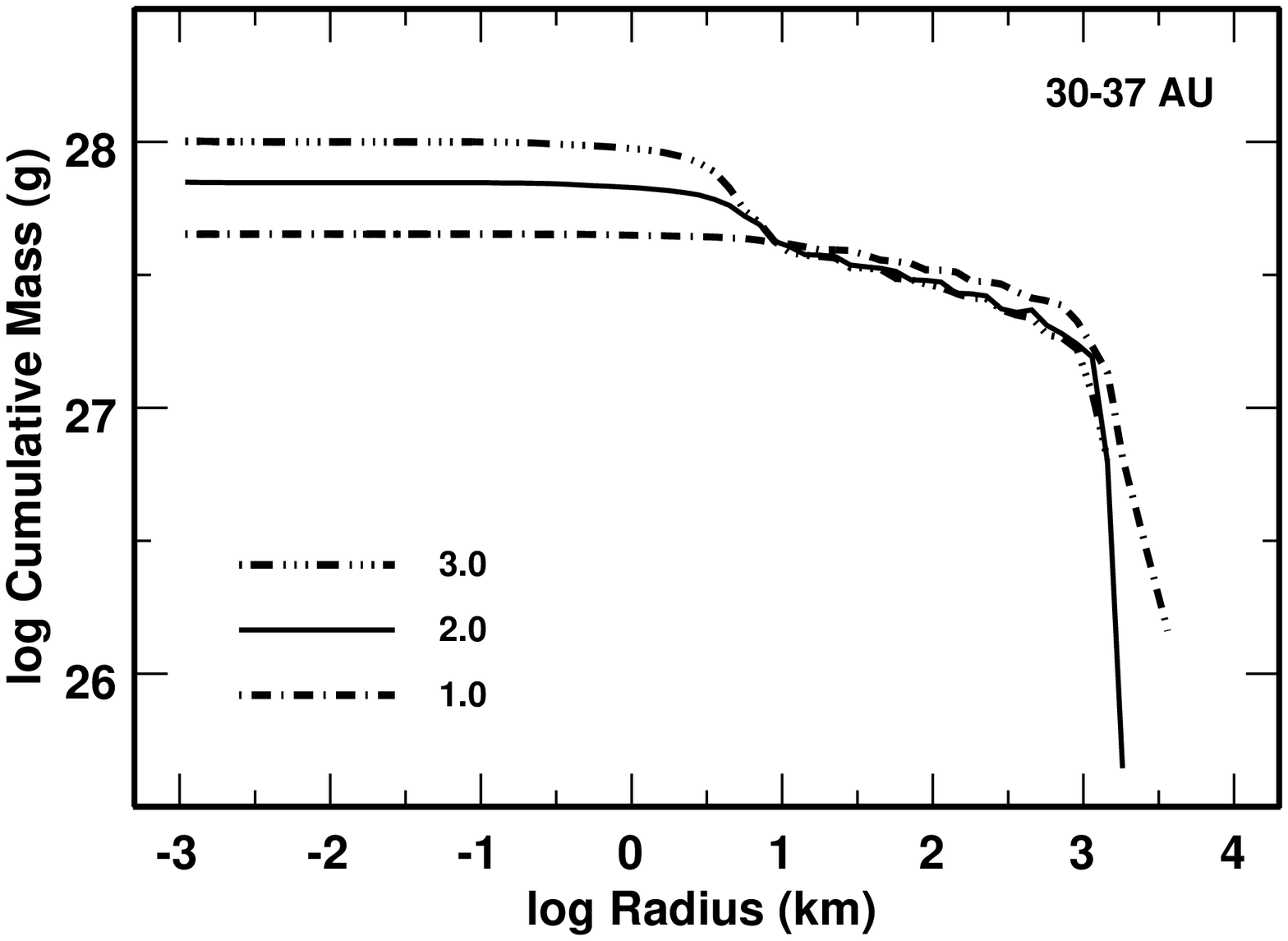}{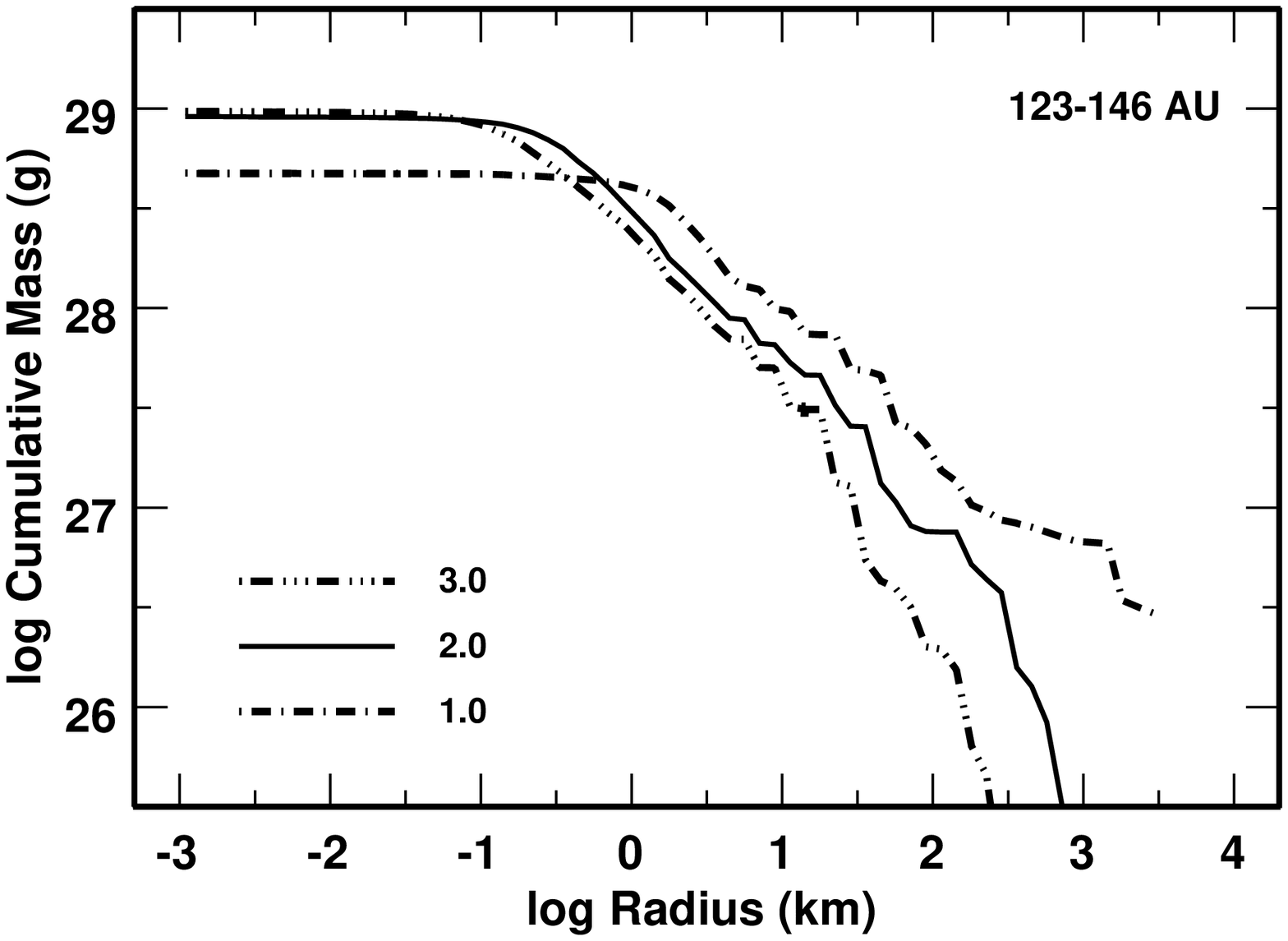}
\figcaption[f11.eps]
{Median cumulative mass distributions at $t = t_{ms}$ for annuli 
at 30--37~AU (left panel) and at 123--146~AU (right panel) for identical 
disks ($\Sigma = 0.18 ~ (a_i/{\rm 30~AU})^{-3/2}$ g cm$^{-2}$) 
around 1--3 \msun\ stars. The legend indicates the stellar mass 
in \msun. In the inner disk, many large planets form and the 
collisional cascade removes a large fraction of the material in 
objects with $r \lesssim$ 1--10~km. In the outer disk, few large 
planets form; collisions are inefficient at removing material in 
small objects.
\label{fig:sd4}}
\end{figure}

Although planet formation proceeds faster with increasing stellar 
mass, stellar evolution halts the collisional cascade and the growth 
of planets in more massive stars (Fig. \ref{fig:sd4}). Planets reach 
their maximum radii in the inner disks for all 1--3~\msun\ stars;
however, the timescale for the collisional cascade to run to 
completion is long compared to the main sequence lifetimes of
2--3~\msun\ stars. Thus, the collisional cascade removes a larger 
fraction of material from the inner disks around 1~\msun\ stars 
than from the inner disks of 2--3~\msun\ stars (Fig. \ref{fig:sd4};
left panel). In the outer disk, the growth time for 1000~km planets
is large compared to the main sequence lifetime for 2--3 \msun\ stars. 
Thus, planets do not reach their maximum radii of $\sim$ 1000--2000~km 
in the outer disks of 2--3 \msun\ stars (Fig. \ref{fig:sd4}; right panel)

To conclude our comparison of icy planet formation around stars of
different masses, we consider the long-term evolution of all solids
in the disk. For $t = 0.3-1.0~t_{ms}$, the median fraction of solids
remaining in the disk is
\begin{equation}
f_s \approx 0.38 \left ( \frac{a}{\rm 100~AU} \right )^{1.25} \left ( \frac{1}{x_m} \right )^{1/4} ~ .
\label{eq:frac10gyallm}
\end{equation}
for the ensemble of calculations for 1.5 \msun\ stars and
\begin{equation}
f_s \approx 0.6 \left ( \frac{a}{\rm 100~AU} \right )^{1.6} \left ( \frac{1}{x_m} \right )^{1/3} ~ .
\label{eq:fracendallm}
\end{equation}
for calculations for 2--3 \msun\ stars. At 30--50~AU, all stars lose 
roughly the same fraction of mass from the disk. At larger disk radii, more 
massive stars evolve off the main sequence before the collisional cascade  
removes most of the leftover 1--10~km planetesimals. Thus, icy planet
formation around lower mass stars converts a larger fraction of the 
initial solid mass into dusty debris. 

Although planet formation around massive stars converts a smaller 
fraction of the initial mass into dusty debris, icy planet formation is 
equally efficient at producing massive objects around all 1--3 \msun\ stars.
For all disks in our calculations, the median fraction of the initial
disk mass in 1000~km and larger objects is
\begin{equation}
\label{eq:frac1000allm}
f_{1000} = 0.035 \left ( \frac{\rm 30~AU}{a} \right ) ~.
\end{equation}
The median fraction of the mass in 100~km and larger objects is 
$\sim$ 50\% larger,
\begin{equation}
\label{eq:frac100allm}
f_{100} = 0.06 \left ( \frac{\rm 30~AU}{a} \right ) ~.
\end{equation}
Mass distributions for icy planets are top heavy for all 1--3 \msun\ stars. 
As for calculations around 1 \msun\ stars, the typical inter-quartile 
ranges are $\sim$ 0.1 for $f_s$ and $\sim$ 20\% for $f_{100}$ and
$f_{1000}$. Thus, identical starting conditions lead to a modest
range of outcomes.  

\subsubsection{Evolution of Dust}

The evolution of dusty debris in disks around 1.5--3 \msun\ stars
generally follows the evolution for 1 \msun\ stars. As oligarchs
form, stirring leads to a collisional cascade that converts 10~km 
and smaller objects into small dust grains. Because planets form 
more rapidly around more massive stars, disks around massive stars 
produce dust sooner than disks around lower mass stars. In our 
calculations, the disk mass scales with the stellar mass. Thus, 
disks around massive stars also produce more dust than disks around 
lower mass stars. However, massive stars do not live as long as 
lower mass stars, preventing the collisional cascade from removing 
all of the small objects from the disk (Fig. \ref{fig:sd3}). Over
the lifetime of the star, disks around lower mass stars form more 
dust than disks around more massive stars.

\begin{figure}
\includegraphics[width=7.0in]{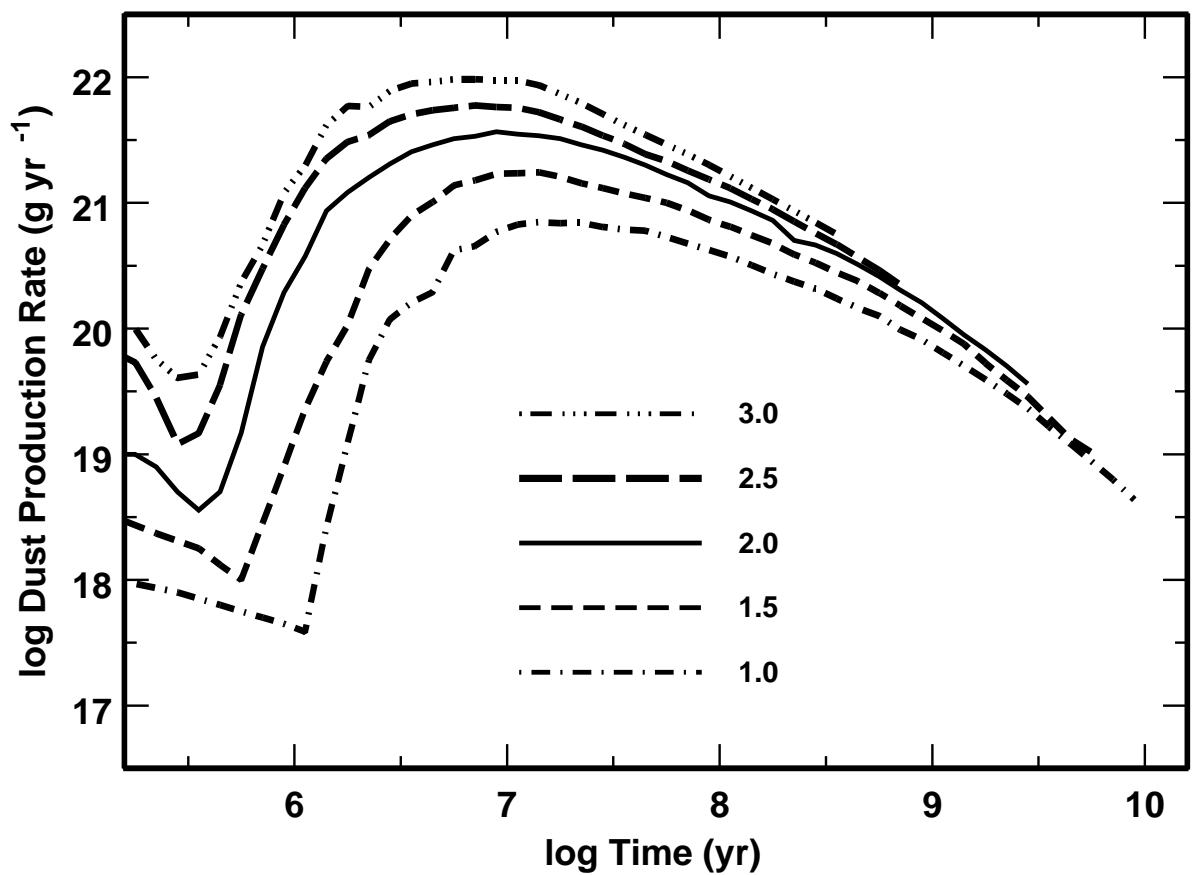}
\figcaption[f12.eps]
{Median production rate of 0.01--1 $\mu$m objects at 30--150~AU
as a function of time for scaled MMSN ($x_m$ = 1) around 1--3 
\msun\ central stars.  The legend indicates the stellar mass 
in \msun. For scaled MMSN, disks around more massive stars 
eject much more material in very small grains at early times 
($t \lesssim$ 10--100~Myr). In an ensemble of stars with a variety 
of disk masses, there is wide range of dust production rates.
\label{fig:dust3}}
\end{figure}

To illustrate these points, Fig. \ref{fig:dust3} shows the time
evolution of the dust production rate for a scaled MMSN ($x_m$ = 1)
around 1 \msun, 2 \msun, and 3 \msun\ stars. During runaway growth,
destructive collisions are rare; thus, the dust production rate 
declines with time. As runaway growth ends, there are three distinct 
phases in dust production: (i) an exponential rise when runaway
objects start to stir leftover planetesimals in the inner disk, 
(ii) a long plateau as oligarchs form farther and farther out 
in the disk, and (iii) a long decline in dust production as the 
collisional cascade depletes the disk of 1--10~km objects. 

Because planets grow more rapidly around more massive stars, the 
exponential growth in dust production occurs first around more 
massive stars.  The timescale for the onset of dust production
also scales inversely with disk mass; thus, more massive disks
produce dust faster than low mass disks. 

When runaway growth ends and oligarchic growth begins, the dust
production rate reaches a clear plateau (Fig. \ref{fig:dust3}).
We define the onset of the plateau phase as the time of maximum
dust production\footnote{Because there are small fluctuations 
in the dust production rate during the plateau phase, we define
the maximum dust production as the time when the derivative
$d \dot{M}(t)/dt$ first changes sign.}. For our calculations,
there is a simple relation between the timescale of maximum
dust production and the masses of the disk and central star,
\begin{equation}
t_{\dot{M}_{max}} \approx 5 ~ x_m^{-1} ~ \left ( \frac{2~M_{\odot}}{M_{\star}} \right )^{1.5} {\rm Myr} ~ .
\label{eq:tmdot1allm}
\end{equation}
At this time, our simulations yield a simple relation between the maximum 
dust production rate and the masses of the disk and the central star,
\begin{equation}
\dot{M}_{max} \approx 3.5 \times 10^{21} ~ x_m^2 ~ \left ( \frac{M_{\star}}{2~M_{\odot}} \right )^{2.5} ~ {\rm g~yr^{-1}} ~ .
\label{eq:mdot1allm}
\end{equation}

Each of these scaling laws has a simple physical origin.  The maximum 
dust production rate, $\dot{M}_{max}$, depends on the collision rate, 
the square of the disk mass divided by the orbital period. Thus, 
$\dot{M}_{max}$ $\propto \Sigma^2 / P$ $\propto \Sigma^2 M_{\star}^{1/2}$.
For disks where the surface density scales with stellar mass 
(Eq. (\ref{eq:sigma})), $\Sigma \propto x_m M_{\star}$. Thus,
$\dot{M}_{max}$ $\propto x_m^2 M_{\star}^{5/2}$. The timescale to 
reach this rate is the orbital period divided by the disk mass. 
Thus, $t_{\dot{M}_{max}}$ $\propto \Sigma^{-1} M_{\star}^{-1/2}$ 
$\propto x_m^{-1} M_{\star}^{-3/2}$.

Once oligarchs form in the outer disk, the dust production rate declines
rapidly. Stellar evolution sets the duration of this decline. In massive
stars, the short main sequence lifetime halts the evolution before the
collisional cascade depletes the disk of 1--10~km objects. Thus, the
dust production rate declines by roughly an order of magnitude before
the central star evolves off the main sequence. For lower mass stars,
the long main sequence lifetime allows the collisional cascade to
remove some material in the outer disk. Thus, the dust production 
rate declines by $\sim$ two orders of magnitude before the central
star evolves into a red giant. We show in \S4 how the long-term
evolution of the dust production rate as a function of stellar mass
produces observable differences in the IR excesses of massive and 
low mass stars.

\begin{figure}
\includegraphics[width=7.0in]{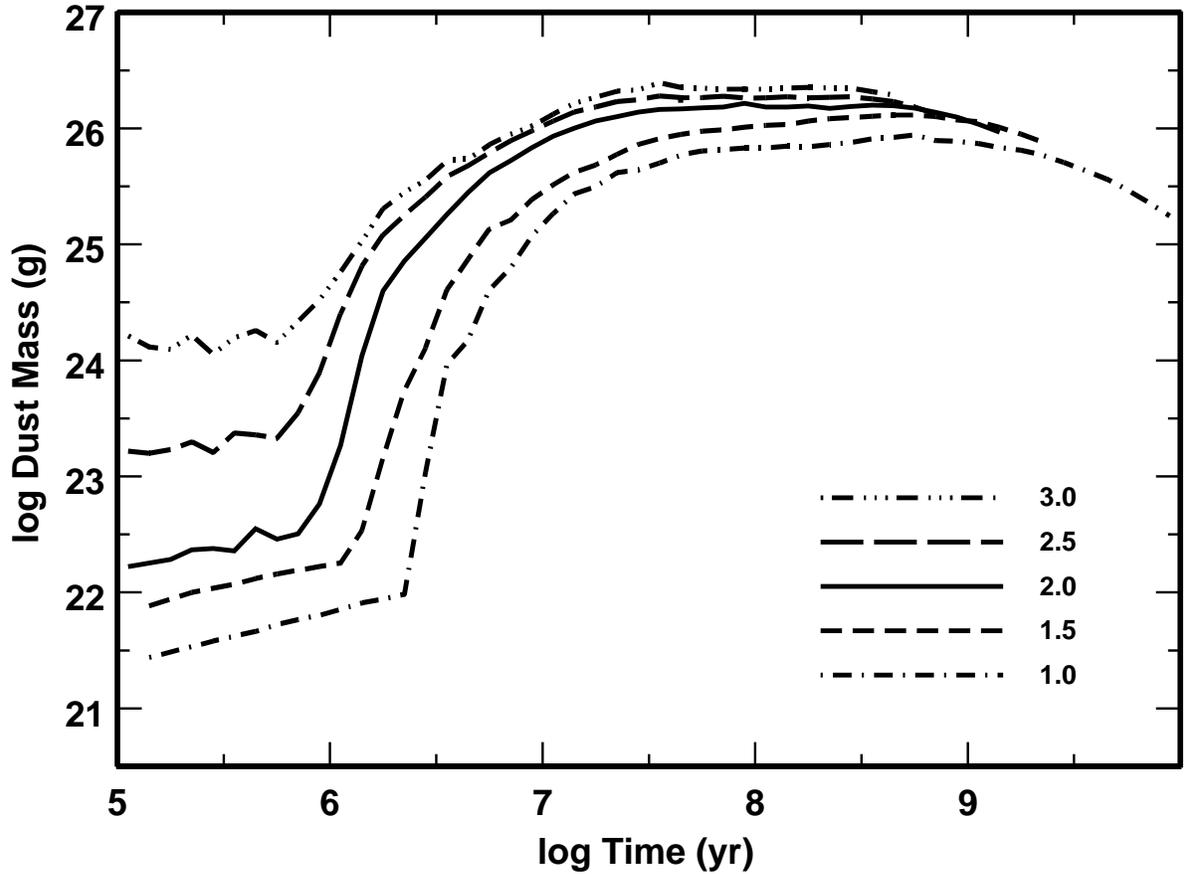}
\figcaption[f13.eps]
{Median mass in 0.001--1 mm objects as a function of time for scaled 
MMSN ($x_m$ = 1) at 30--150~AU around 1--3 \msun\ central stars. The 
legend indicates the stellar mass in \msun.
For $t \lesssim$ 1--3~Myr, icy planet formation produces little dust.
At 10--100~Myr, the mass in small grains is $\sim$ 1 lunar mass for
most disks. At late times, the mass in small grains slowly declines
to currently undetectable levels.
\label{fig:dust4}}
\end{figure}

Despite the large differences in dust production rates, there are
smaller variations in the amount of dust as a function of disk mass
and stellar mass. Fig. \ref{fig:dust4} shows the time evolution of the 
median mass in small grains for scaled MMSN around 1--3 \msun\ stars.
Once the exponential rise in dust production begins, the dust masses
rapidly evolve to similar configurations with $\sim 10^{26}$ g in
small grains. For our set of calculations, the maximum mass in small 
grains is
\begin{equation}
M_{max,small} \approx 0.026 ~ x_m ~ \left ( \frac{M_{\star}}{2~M_{\odot}} \right ) M_{\oplus} ~ .
\label{eq:msmallallm}
\end{equation}
The coefficient in this equation, $0.026 M_{\oplus}$, is roughly twice the
mass of the Moon.  The maximum mass in large particles is
\begin{equation}
M_{max,large} \approx 1.0 ~ x_m ~ \left ( \frac{M_{\star}}{2~M_{\odot}} \right ) M_{\oplus} ~ .
\label{eq:mlargeallm}
\end{equation}

The timescale to reach the maximum dust mass is a function of the particle
size. For the small grains,
\begin{equation}
t_{max,small} \approx 135 ~ x_m^{-1/2} ~ \left ( \frac{M_{\star}}{2~M_{\odot}} \right )^{-1} {\rm Myr} ~ .
\label{eq:tdustsmallm}
\end{equation}
For the large grains,
\begin{equation}
t_{max,large} \approx 300 ~ x_m^{-1/2} ~ \left ( \frac{M_{\star}}{2~M_{\odot}} \right )^{-1} {\rm Myr} ~ .
\label{eq:tdustlgallm}
\end{equation}
As described in \S3.1.2, the collision rate, the dynamical timescale,
and Poynting-Robertson drag combine to produce the shorter timescale 
for smaller dust grains.

\subsection{Limitations of the Calculations}

In previous papers, we have described limitations to multiannulus
\citep[][2002a, 2004b, 2005, 2006]{bk06,kb01} and single annulus 
\citep[][1999]{kl98} coagulation calculations.  Here, we review
how several of these limitations affect results for the simulations
described above.

As long as the statistical assumptions underlying the formalism are
met, coagulation calculations provide a reasonable representation of 
real collision evolution
\citep{wet80,gre84,dav85,bar91,spa91,lis93,wet93,st97,wei97,kl98,ina01}.
For calculations at 30--150~AU around 1--3~\msun\ stars, the spacing 
of mass bins in an annulus and the spacing of annuli in the disk limit 
the accuracy of the results. Our standard mass spacing, $\delta = 2$, 
lengthens the evolution time by 10\% to 20\% relative to more accurate 
calculations with $\delta \lesssim$ 1.4 \citep[see][and references 
therein]{kl98}. The radial resolution, $\Delta a_i/a_i$ = 0.025, also 
lengthens the evolution time. Compared to calculations described in
\citet{kb04b}, improvements in our treatment of interactions among 
particles in neighboring annuli reduce lags by a factor of two, 
from $\sim$ 20\% to $\sim$ 10\%. Combining the lags for mass
spacing and radial resolution, our evolution timescales are $\sim$ 
20\% to 30\% longer than the actual evolution times. This lag is
comparable to the dispersion in timescales derived from multiple 
calculations with identical starting conditions. Thus, improvements 
in resolution are unlikely to alter our results significantly.

\subsubsection{Dynamical interactions}

The coagulation algorithm begins to break down when (i) a few large objects
contain most of mass in the grid and (ii) the gravity of these objects 
dominates the stirring. For $r \lesssim$ 500--1000~km, the largest
objects contain a small fraction of the mass in an annulus; individual 
dynamical interactions are much smaller than the Fokker-Planck stirring 
rates. Thus, kinetic theory yields good estimates for collisions and 
stirring among small objects. As objects grow beyond $\sim$ 1000~km, 
however, both assumptions of our statistical approach begin to fail:
(i) the collisional cascade removes leftover planetesimals, increasing
the fraction of mass in the largest objects and (ii) individual 
interactions among the largest objects in an annulus can deflect large 
objects into neighboring annuli, raising collision and stirring rates 
significantly. With $\sim$ 100--1000 Pluto-mass objects at 30--150~AU 
(see Tables \ref{tab:rad1000.1}--\ref{tab:rad1000.2}), interactions among
the largest objects could play a significant role in the late-time 
evolution of our models.

Dynamical interactions among an ensemble of Pluto-mass planets occur 
when the radial spacing is $\Delta a \sim 2 \sqrt{3} r_H$, where $r_H$
is the Hill radius in Eq. (\ref{eq:rhill}).  For planets with mass
$M_p \sim 6 \times 10^{24}$ g and $M_{\star}$ = 1 \msun, $r_H$ = 
0.001 $a$. Thus, dynamical interactions among the largest objects 
are inevitable when $n_P \approx$ 50--100 in a region with a radial 
extent $\Delta a / a \approx$ 0.2 \citep{gol04,kb06}. Many of our 
calculations yield such large numbers of Pluto-mass objects.

To save computer time, we did not calculate the typical long-term evolution 
of hundreds of Plutos using our hybrid evolution code \citep{bk06}. 
However, we can infer the outcome from scaling the results of 
calculations for the formation of the Earth at 1~AU 
\citep[e.g.,][]{cha01,bk06,kb06}. At 0.4--2~AU, dynamical evolution of
100--200 lunar mass objects produces several Earth-mass planets in 
10--30~Myr. The spacing of lunar mass objects in these calculations 
is $\sim$ 70\% of the critical spacing $\Delta a \sim 2 \sqrt{3} r_H$,
similar to the spacing of Pluto-mass objects at late times in our 
calculations at 30--150~AU.  Scaling the evolution times by the ratio 
of orbital periods suggests that 100--200 Pluto mass objects collide 
to form planets with masses $\sim$ 0.1 \mearth\ on 1--3 Gyr timescales
\citep[see also][]{lev01,gol04}.

This analysis suggests that dynamical interactions between large numbers 
of Plutos at 30--150~AU are interesting only for low mass stars. For 
2--3 \msun\ stars, the main sequence lifetimes are shorter than the time 
needed for Plutos to interact, to collide, and to grow into Mars-mass 
planets. For lower mass stars, several test calculations with our hybrid 
code confirm that ensembles of 100--200 Plutos can grow into several 
Mars-mass planets on timescales of 2--5 Gyr\footnote{For $a \gtrsim$ 75~AU,
the escape velocity of Mars-mass planets exceeds the orbital velocity.
Although dynamical interactions among Mars-mass objects can produce 
ejections in these circumstances \citep{gol04}, damping by leftover 
planetesimals limits ejections in our simulations.}. Although this evolution
leads to some extra stirring of leftover low mass planetesimals, there 
are only small changes in the dust production rate and the total mass
in small grains. Thus, dynamical interactions among Plutos have little
impact on our general results.

\subsubsection{Fragmentation parameters}

Fragmentation is another uncertainty in our calculations. We treat 
destructive collisions with an energy-scaling algorithm that uses
(i) the ratio of the center of mass collision energy $Q_c$ to the critical
disruption energy $Q_d^*$ and (ii) a simple power-law size distribution
to apportion ejected material into lower mass bins. Although the 
energy-scaling algorithm yields a reasonable treatment of collisions
in coagulation calculations, the disruption energy $Q_d^*$ sets the 
strength of the collisional cascade. Large $Q_d^*$ leads to a weak
cascade with little debris; small $Q_d^*$ allows a strong cascade 
with significant debris. Because $Q_d^*$ and the size distribution 
of the ejecta set the amount of material in small grains, we now
discuss how our choices for these input parameters affect our results. 

Detailed comparisons of various approaches suggest that the size 
distribution of the ejected mass has little impact on our results. For 
the large collision rates in our calculations, all methods for dividing 
ejected mass among lower mass bins -- including dividing the ejected 
mass equally among 2--3 lower mass bins -- leads to a power-law mass 
distribution with a characteristic slope of $dn/dm \approx -0.8$ 
\citep{doh69,wil94}. Thus, the adopted mass distribution for the ejecta 
is relatively unimportant.

\citet{kb05} and \citet{kbod08} describe how the form of $Q_d^*$ in Eq. 
(\ref{eq:Qd}) impacts collisional evolution of icy objects at 30--150~AU.
Here, we divide $Q_d^*$ into a bulk component (the first term of Eq. 
(\ref{eq:Qd})) and a gravity component (the second term of Eq. (\ref{eq:Qd})).
Gravity provides nearly all of the binding energy for large objects with
$r \gtrsim$ 10~km; the bulk component of $Q_d^*$ provides most of the
binding energy for small objects with $r \lesssim$ 1--10~km.  For icy 
objects with maximum sizes $r_{max} \sim$ 2000~km, stirring never leads to 
orbital motions large enough to disrupt objects with $r \gtrsim$ 10--20~km.
Thus, our choices for the gravity component of $Q_d^*$ have little impact 
on our results. Although both components of the bulk strength -- $Q_b$
and $\beta_b$ -- contribute to $Q_d^*$, quoted uncertainties in $\beta_b$
derived from theoretical simulations lead to unimportant variations in 
$Q_d^*$ as a function of $r$. Thus, we concentrate on $Q_b$.

To quantify the impact of $Q_b$ on our results, we first consider the 
evolution of the dust production rate and the amount of material in 
large and small grains. During runaway growth, the dust production 
rates for models with $Q_b \le 10^4$ erg g$^{-1}$ are 10\%--20\% 
larger than dust production rates for models with $Q_b \ge 10^5$ 
erg g$^{-1}$. At the same time, the total mass in large and small grains 
is $\sim$ 10 times larger for models with small $Q_b$ than for models 
with large $Q_b$. In both cases, models with the smallest initial disk
mass have the largest differences as a function of $Q_b$. During 
oligarchic growth, these differences disappear.  For models with 
$Q_b = 1 - 10^6$ erg g$^{-1}$, the dispersion in dust production rates 
near the time of maximum dust production is $\sim$ 5\% or less for all 
disks around 1--3 \msun\ stars. Although the dispersion in the total mass 
in large and small grains is a factor of $\sim$ 3 during the early 
stages of oligarchic growth, the dispersion in dust masses declines
to 10\% or less at late times when the dust masses reach their
maximum values \citep{kb04b}. 

The time variation in dust production rate and total dust mass as a 
function of $Q_b$ has a simple physical origin \citep[see also][]{kbod08}. 
During runaway and oligarchic growth, the collision energies of small
objects scale with the mass of the largest objects in the grid 
\citep[see also][]{gol04}. Thus, small objects have larger and larger 
collision energies $Q_c$ at later and later evolution times. Because
this feature of the evolution depends only on gravitational stirring,
it is independent of $Q_b$. Throughout the evolution, the mass ejected
in a collision scales with $Q_c / Q_b$ (Eq. (\ref{eq:mej})).  Thus, 
calculations with small $Q_b$ eject more material at early times 
compared to models with large $Q_b$, leading to a large dispersion
in the dust production rate and total dust masses early in the evolution. 
At late times, all calculations produce objects with $r_{max} \approx$ 
1500--2000~km. These large objects stir all leftover small planetesimals 
to large random velocities, where the collision energies $Q_c \gg Q_d^*$
for all $Q_b$. All collisions then lead to copious mass loss, which
eliminates the dispersion in dust production rates and total dust
masses at late times \citep[see also][]{kb04b}.

In addition to the small late-time dispersion in dust production rates
and total dust masses, our results yield negligible differences in the 
late-time fractions of mass in large objects ($r \gtrsim$ 100~km) as 
a function of $Q_b$.  The median radius of the largest object and the 
median number of Pluto mass objects are also independent of $Q_b$. 
Thus, our analysis suggests that the fragmentation parameters have a 
small impact on observable quantities. For low mass disks at $t \lesssim$ 
10~Myr, destructive collisions between planetesimals with small $Q_b$ 
produce more dust than objects with large $Q_b$. Although these 
differences are probably large enough to be observable, they disappear
at late times when planets reach their maximum sizes.

\subsubsection{Treatment of small particles}

Our algorithm for deriving the evolution of small particles with
$r_{min} \lesssim$ 1~m is a final uncertainty in our calculations. To follow 
the evolution of sizes and orbits for large objects in a reasonable amount 
of computer time, we do not calculate the evolution of small particles
directly. Instead, we use the known production rate of small particles
from the detailed calculation $\dot{M}(t)$, an adopted power-law size
distribution, and a simple collision algorithm to evolve the small
particle size distribution with time. 

Because we include radiation pressure and Poynting-Robertson drag in this 
simple treatment of collisional evolution, our predicted size distributions 
consist of three distinct pieces. For particle sizes where the collisional 
timescale is shorter than the timescale for Poynting-Robertson drag, 
$n \propto r^{-0.8}$. For very small sizes where radiation pressure ejects 
grains, we adopt $n \propto r^{-0.8}$ for grains in a constant velocity
outflowing wind (see the Appendix). For intermediate sizes, 
Poynting-Robertson drag can remove grains faster than collisions replenish 
them. Thus, the particle number $n \rightarrow$ 0. To conserve mass,
we solve a continuity equation to derive the number density of grains
dominated by Poynting-Robertson drag.

Although our solution for the evolution of small particles is efficient,
it does not consider how fluctuations in the collision and fragmentation
rates might modify the size distribution. \citet{cam94} note that 
size-dependent fluctuations can produce wavy size distributions for
0.1--10 mm particles. In their simulations of the $\beta$ Pic disk,
\citet{the03} derive steady-state size distributions with substantial
deficits of 0.1--10 mm particles compared to a standard $n \propto r^{-0.8}$
power-law \citep[see also][]{kri06,the07,loh08}. If these deficits are
typical, then our algorithm seriously overestimates the mass in small 
dust grains and thus the infrared fluxes of debris disks.

To check for this possibility, we computed several models with a simple 
version of our multiannulus coagulation code. In these tests, we 
extracted a complete disk model near the peak of the collisional cascade, 
extended the lower end of the size distribution from $r_{min}$ = 1~m to 
1~$\mu$m using a power law $n \propto r^{-\alpha}$, and continued the 
calculation for $\sim$ 100~Myr with collisions and Poynting-Robertson 
drag but without our Fokker-Planck velocity evolution. To estimate the 
range of errors in our simple algorithm, we varied the power law exponent 
for the size distribution, $\alpha \approx$ 0.6--1, the power law exponent 
for the fragmentation law, $\beta_b \approx$ -0.5--0, and the magnitude of
the bulk strength $Q_b$ = 1--$10^5$ erg g$^{-1}$. For a range of disk
masses around a 2 \msun\ star, this approach provides a straightforward
estimate for the accuracy of our results for small particles.

These tests confirm that the simple collision algorithm yields results
reasonably close to more detailed coagulation calculations. For models 
with $\beta_b \approx$ 0, $Q_b \gtrsim$ $10^3$ erg g$^{-1}$, and 
$\alpha \approx$ 0.6--1, the derived size distributions are within 
$\sim$ 20\% of those predicted by the simple model for all particles 
with $r \approx$ 0.01--100 mm.  Although calculations with 
$Q_b \lesssim 10^2$ erg g$^{-1}$ yield larger deviations from the 
simple model, these are small compared to those quoted by \citet{the07}. 
Because particles with small $\beta_b$ are harder to fragment, 
calculations with $\beta_b \lesssim -0.25$ tend to produce smaller 
departures for a power law size distribution than those with 
$\beta_b \gtrsim -0.25$. 

Several features of our calculations combine to minimize wavy size
distributions for small particles in disks at 30--150 AU. Because 
icy planet formation is inefficient, the collisional cascade begins
when most of the initial disk mass is in 1--10 km planetesimals.  
Fragmentation of the leftovers leads to a very large production 
rate of 1~m and smaller objects. Continued fragmentation of these
objects tends to wash out wavy size distributions produced by a
low mass cutoff \citep{cam94,the03}. In our Fokker-Planck treatment 
of velocity evolution, leftover planetesimals are also in dynamical 
equilibrium with larger protoplanets that are `safe' from fragmentation.
Thus, the dust production rate from the collisional cascade is 
well-matched to the dynamical state of the system and tends to 
sustain a power-law size distribution for the smallest objects.

\subsection{Highlights of Icy Planet Formation Around 1--3 \msun\ Stars}

Starting with a disk of 1~km planetesimals, icy planet formation at 
30--150~AU follows the same path for all 1--3 \msun\ stars. This
evolution has six main features.

\begin{itemize}

\item It takes 5--30 Myr for runaway growth to produce an ensemble 
of oligarchs with radii of 500--1000 km. Throughout runaway growth, 
oligarchs stir up the orbits of leftover planetesimals. Collisions 
between leftover planetesimals produce more and more debris.

\item From $\sim$ 10 Myr to the main sequence turnoff, planets slowly 
grow to a characteristic radius.  For a broad range of input parameters, 
the maximum size of an icy planet is $\sim$ 1750~km at 30--150 AU. 
Because the timescale for planet formation at 100--150 AU is 
similar to the main sequence lifetime of a 1--3 \msun\ star, the 
inner disk contains more 1500--2000~km planets than the outer disk.

\item As planets grow slowly, a collisional cascade grinds leftover 
planetesimals to dust. Early on, radiation pressure ejects the 
smallest grains in an outflowing wind. Later, Poynting-Robertson 
drag also removes larger grains from the disk. In our calculations,
radiation pressure removes roughly twice as much mass from the
disk as Poynting-Robertson drag.  The timescale for the collisional 
cascade to remove leftover planetesimals is close to the main sequence 
lifetime of the central star. Thus, the cascade removes more material 
from the inner disk than from the outer disk.

\item Icy planet formation is inefficient. In our calculations,
icy planets with radii exceeding 1000~km contain $\lesssim$ 3--4\% 
of the initial mass in solid material. Objects with radii $\sim$ 
100--1000~km contain $\sim$ 2--3\% of the initial mass. Because
short stellar lifetimes limit the growth of planets in the outer
disk, the mass in large objects declines linearly with increasing 
distance from the central star. Thus, the inner region of the disk 
contains many more Pluto-mass objects than the outer region.

\item The dust produced by the collisional cascade is observable.
For disks around 1--3 \msun\ stars, the maximum mass in small dust
grains with radii of 1~$\mu$m to 1~mm is $\sim$ 1--2 lunar masses.
This mass is comparable to the masses derived for the most luminous 
debris disks around A-type and G-type stars. The time evolution of 
the dust production rate and the mass in small dust grains suggest 
the dust luminosity declines with time.

\item Dusty debris is a signature of the formation of a planetary
system. This debris is present throughout the lifetime of the 
central star.

\end{itemize}

\section{DEBRIS DISK EVOLUTION}

To convert our derived size distributions into observable quantities, 
we perform a radiative transfer calculation.  For each evolution time 
$t$, we derive the luminosity $L_{\star}$ and effective temperature
$T_{\star}$ of the central star from the $Y^2$ stellar evolution models
\citep{dem04}.  We then compute the optical depth $\tau (a)$ of each 
annulus in our model grid. The optical depth allows us to derive the 
fraction of the stellar luminosity $L_{\star}$ absorbed by each annulus. 
For each grain size in each annulus, we derive an equilibrium grain 
temperature $T(r,a)$ and an emitted spectrum. Summing the emitted 
spectra over $r$ and $a$ yields the predicted spectral energy 
distribution (SED) and the total dust luminosity $L_d$ as a function 
of time. The Appendix describes this calculation in more detail 
\citep[see also][]{kb04b}.

In our calculation of observable quantities, the most important input
parameters are the smallest stable grain size $r_2$ (also known as the
`blowout' radius; see Backman \& Paresce 1993) and the slope $q$ of the 
emissivity law for small grains.  Although several estimates for the 
minimum grain size suggest $r_2 \approx 0.5-2 M_{\star}^3$~$\mu$m 
for 1--3 \msun\ stars \citep[e.g.][]{bur79,art88,bac93,kim02}, the 
coefficient and the scaling relation are sensitive to the composition, 
internal structure, and radiative properties of the grains. Because 
observations allow few tests of this relation, we adopt $r_2$ = 1~$\mu$m 
for all stars. If more luminous stars have larger $r_2$, our calculations 
overestimate the optical depth in small grains. Thus, we overestimate 
the mid-IR and submm excesses. For the emissivity, submm data suggest 
$q \approx$ 0.6--1 from a handful of debris disks \citep{naj05,wil06}.  
To provide some balance for our likely underestimate of $r_2$, we adopt 
$q$ = 1. Grains with smaller $q$ emit more efficiently at longer 
wavelengths; our models then underestimate mid-IR and submm excesses.

To describe the evolution of observable quantities with time, we focus on 
the dust luminosity $L_d$ and the excesses at IR and submm wavelengths.
The fractional dust luminosity $L_d / L_{\star}$ provides a clear measure 
of the relative luminosity of the debris disk. For excesses at specific
wavelengths, we quote the total emission of the disk and the central 
star relative to the emission from the stellar photosphere, 
$F_{\lambda} / F_{\lambda,0}$. With this definition, disks that produce
no excess have $F_{\lambda} / F_{\lambda,0}$ = 1; disks where the excess
emission is comparable to the emission from the central star have
$F_{\lambda} / F_{\lambda,0}$ = 2.

We begin this section with a discussion of excess emission for 
1 \msun\ stars.  After discussing results for 1.5--3 \msun\ stars, we 
conclude this section with a brief summary. To facilitate comparisons of 
our results with observations, Tables \ref{tab:mod-1p0}--\ref{tab:mod-3p0} 
list results for the fractional dust luminosity and excesses at 
24--850~$\mu$m. The paper version lists the first five lines of results for 
$x_m$ = 1/3, 1, and 3. The electronic version includes all results for 
these $x_m$.

\subsection{Evolution for 1 \msun\ stars}

\begin{figure}
\includegraphics[width=7.0in]{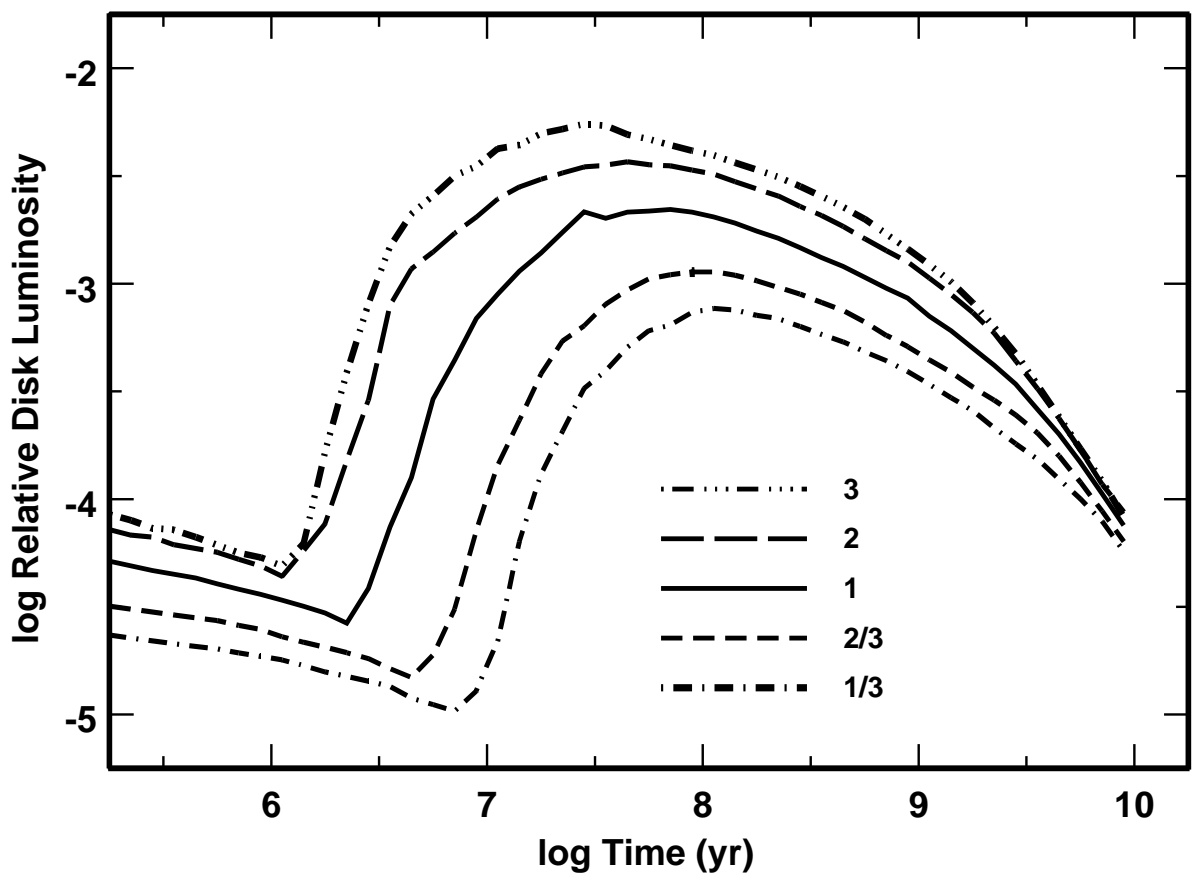}
\figcaption[f14.eps]
{Time evolution of the median $L_d/L_{\star}$ (dust luminosity 
relative to the luminosity of the central star) for disks 
surrounding a 1 $M_{\odot}$ star. The legend indicates
the disk mass in units of the MMSN. More massive disks reach
larger peak dust luminosities earlier than less massive disks.
The typical peak dust luminosity is comparable to the dust 
luminosity of the most luminous debris disks associated with 
solar-type stars.
\label{fig:ldust-1}}
\end{figure}

Fig. \ref{fig:ldust-1} shows the evolution of the fractional dust 
luminosity $L_d/L_{\star}$ for disks surrounding a 1 \msun\ star. 
Early in the evolution, collisions produce mergers instead of debris.
For an ensemble of growing planetesimals, the surface area per unit
mass (and hence the opacity) decreases with time. Thus, $L_d/L_{\star}$
declines with time. Less massive disks have smaller dust masses and 
smaller dust luminosities.  As oligarchic growth begins, the dust 
luminosity rises rapidly and reaches a peak 
$L_d/L_{\star} \approx 2 \times 10^{-3}$ in 30--100~Myr. More massive 
disks reach larger peak luminosities earlier than less massive disks. 
At late times, all disks converge to the same dust luminosity, 
$L_d/L_{\star} \approx 10^{-4}$ at $\sim$ 10 Gyr.

\begin{figure}
\epsscale{1.1}
\plottwo{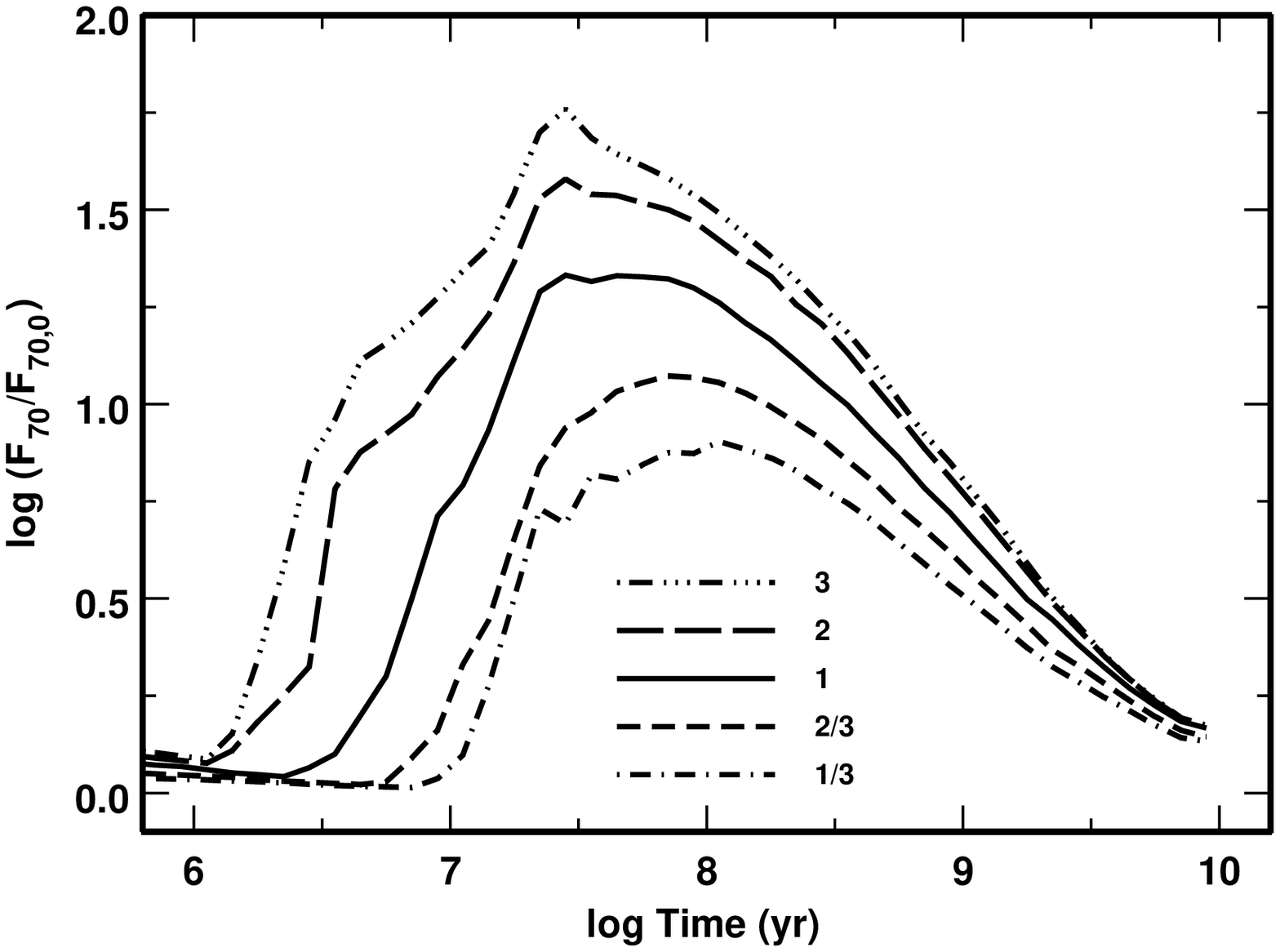}{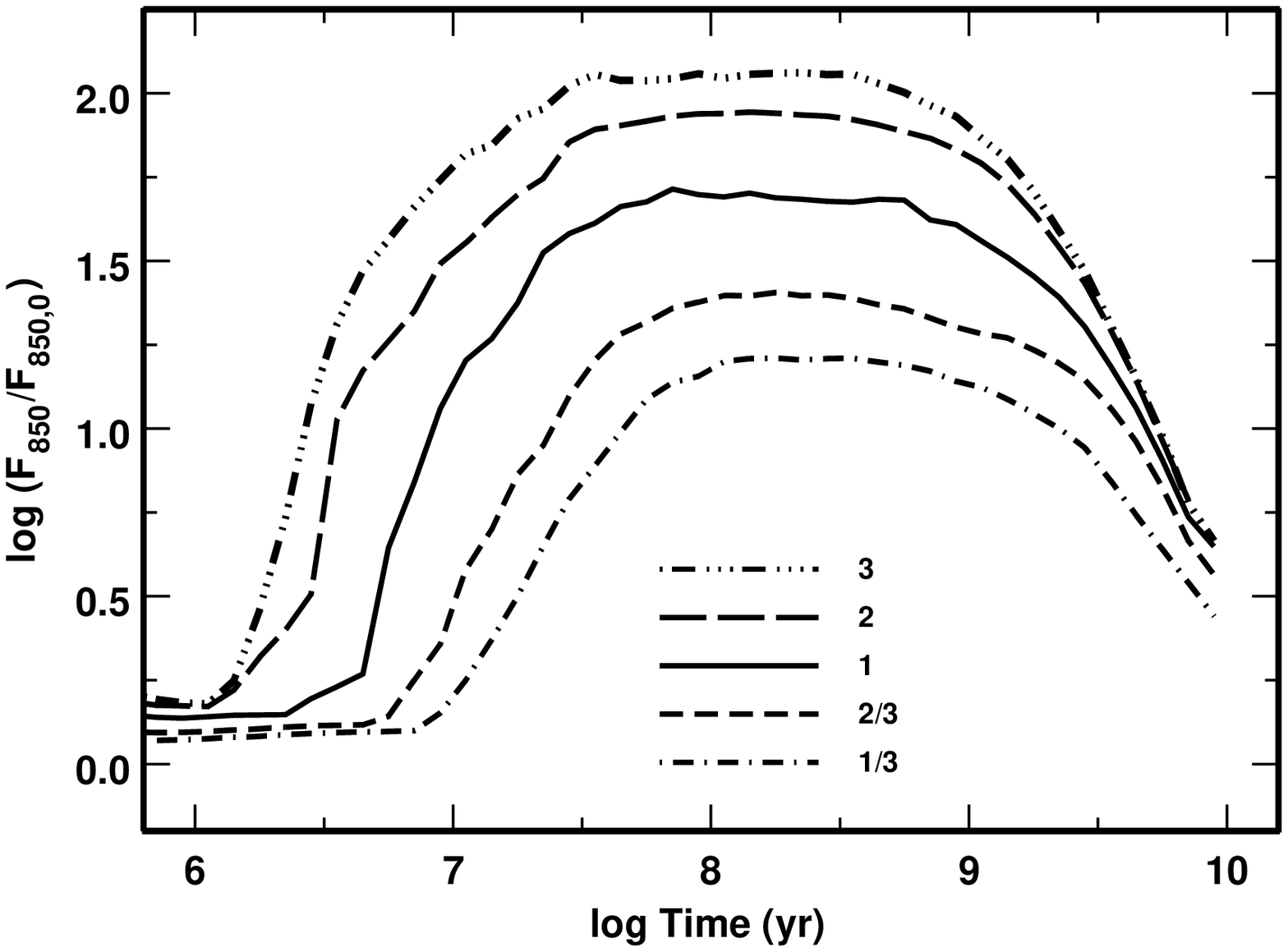}
\figcaption[f15.eps]
{As in Fig. \ref{fig:ldust-1} for the median 70 $\mu$m excess
(left panel) and the median 850~$\mu$m excess (right panel).
At both wavelengths, dust emission begins to increase at 5--10~Myr.
Peak dust emission occurs at 30--100~Myr (70~$\mu$m) and 
100--300~Myr (850~$\mu$m). When the central star evolves off the
main sequence, the typical excess at 70~$\mu$m (850~$\mu$m) is 
$\sim$ 2--3 (3--10) times the flux from the central star.
\label{fig:f70x850-1}}
\end{figure}

Despite their small fractional dust luminosities, these disks produce 
large excesses at 70~$\mu$m (Fig. \ref{fig:f70x850-1}; left panel). 
For massive disks with $x_m$ = 2--3, the 70~$\mu$m excess rises from 
$F_{70}/F_{70,0}$
$\sim$ 2--3 at 3~Myr to $F_{70}/F_{70,0}$ $\sim$ 30--50 at 30~Myr.
Lower mass disks with $x_m$ = 1/3 to 1/2 produce smaller peak excesses
at later times, $F_{70}/F_{70,0}$ $\sim$ 10 at $\sim$ 100~Myr. For all
disk masses, the 70~$\mu$m excess is close to its maximum value for
a short period when planet formation peaks in the inner disk. The
excess then declines with time. The rapid decline leads to modest 
excesses at late times, $F_{70}/F_{70,0}$ $\sim$ 3--5 at $\sim$ 1 Gyr 
and $F_{70}/F_{70,0}$ $\sim$ 2 at $\sim$ 3--10 Gyr.

The large excesses at 70~$\mu$m are a simple consequence of blackbody 
radiation from small grains at 30--50 AU around a solar-type star. 
With typical temperatures $\sim$ 40--60~K, these grains emit most of
their radiation at $\sim$ 50--70~$\mu$m.  The peak flux from a blackbody 
grain at temperature T is $F_{\lambda,max} \propto T^5$ \citep{all76}. 
Setting the total disk luminosity $L_d \propto T^4$ yields 
$F_{\lambda,max} \propto L_d T$.  At this wavelength, the stellar flux 
follows a Rayleigh-Jeans law, $F_{\lambda} \propto T_{\star} \lambda^{-4}$. 
Combining these relations and including a correction factor for 
inefficient radiation from small grains yields a simple relation for the 
70~$\mu$m flux from the disk and central star,
\begin{equation}
F_{70}/F_{70,0} \approx 1 + 10 \left ( \frac{L_d/L_{\star}}{10^{-3}} \right ) ~ .
\end{equation}
For the luminosities in Fig. \ref{fig:ldust-1}, this relation accounts 
for the 70~$\mu$m excesses in Fig. \ref{fig:f70x850-1} at all times.

At longer wavelengths, the disks in our calculations achieve larger peak 
excesses and stay close to the peak excess for longer periods of time 
(Fig. \ref{fig:f70x850-1}; right panel). 
Disks reach their peak excesses at 850~$\mu$m on timescales similar to 
those at 70~$\mu$m, $\sim$ 30~Myr for disks with $x_m$ = 2--3 and 
$\sim$ 100~Myr for disks with $x_m$ = 1/3. The fractional excesses at 
850~$\mu$m are a factor of $\sim$ 2 larger than the excesses at 70~$\mu$m. 
Because the emitting region evolves more slowly, these disks are 
luminous for $\sim$ 1 Gyr and then decline with time. Despite the 
rapid decline, the excesses are significant at late times, with 
$F_{850}/F_{850,0}$ $\sim$ 3--10 at $\sim$ 3--10 Gyr.

\begin{figure}
\epsscale{1.1}
\plottwo{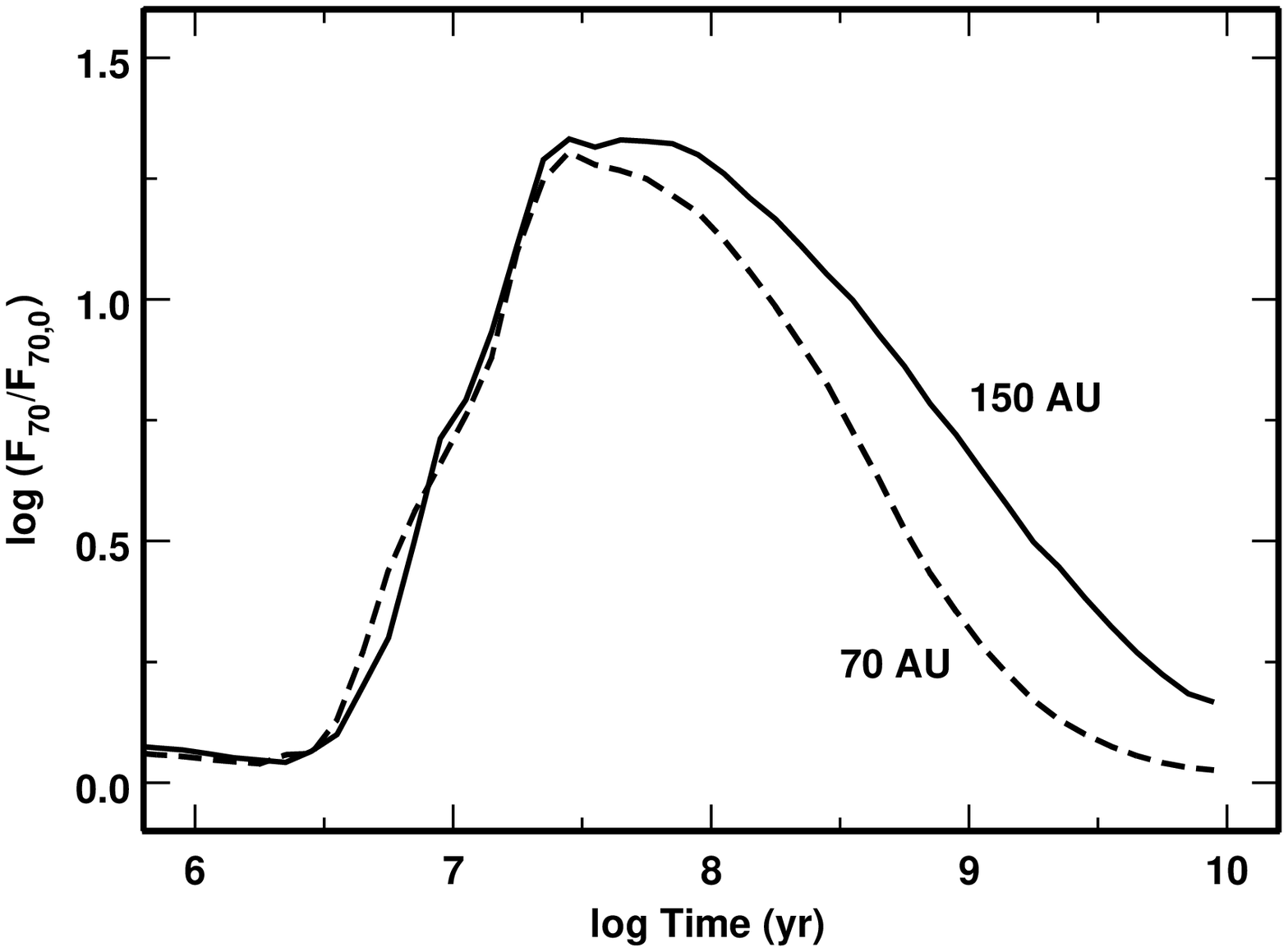}{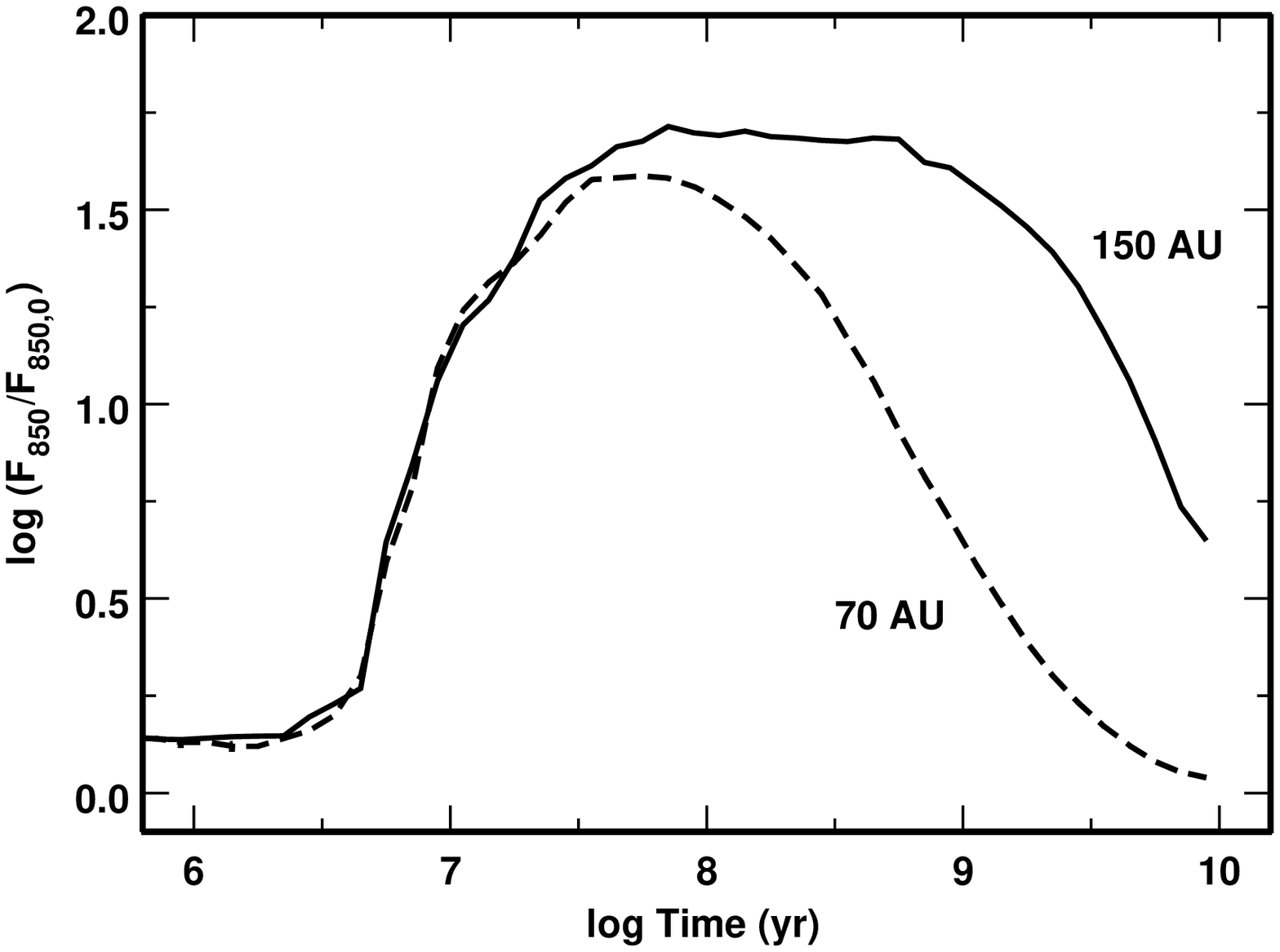}
\figcaption[f16.eps]
{Variation of dust excess with disk size.
{\it Left panel:} time evolution of the median 70 $\mu$m excess for 
MMSN disks with outer radii of 70~AU (dashed line) and 150~AU (solid line).
At late times, smaller disks produce smaller IR excesses.
{it Right panel:} as in the left panel for the median 850 $\mu$m excess.
\label{fig:f70x850-2}}
\end{figure}

The time variation of IR excess also depends on the outer radius of the 
disk.  For solar mass stars, grains at 30--50~AU in the inner disk produce 
most of the flux at 50--100~$\mu$m. Thus, disks with outer radii of 
70~AU and 150~AU produce similar 70~$\mu$m excesses for $t \lesssim$ 
30~Myr (Fig. \ref{fig:f70x850-2}; left panel). Once the smaller disk 
reaches peak 
emission, the 70~$\mu$m excess begins a dramatic decline. The larger 
disk maintains the peak excess for $\sim$ 30~Myr and then declines
more slowly with time. For $t \gtrsim$ 100~Myr, the smaller disk is 
a factor of 2--3 fainter at 70~$\mu$m than the larger disk.

The evolution of disks with different sizes is more dramatic at 
850~$\mu$m (Fig. \ref{fig:f70x850-2}; right panel). For typical 
grain temperatures $\sim$ 20--60 K, long wavelength emission from 
the disk follows the Rayleigh-Jeans tail of a set of blackbodies. 
The radiation from each disk annulus is then $\propto a^2T$. Because 
the outer disk produces more long wavelength 
emission than the inner disk, the 850~$\mu$m excess scales with the outer 
disk radius. For $t \gtrsim$ 1 Gyr, we derive $F_{850}/F_{850,0} \propto$
$a_{out}^n$, where $a_{out}$ is the outer radius of the disk and 
$n \approx$ 3--4. Thus, doubling the outer disk radius increases the 
predicted 850~$\mu$m excess by a factor of $\sim$ 10 at late times.

\subsection{Evolution for 1.5-3 \msun\ stars}

Several factors change the evolution of the dust luminosity and the IR/submm 
excesses in stars more massive than 1 \msun. More massive stars are hotter; 
for the Y$^2$ stellar evolution isochrones $T_{\star} \propto M_{\star}$
\citep{dem04}. Thus, grains in the inner disks around massive stars are 
warmer, emit more short wavelength radiation, and produce bluer colors than 
disks around less massive stars. More massive stars also evolve faster, 
$t_{ms} \propto M_{\star}^{-3}$.  Because the evolutionary timescales for 
solids in the disk are much less sensitive to stellar mass, 
$t \propto  M_{\star}^{-1}$, massive stars have more dust at the end
of their main sequence lifetime than low mass stars 
(e.g., Fig. \ref{fig:dust4}). Thus, these systems have relatively large 
IR excesses when their central stars evolve off the main sequence.

\begin{figure}
\epsscale{1.1}
\plottwo{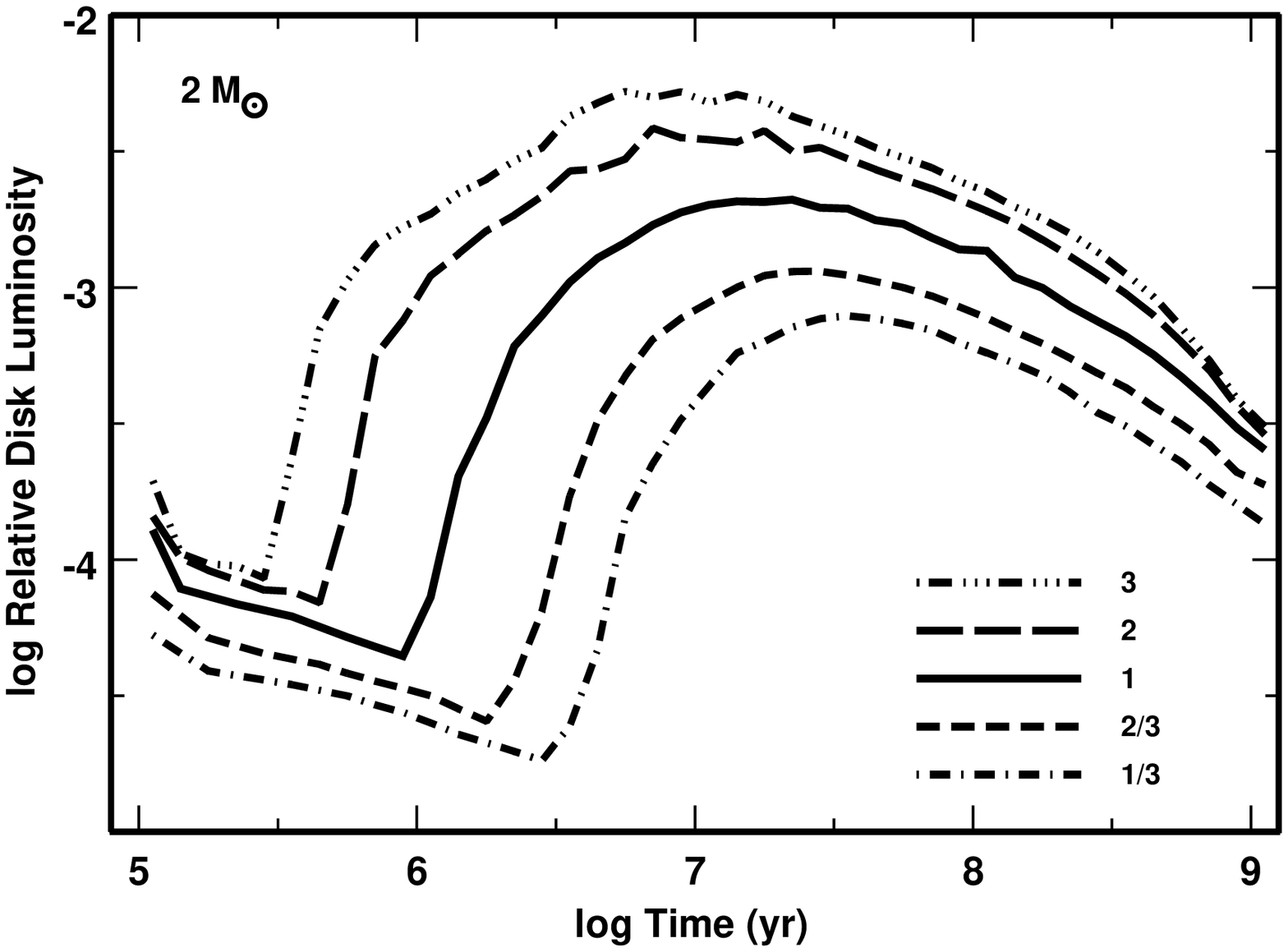}{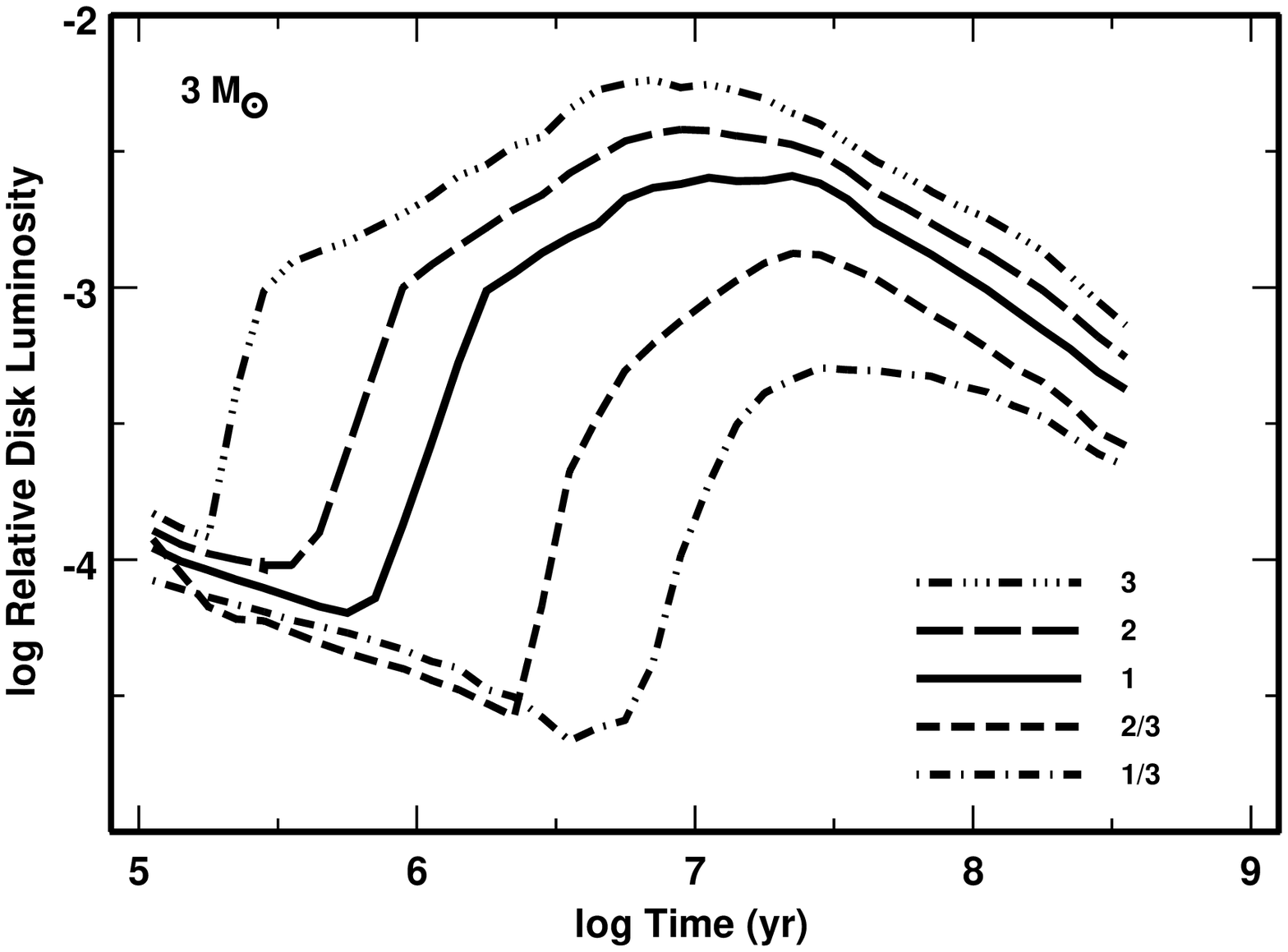}
\figcaption[f17.eps]
{Time evolution of the median $L_d/L_{\star}$ for MMSN disks surrounding 
2 $M_{\odot}$ stars (left panel) and 3 $M_{\odot}$ stars (right panel). 
The legend indicates the initial disk mass in units of the scaled MMSN. 
The typical maximum dust luminosity, $L_d/L_{\star} \sim 10^{-3}$ is
comparable to the dust luminosity of the brightest debris
disks around A-type stars.
\label{fig:ldust-2}}
\end{figure}

To compare the evolution of dust emission in debris disks around 
1--3~\msun\ stars, we begin with the evolution of the dust luminosity
(Fig. \ref{fig:ldust-2}). Planets grow faster around more massive stars; 
thus, the dust luminosity rises earlier for more massive stars. Once 
the collisional cascade begins, the timescale to reach the peak 
luminosity depends only on the initial disk mass and the stellar mass,
\begin{equation}
t_{d,max} \approx 25~x_m^{-2/3}~\left ( \frac{2 ~ M_{\odot}}{M_{\star}} \right ) {\rm Myr} .
\label{eq:tlmax}
\end{equation}
This timescale is similar to the timescale required to produce the 
first Pluto-mass object in the inner disk (Eq. (\ref{eq:t1000allm})).
The peak luminosity depends only on the initial disk mass
\begin{equation}
L_{d,max}/L_{\star} \approx 2 \times 10^{-3} x_m ~ .
\label{eq:ldisk}
\end{equation}
The luminosity remains close to the peak for $\sim$ 10--30~Myr and 
then declines with time.

Stellar evolution has a clear impact on the evolution of the dust luminosity. 
For 1 \msun\ stars, the dust luminosity declines by a factor of $\sim$ 20 
before the star evolves off the main sequence (Fig. \ref{fig:ldust-1}).
For 3 \msun\ stars, the typical decline in $L_d/L_{\star}$ is only a 
factor of $\sim$ 4. Because debris disks have roughly the same peak 
luminosities, an ensemble of debris disks around middle-aged low mass 
stars should be systematically less luminous than disks around 
middle-aged high mass stars.

\begin{figure}
\includegraphics[width=7.0in]{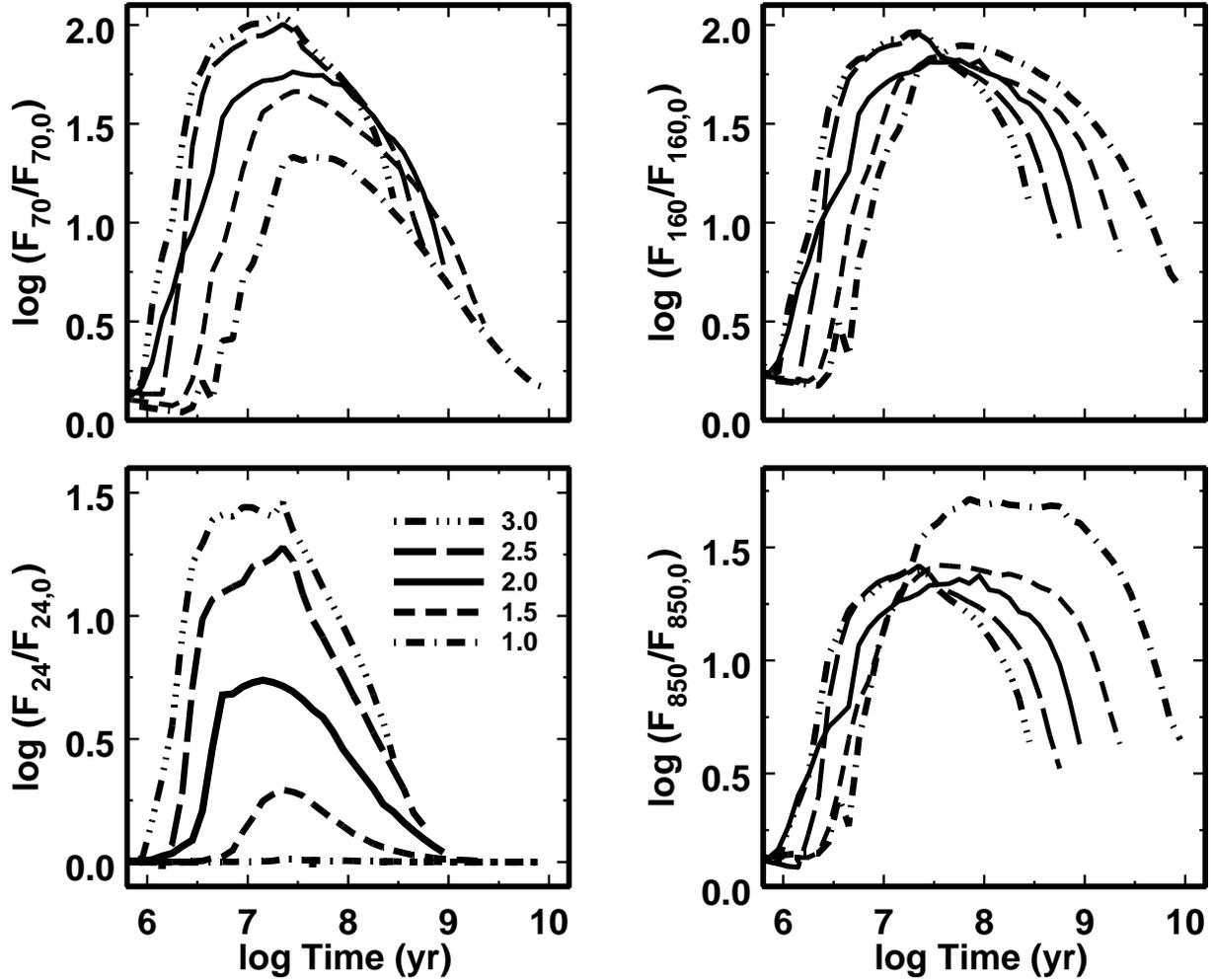}
\figcaption[f18.eps]
{Time evolution of median IR excesses for MMSN disks around 
1--3~$M_{\odot}$ stars. The legend in the lower left panel 
indicates the stellar mass in solar masses for each curve in 
all panels.
{\it Lower left panel:} 24~$\mu$m excess. 
{\it Upper left panel:} 70~$\mu$m excess. 
{\it Upper right panel:} 160~$\mu$m excess.
{\it Lower right panel:} 850~$\mu$m excess.
At 24~$\mu$m, the peak excess increases dramatically with the
temperature of the central star. Thus, hotter stars produce 
much larger 24 $\mu$m excesses. At longer wavelengths, the
magnitude of the excess is correlated with the mass of the 
central star.  Roughly independent of stellar mass, the magnitude 
of the excess at 24--850~$\mu$m peaks at 10--30~Myr as observed 
in debris disks around A-type stars \citep{cur08a}. At late times,
the 160--850~$\mu$m excesses for all stars are $\sim$ 3--5 times 
the flux from the stellar photosphere. 
\label{fig:f24-850}}
\end{figure}

Stellar physics also produces dramatic differences in the behavior
of the 24~$\mu$m excess with stellar mass (Fig. \ref{fig:f24-850};
lower left panel).
At 30--50~AU, the grain temperatures range from $\sim$ 40--60~K 
for 1 \msun\ stars to $\sim$ 80--120~K for 3 \msun\ stars. For these
temperatures, radiation at 24~$\mu$m is on the Wien side of the 
blackbody peak and thus varies exponentially with temperature.  Our 
calculations for 1 \msun\ stars produce very little 24~$\mu$m radiation 
from material at 30--150 AU. However, the peak 24~$\mu$m excesses reach 
$F_{24}/F_{24,0} \sim$ 20 for disks around 3 \msun\ stars. For all
1--3 \msun\ stars, our results yield 
\begin{equation}
{\rm log}~F_{24,max}/F_{24,0} \approx 0.74~(M_{\star} - 1 M_{\odot}) + 0.27~(M_{\star}/M_{\odot})~{\rm log}~x_m ~ .
\label{eq:f24}
\end{equation}
This maximum flux occurs at roughly the same time as the peak dust luminosity.

At longer wavelengths, the excesses are less sensitive to stellar mass.
Radiation at 70~$\mu$m is at the blackbody peak for grains in the inner 
disk. Thus, the inner disk produces most of the 70~$\mu$m excess.  The 
peak excess is then independent of the stellar luminosity and depends 
only on the total disk mass (Fig. \ref{fig:f24-850}; upper left panel),
\begin{equation}
F_{70,max}/F_{70,0} \approx 55~x_m^{0.90} \left ( \frac{M_{\star}}{2~M_{\odot}}\right )  ~ .
\label{eq:f70}
\end{equation}
For grains at 30--150~AU, radiation at longer wavelengths is on the 
Rayleigh-Jeans tail of the blackbody. Thus, observations at 160--850~$\mu$m
probe material throughout the disk. At 160~$\mu$m, extra emission from 
hotter grains in disks around more massive stars is balanced by more flux 
from the hotter central star. Thus, the excess is independent of stellar
mass and depends only on $x_m$ (Fig. \ref{fig:f24-850}; upper right panel), 
\begin{equation}
F_{160,max}/F_{160,0} \approx 65~x_m^{0.90}  ~ .
\label{eq:f160}
\end{equation}
At 850~$\mu$m, grains in disks around 1 \msun\ stars are closer to their
blackbody peaks than grains in disks around more massive stars. Thus,
the 850~$\mu$m excesses are larger for 1 \msun\ stars 
(Fig. \ref{fig:f24-850}; lower right panel),
\begin{equation}
F_{850,max}/F_{850,0} \approx \left \{
\begin{array}{l l}
40~x_m^{0.9} & ~~~~~M_{\star} = 1~M_{\odot} \\
\\
25~x_m^{0.9} & ~~~~~M_{\star} = 1.5-3.0~M_{\odot} \\
\end{array}
\right \}
\label{eq:f850}
\end{equation}
At 70--850~$\mu$m, the time of peak excess is similar to the maximum
in the dust luminosity. Thus, all excesses at 24--850~$\mu$m peak at 
$\sim$ 20--30~Myr for 1--3 \msun\ stars.

Following the peak in the excess at 20--30~Myr, the relative disk 
luminosity and the excesses at 24--850~$\mu$m decrease monotonically 
with time. For this evolution, simple debris disk models predict a 
power law decline, $L_d/L_{\star} \propto t^{-n}$ with $n \approx$ 
1--2 \citep[e.g.][]{dom03,wya07a,wya07b}. To compare our results 
for $t \gtrsim t_{d,max}$ with these predictions, we adopt
\begin{equation}
f_d \equiv L_d/L_{\star} \propto t^{-n_d}
\end{equation}
and
\begin{equation}
f_{\lambda} \equiv F_{\lambda}/F_{\lambda,0} \propto t^{-n_{\lambda}}
\end{equation}
and derive the power law exponents $n_d = d~{\rm log}~f_d/d~{\rm log}~t$ 
and $n_{\lambda} = d~{\rm log}~f_{\lambda}/d~{\rm log}~t$ from all of
our calculations as a function of disk mass, stellar mass, and time.

Throughout the evolution of all our debris disk models, $n_d$ changes 
continuously with time. For $t \gtrsim t_{d,max}$, collisions and 
radiation pressure dominate the removal of small grains.  As collision 
rates slowly decline with time, the exponent increases slowly from 
$n_d \approx$ 0 to $n_d \approx$ 1. When the central star approaches
the end of its main sequence lifetime, Poynting-Robertson drag starts 
to dominate collisions. The disk luminosity then decreases rapidly; 
$n_d$ increases from $\sim$ 1 to $\sim$ 2. Because most systems are 
collisionally-dominated, our calculations yield a typical 
$n_d \approx$ 0.6--0.8.

For $\lambda \approx$ 24--850~$\mu$m, the exponents $n_{\lambda}$ 
follow the evolution of $n_d$. Because collision rates are larger 
in the warmer, inner disk than in the colder outer disk, $n_{\lambda}$ 
increases slowly with $\lambda$. Thus, the typical $n_{24} \approx$ 
0.6--0.8 is smaller than the typical $n_{850} \approx$ 0.8--1.0.

The exponents $n_d$ and $n_{\lambda}$ are somewhat sensitive to the 
disk mass and the stellar mass. At fixed stellar mass, more massive disks 
evolve faster. Thus, $n_d$ changes faster for more massive disks and 
is larger at the main sequence turnoff. For fixed disk mass, lower
mass stars live longer and have more time to reach the Poynting-Robertson
drag-dominated regime. Our results suggest a 0.1--0.2 range in $n_d$ and 
$n_{\lambda}$ for a factor of 10 range in $x_m$ and a factor of 3 range
in stellar mass. 

\begin{figure}
\includegraphics[width=7.0in]{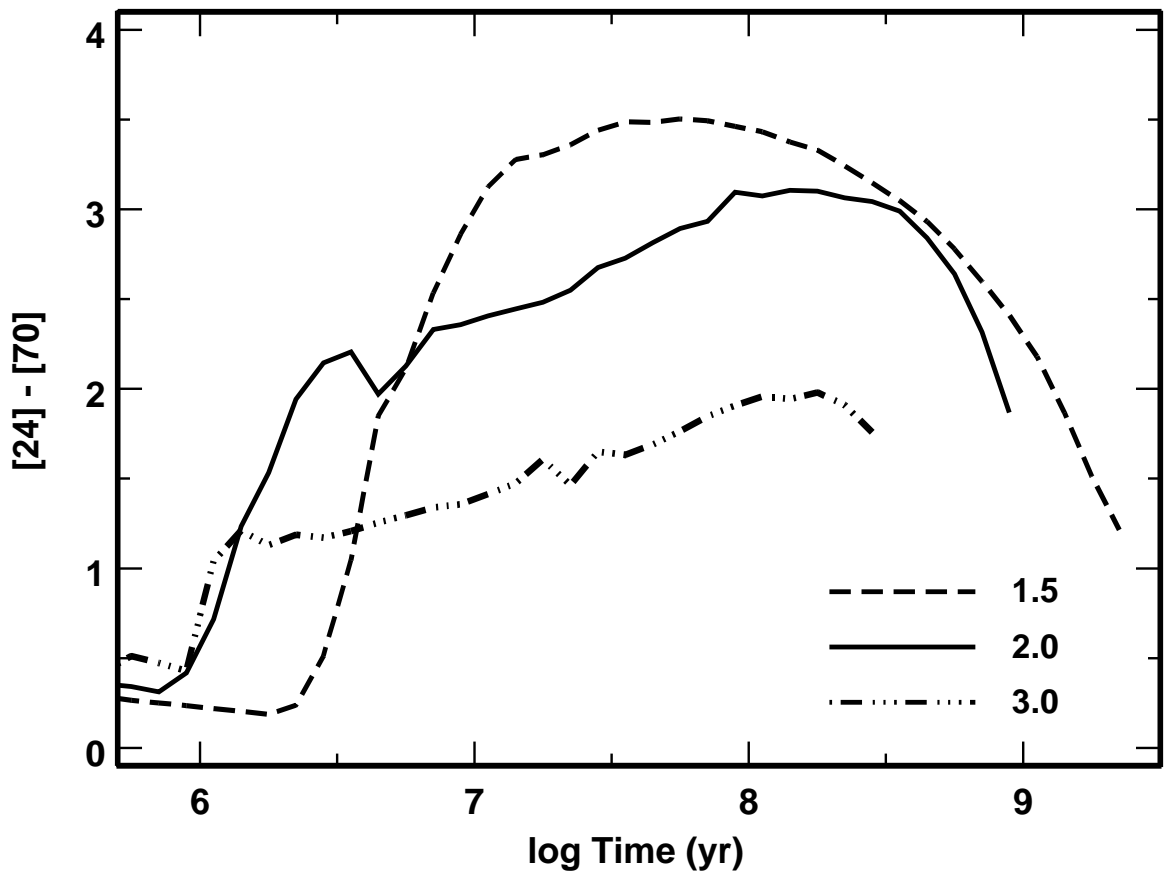}
\figcaption[f19.eps]
{Time evolution of the median [24]--[70] color as a function 
of time for 1.5 \msun\ (dashed line), 2.0 \msun\ (solid line), 
and 3.0 (triple dot-dashed line) \msun\ stars. 
Debris disks around lower mass stars have redder [24]--[70] 
colors than disks around more massive stars.
For massive stars ($\gtrsim$ 2 \msun),
the [24]--[70] color increases slowly throughout the
main sequence lifetime and then declines just before 
central star evolves off the main sequence.
For lower mass stars, the [24]--[70] color reaches a
broad maximum at 300~Myr to 1 Gyr and then declines.
\label{fig:color1}}
\end{figure}

In addition to excesses at specific wavelengths, the evolution of color 
excesses yield interesting trends with stellar mass and time.  Because 
the 24~$\mu$m excess is sensitive to stellar mass, the [24]--[70] color 
cleanly distinguishes debris disks around stars of different masses (Fig. 
\ref{fig:color1}). For 2--3 \msun\ stars, \2470\ rises rapidly to
\2470\ $\approx$ 1--2 at $\sim$ 1~Myr and then rises slowly throughout
the main sequence lifetime of the central star. For 1--1.5 \msun\ stars,
the color rises later, reaches \2470\ $\approx$ 3--4 at 30--100~Myr, and 
then declines slowly. 

For disks at 30--150~AU, the variation of \2470\ with $M_{\star}$ 
depends solely on the properties of the central star. Because more
massive stars are hotter, their disks are warmer. Warmer disks 
produce bluer colors. Thus, the peak \2470\ scales with $M_{\star}$.

\begin{figure}
\includegraphics[width=7.0in]{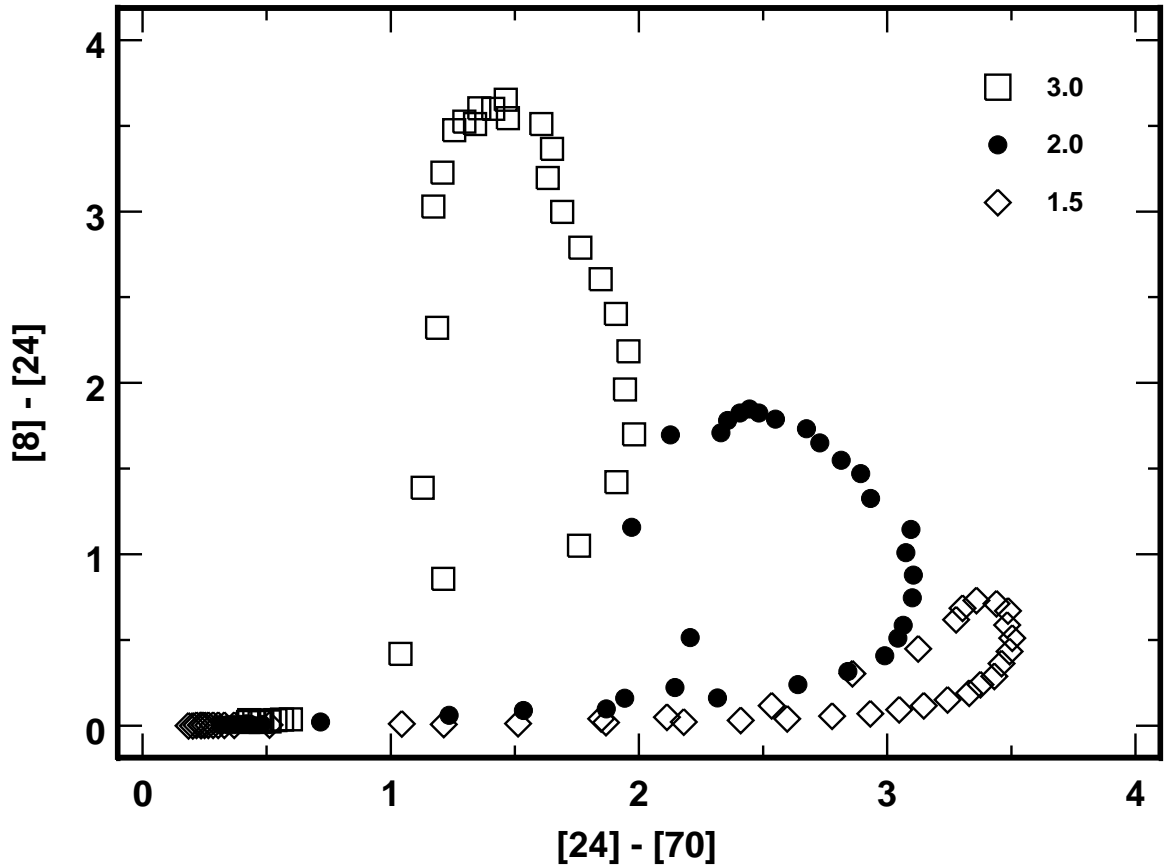}
\figcaption[f20.eps]
{Evolution of debris disks with $x_m$ =  1 in color-color space.  
For icy planet formation at 30--150~AU, debris disks around massive 
stars are hotter than debris disks around less massive stars. Thus, 
debris disks around stars of different masses occupy specific regions 
of the [8]--[24] {\it vs} [24]--[70] color-color diagram.
\label{fig:cc-1}}
\end{figure}

\begin{figure}
\includegraphics[width=7.0in]{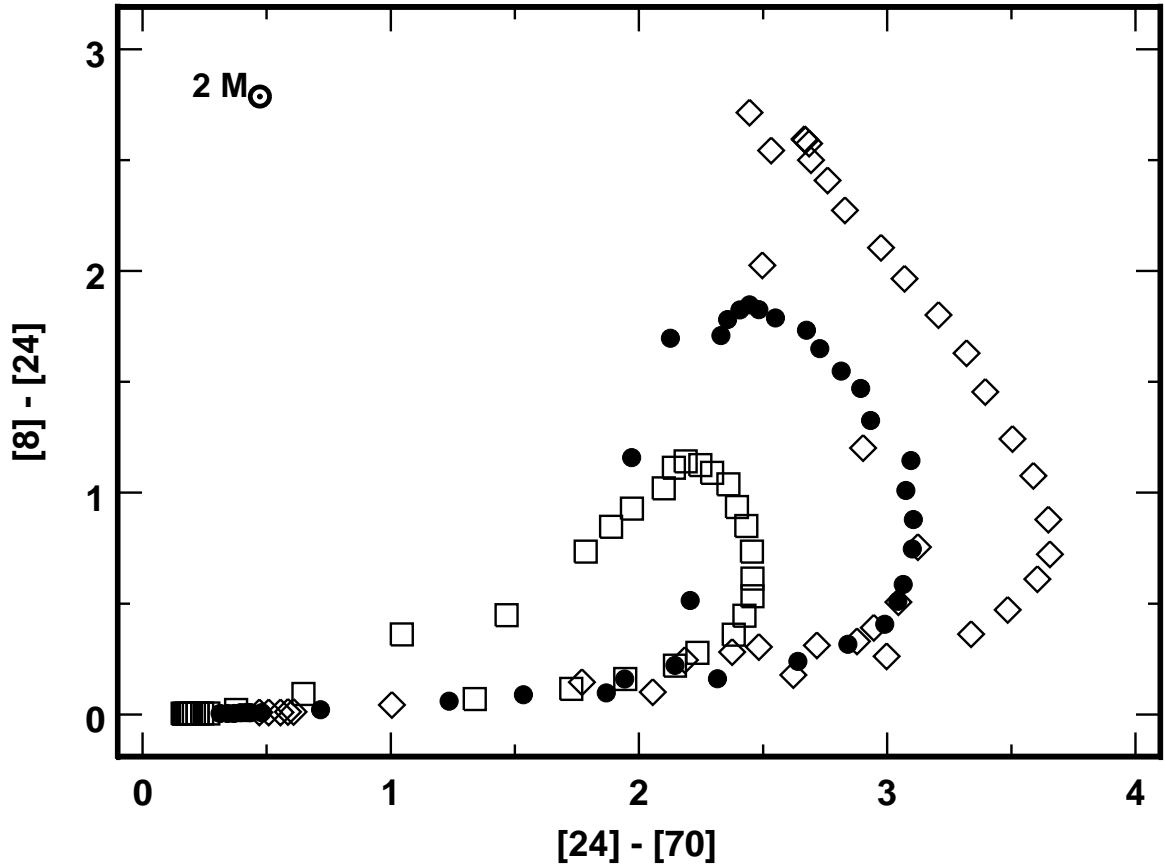}
\figcaption[f21.eps]
{As in Fig. \ref{fig:cc-1} for disks with different initial masses
around 2 \msun\ stars. Boxes: $x_m$ = 1/3, filled circles: $x_m$ = 1,
diamonds: $x_m$ = 3.  Although more massive disks have redder 
[8]--[24] and [24]--[70], the shape of the color-color track is
independent of mass. Thus, the color-color diagram isolates stars
of different masses.
\label{fig:cc-2}}
\end{figure}

The mass-dependent color evolution of debris disks at 30--150 AU 
suggests that color-color diagrams can discriminate masses of the
central star. 
In Fig. \ref{fig:cc-1}, color-color tracks for scaled MMSN around 
1.5~\msun\ stars are clearly distinct from tracks for scaled MMSN
around 2~\msun\ and 3~\msun\ stars. In Fig. \ref{fig:cc-2}, tracks 
for a range of disks around 2~\msun\ stars define a triangle-shaped
locus distinct from the tracks for 1.5~\msun\ and 3~\msun\ stars. 

To establish a triangular debris disk locus for each stellar mass, 
we define two vectors. Adopting a vertex, x$_0$,y$_0$, the upper
boundary of the locus is a vector connecting the vertex with an
upper point, x$_u$,y$_u$. The lower boundary is a second vector
connecting the vertex with a lower point, x$_l$,y$_l$. 
Table \ref{tab:cc-loci} lists our results for the vertex and the
upper/lower points as a function of stellar mass. For each stellar 
mass, colors for debris disks at 30--150~AU lie within the area 
defined by the two vectors. More massive disks produce redder 
colors. Within each locus, the initial disk mass scales with 
distance from the vertex.

When dust inside $\sim$ 30~AU produces a small IR excess, this color-color 
diagram provides a useful discriminant of stellar mass. For disks at 
30--150 AU around 1--3 \msun\ stars, the typical [5.8]--[8] color is 
small, with [5.8]--[8] $\lesssim$ 0.1 at all times. Predicted colors
for terrestrial debris disks are much larger. For 3 \msun\ (1.5 \msun)
stars, we predict maximum colors [5.8]--[8] $\sim$ 0.5--1 (0.2--0.5)
\citep[e.g.,][2005]{kb04a}.  Thus, mid-IR color-color diagrams are 
useful diagnostics of the outer disk for [5.8]--[8] $\lesssim$ 0.1.

\subsection{Summary}

Planet formation and stellar evolution combine to produce several robust 
trends in the time evolution of the dust luminosity and IR/submm excesses
from debris disks around 1--3 \msun\ stars.

For scaled MMSN, the maximum dust luminosity is 
$L_{d,max} \sim 2 \times 10^{-3}$. For an ensemble of debris disks,
the range in the peak dust luminosity scales with the initial mass of 
solid material in the disk. The dust luminosity reaches this peak at 
roughly the time when the first Pluto-mass objects form at 30--50~AU.
Following this peak, the luminosity declines as $t^{-n_d}$ with 
$n_d \approx$ 0.6--0.8. Because lower mass stars have longer main 
sequence lifetimes, debris disks around lower mass stars reach smaller 
fractional dust luminosities at late times.

The IR/submm excesses from debris disks at 30--150~AU are sensitive to 
the mass of the central star. At 24~$\mu$m, disks around more massive 
stars produce larger excesses; disks around stars with $M_{\star} \lesssim$ 
1~\msun\ produce negligible excesses at 24~$\mu$m. At 70~$\mu$m, the
excess is a simple function of the total dust luminosity, 
$F_{70}/F_{70,0} \approx$ 1 + $10^4 L_d/L_{\star}$. At 850~$\mu$m, debris
disks around 1 \msun\ stars produce larger peak excesses than disks around
more massive stars. At late times, however, the typical 850~$\mu$m excess
is fairly independent of stellar mass, with $F_{850}/F_{850,0} \approx$
3--5 for stars with ages $t \sim t_{ms}$.

Among stars with different masses, mid-IR colors provide a sensitive 
discriminant of debris disk evolution when the [5.8]--[8] color is
small (Fig. \ref{fig:color1}--\ref{fig:cc-2}).
For 2--3 \msun\ stars, [24]--[70] slowly becomes redder with the age 
of the central star; for 1--1.5 \msun\ stars, [24]--[70] rises more
rapidly, remains at peak color for 300~Myr to 1 Gyr, and then declines
rapidly with time. For all stars, [8]--[24] and [24]--[70] correlate
with stellar mass. Debris disks around 2--3 \msun\ (1--2 \msun) stars 
have redder (bluer) [8]--[24] colors and bluer (redder) [24]--[70] colors.
Thus, an [8]--[24] {\it vs} [24]--[70] color-color diagram provides a
way to analyze debris disks around stars with different masses (Table
\ref{tab:cc-loci}). 

\section{APPLICATIONS}

To test whether our predictions provide a reasonable match to observations,
we now consider several applications of our models to real systems. For 
these calculations, the broad trends in the evolution of IR excesses and 
colors are sensitive to the physics of planet formation and the collisional 
cascade. Thus, our main goal is to compare our results with observed trends 
of excesses and colors for large samples of main sequence stars observed
with the {\it IRAS}, {\it ISO}, and {\it Spitzer} satellites. In addition
to long-term trends, the absolute level of the excesses depends on $r_2$ 
and $q$. Thus, our second goal is to learn whether our assumptions yield 
mid-IR and submm excesses similar to those observed. 

We begin with an analysis of {\it Spitzer} data for the prototypical
debris disk, Vega. After demonstrating that our models can explain 
the mid-IR fluxes and morphology of {\it Spitzer} images for this system, 
we show that our predictions provide a good match to observations of 
mid-IR excesses for a sample of A-type stars \citep{rie05,su06} and 
a sample of solar-type stars \citep{bei06,hil08}.

\subsection{The Vega Disk}

Observations of Vega with {\it IRAS} first revealed a large excess of 
emission above the A-type photosphere for wavelengths exceeding 
12~$\mu$m \citep{aum84}. The best-fitting single temperature blackbody 
to the {\it IRAS} data yields a temperature of $\sim$ 85 K, a fractional 
luminosity of $\sim 2.5 \times 10^{-5}$ relative to the central star, 
and a radius of $\sim$ 150--200~AU for the emitting material. Because 
the lifetime for small grains at 150--200~AU is much shorter than the 
age of Vega, \citet{aum84} concluded that the grains have sizes larger 
than 1~mm. Thus, Vega provided the first direct evidence for grain 
growth outside the Solar System.

Since the \citet{aum84} discovery, Vega has become the prototypical 
debris disk \citep[e.g.,][]{bac93,art97,lag00}. The debris consists of a 
bright torus with small-scale clumps at 80--1000~AU from the central star 
\citep{hol98,wil02,wliu04,su05} and a smaller disk of debris at $\sim$ 
1~AU from the central star \citep{abs06}. Dust in the small disk is 
hot ($\sim$ 1500~K), luminous ($L_d/L_{\star}$ $\sim 5 \times 10^{-4}$), 
and mostly confined to a narrow ring with a diameter of $\sim$ 0.5--1~AU
\citep{abs06}. This dust might be a result of collisions between larger 
objects at 1~AU or grains lost from icy comets at 80--100~AU in the 
outer disk.

Recently, \citet{su05} analyzed high quality {\it Spitzer} images at 
24, 70, and 160~$\mu$m. Their results demonstrate that the large-scale
debris consists of a bright ring at 80--200~AU and a smooth `halo' that 
extends to $\sim$ 1000~AU at 160 $\mu$m. The halo has an $a^{-2}$ radial 
density profile, consistent with a wind of small grains ejected by radiation
pressure. Fits to the radial surface brightness profiles and the spectral
energy distribution suggest the grains in the wind have sizes of 
1--50~$\mu$m and a total mass of 
$M_{d,1-50} \sim 3 \times 10^{-3}$~\mearth. The grains in the bright ring 
are larger, with typical sizes of $\sim$ 240~$\mu$m, and have a total mass 
of $M_{d,240} \sim 2 \times 10^{-3}$~\mearth\ \citep[see also][]{mar06}.  
For an adopted residence
time of $\sim 10^3$~yr in the wind, the mass in small grains implies
that larger grains in the ring produce smaller dust particles at a rate 
of $\sim 10^{15}$~g~s$^{-1}$. 


To check whether our model predictions can match the MIPS data for Vega, 
we make a simple comparison with median results from several calculations. 
In addition to the observed fluxes at 24, 70, 160, and 850 $\mu$m 
\citep{su05}, we adopt published values for the age \citep[200~Myr;][]{su06}, 
luminosity \citep[37~\lsun;][]{auf06}, and mass \citep[2.3~\msun;][]{auf06} 
of the central star. \citet{su05} separate the observed fluxes into 
contributions from the debris disk and the central star.  The first 
row of Table \ref{tab:Vega} lists these results, along with their 
derived values for the mass in 1--50~$\mu$m dust grains and the dust 
production rate. The rest of Table \ref{tab:Vega} lists predictions 
for four of our debris disk models around 2--2.5~\msun\ stars.

The comparison in Table \ref{tab:Vega} suggests a reasonable match to 
the data. Predictions for the total mass in 1--50~$\mu$m dust grains 
and the fluxes at 160~$\mu$m and at 850~$\mu$m bracket the observed 
values.  Our models also predict a bright ring in the 70--160~$\mu$m 
dust emission at 80--130 AU, close to the observed position of the 
bright ring inferred from Spitzer images \citep[85--200~AU;][]{su05}. The 
mass of 0.1--1~mm particles in this ring, $\sim 3-5 \times 10^{-3}$~\mearth, 
also agrees with the mass in 240~$\mu$m grains derived from the 
Spitzer data.  However, our models overpredict the fluxes at 24~$\mu$m 
and at 70~$\mu$m by a factor of 2--10 and underpredict the dust 
production rate by a similar factor. 

To understand possible origins for the mismatches between the data and 
the models, we consider the evolution of small grains in our calculations. 
When the collisional cascade starts removing material from the disk, the 
most destructive collisions involve grains with comparable masses.  These 
collisions gradually erode the parent objects and produce modest amounts 
of debris in smaller particles. Because (i) the collision timescale is 
much shorter than the timescale for Poynting-Robertson drag and (ii) the
ratio of the radiation force to the gravitational force is 
$\beta_{rad} \equiv F_{rad}/F_{grav} \propto r^{-1}$, erosion continues 
until particles reach a size $r_2$ where $\beta_{rad} \gtrsim$ 0.5--1
\citep{bur79}.  Particles with $r \lesssim r_2$ are ejected. For simplicity, 
we assume $r_2 \approx$ 1~$\mu$m for all of our calculations.

In this picture for the collisional cascade, the \citet{su05} results 
provide a simple solution for the overprediction of the 24--70~$\mu$m 
fluxes in our calculations.  If grains with $r_2 \gg$ 1~$\mu$m are in 
the wind, our `Vega models' underestimate the mass in the wind and 
overestimate the mass left behind in the disk.  More mass at larger 
distances from the central star lowers the optical depth in the inner 
disk, reducing the predicted fluxes at short wavelengths. Thus, increasing
our adopted $r_2$ for Vega models should provide a better match between 
observed and predicted fluxes at 24~$\mu$m and at 70~$\mu$m. For an adopted
$L_{\star}$ = 60 \lsun, \citet{bac93} estimated $r_2$ = 14 $\mu$m.  Scaling 
this result for our adopted $L_{\star}$ = 37 \lsun, $r_2 \approx$ 8.5 $\mu$m.  
Several test calculations with $r_2 =$ 10~$\mu$m yield predicted 24~$\mu$m 
and 70~$\mu$m fluxes close to the observed values.

Reconciling the estimated dust production rate with our predictions requires
a more rigorous analysis of dust production and ejection in the Vega debris 
disk. To derive the dust production rate, \citet{su05} assume that (i) the 
240~$\mu$m grains in the bright ring are bound, with $\beta_{rad} \sim$ 0 
and residence times $\gg 10^3$ yr, and (ii) the 1--50 $\mu$m grains in the 
wind are unbound, with $\beta_{rad} \ge$ 1 and residence times $\sim 10^3$ yr. 
If the collisional cascade proceeds as a gradual erosion of larger objects 
into smaller objects, however, we expect a more gradual transition from
grains with $\beta_{rad} \sim$ 0 to grains with $\beta_{rad} \sim$ 1
\citep[see also][]{bur79,art88}. Allowing the residence time to change
gradually from the bound 240~$\mu$m grains to the unbound 1~$\mu$m grains 
provides a way to lower the apparent dust production rate and to resolve 
the mismatch between our models and the observations.

To provide an alternate estimate for the residence time of grains in the Vega 
disk, we consider the collision times in the ring and the wind.  We adopt the 
dust masses derived from the Spitzer images ($M_{d,1-50}$ and $M_{d,240}$) and 
a typical particle size $\langle r \rangle$.  For a ring at $a \sim$ 150~AU 
with a width 
$\Delta a \sim$ 50~AU, the collision time for a single grain is 
\begin{equation}
t_c \sim 10^2~P~\left ( \frac{\langle r \rangle}{10~\mu{\rm m}} \right ) ~ , 
\end{equation}
where $P$ is the local orbital period in yr \citep{lis87,wet93,kl98}. 
Collision times in the wind are similar. For $P \sim 10^3$ yr at 150 AU, 
the collision times range from $\sim$ $10^4$~yr for 1~$\mu$m grains to 
$2-3 \times 10^6$~yr for 240~$\mu$m grains. 

If we assume that the residence times are comparable to the collision times,
we can construct a self-consistent picture for the collisional cascade in the 
Vega disk. Collisions in the bright ring gradually erode 200--300 $\mu$m grains 
until they reach sizes $\lesssim$ 100~$\mu$m, when they become incorporated 
into the wind. Collisions in the wind gradually erode the smaller grains until 
they reach sizes $\sim$ 1~$\mu$m, when they are ejected rapidly from the system. 

As long as the 200--300 $\mu$m grains are replenished from a reservoir of larger 
grains, this cascade can remain in a quasi-steady state over the main sequence 
lifetime of Vega. The required mass for the reservoir of larger objects is 
$\sim$ 100--1000 times the current mass in 240~$\mu$m grains, $\sim$ 
1--5~\mearth. This mass is small compared to the initial mass of solid 
material in a torus at 80--200 AU in a scaled MMSN, $\sim$ 
50--100~\mearth\ (Eq. (\ref{eq:sigma})).  Because the optical depth of 
this reservoir is small, it produces a small IR excess compared to the 
emission from smaller grains. 

This picture relies on two features of the collisional cascade. We need an
approximate equivalence in mass between the large grains in the ring and
the small grains in the wind. The \cite{su05} mass estimates support this 
feature. We also need a gradual change in grain lifetime from the 
$\sim 10^6$~yr collision timescale of the large grains to the $10^3-10^4$~yr 
dynamical lifetime of the smallest grains. Otherwise, collisions in the
broad torus cannot occur fast enough to maintain the current smooth structure 
of the wind for timescales longer than $\sim 10^3-10^4$ yr. Current 
theoretical analyses support this idea \citep{bur79,art88,tak01,gri07}. 
Numerical simulations of a collisional cascade with a careful treatment 
of the interactions between the radiation field and the small grains 
could test this proposal in detail \citep[e.g.,][]{gri07}.

We conclude that our calculations provide a reasonable match to observations 
of Vega. If we adopt $r_2 \sim$ 10~$\mu$m, the data are consistent with a 
standard collisional cascade within a broad torus at 80--200 AU. The cascade
feeds an outflowing wind of small grains with sizes 1--50~$\mu$m. If the grain 
lifetime changes smoothly from $\sim 10^6$ yr for large grains to $\sim$ 
$10^3-10^4$ yr for small grains, the cascade can maintain the wind 
indefinitely.

\subsection{Debris Disks Around A-type Stars}

Since the discovery of the Vega debris disk, {\it IRAS}, {\it ISO}, and 
{\it Spitzer} observations have revealed debris around dozens of nearby 
A-type stars \citep{bac93,lag00,rie05,su06}. Like Vega, several of these disks 
are resolved and thus provide important information on the radial structure
of the dusty disk \citep[e.g.,][]{smi84,sta04,kal05,mey07,su08}.  Although 
most A-type stars with debris disks are unresolved, the sample is large 
enough to probe the time evolution of debris around 1.5--3 \msun\ stars.  
We now consider whether our calculations can explain this evolution.

To compare our model predictions with observations, we examine data for 
nearby A-type stars from \citet{rie05} and \citet{su06}. \citet{rie05}
combined 24--25~$\mu$m data from {\it IRAS} and {\it ISO} with new 24~$\mu$m
photometry from {\it Spitzer} to investigate the decay of planetary debris 
disks around 266 A-type stars.  \citet{su06} analyze a sample of $\sim$ 160 
A-type stars with high quality 24 $\mu$m and/or 70 $\mu$m data acquired with 
MIPS on {\it Spitzer}. The combined sample has 319 (160) stars with 24~$\mu$m 
(70~$\mu$m) observations, spectral types B7--A6, and ages 5--850~Myr. 
From the \citet{kh95} table of stellar effective temperatures and spectral 
types and the \citet{dem04} stellar evolution tracks, 
$\sim$ 75\% ($\sim$ 85\%) of the stars in the \citet{rie05} \citep{su06} 
sample have masses of 1.7--2.5 \msun. Thus, we compare these data with 
our results for debris disk evolution around 2~\msun\ stars.

\begin{figure}
\includegraphics[width=7.0in]{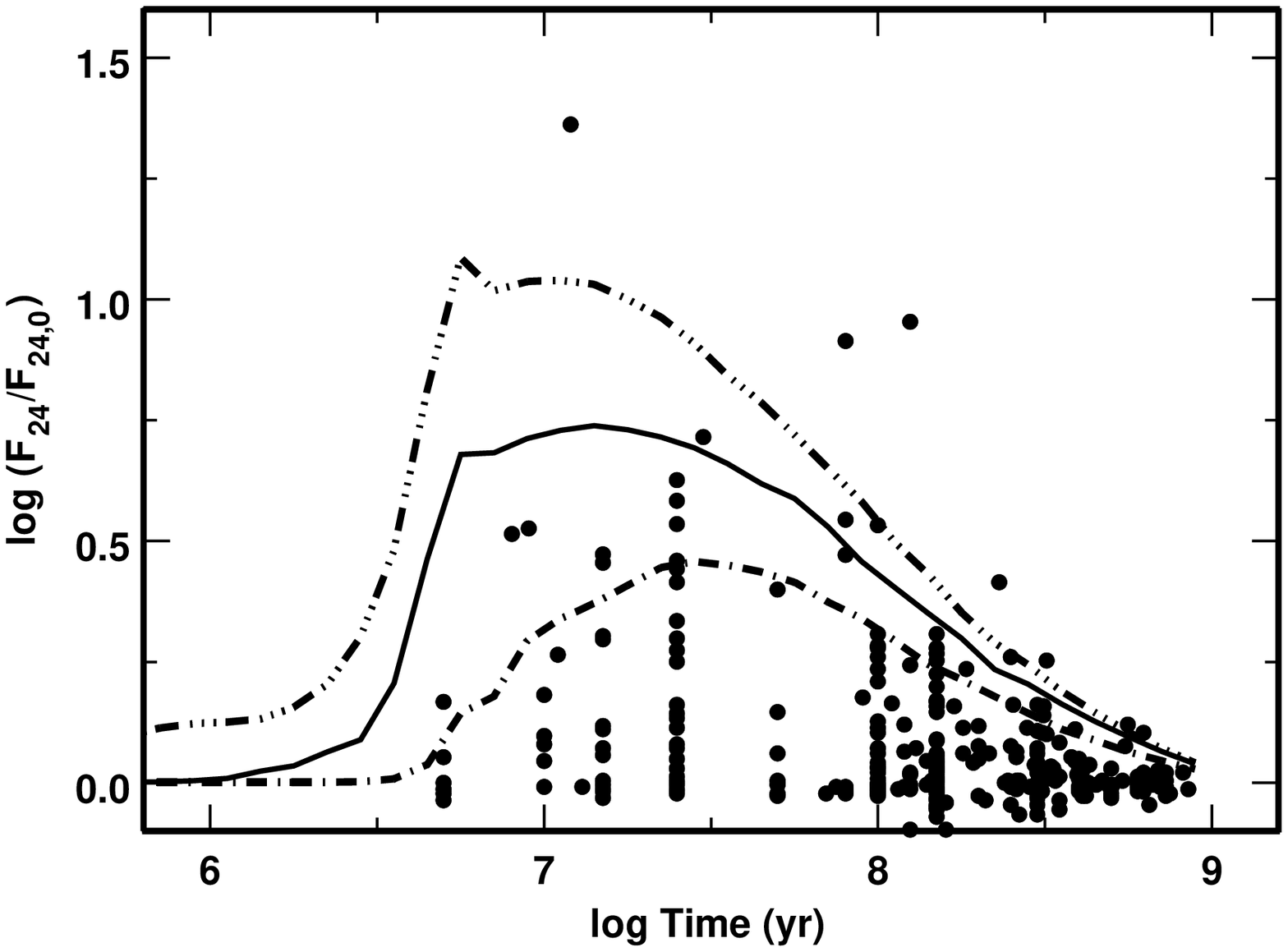}
\figcaption[f22.eps]
{Observations of the 24~$\mu$m excess for nearby A-type stars
with known ages \citep{rie05,su06}. 
The lines show the predicted evolution of the excess for
debris disk models around 2 \msun\ stars
(dot-dashed line: $x_m$ = 1/3; solid line: $x_m$ = 1,
triple dot-dashed line: $x_m$ = 3).
Observations for all but four stars (including one 8 Myr old 
star with log $F_{24}/F_{24,0} \approx$ 2) fall within loci 
defined by our debris disk calculations.  Model predictions 
are also consistent with observational evidence for a peak in 
the 24~$\mu$m excess at 10--20~Myr \citep[see also][]{cur08a}.
\label{fig:su1}}
\end{figure}

\begin{figure}
\includegraphics[width=7.0in]{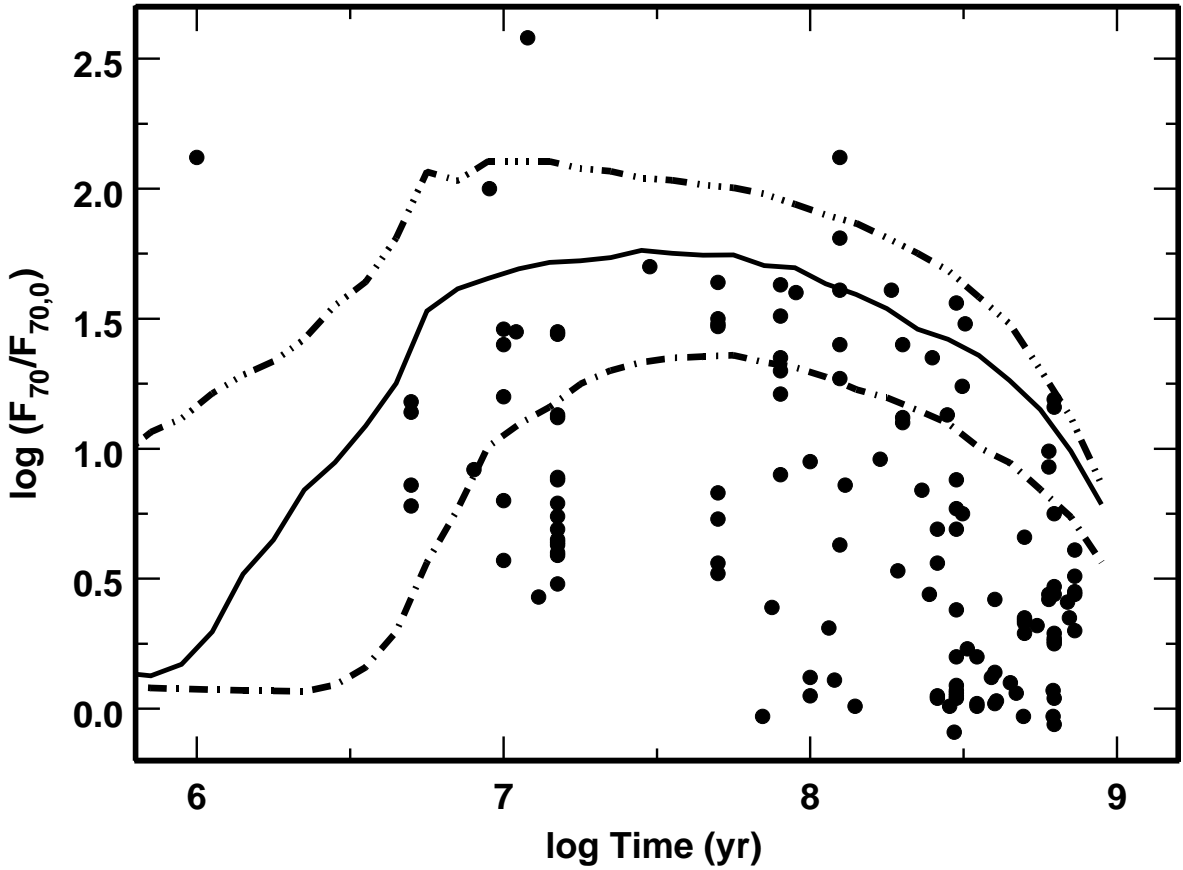}
\figcaption[f23.eps]
{As in Fig. \ref{fig:su1} for the 70 $\mu$m excess.
Observations for all but 2--3 stars fall within the
loci defined by the model tracks.
Consistent with model predictions, the data suggest 
a peak in the 70~$\mu$m excess at 10--20~Myr.
\label{fig:su2}}
\end{figure}

The observed 24--70~$\mu$m excesses of A-type stars show a clear trend
with the age of the star (Fig. \ref{fig:su1}--\ref{fig:su2}). Although 
the statistics are poor, the data suggest a rise in the 24 $\mu$m excess 
at 5--10~Myr. The larger sample of young stars with 70~$\mu$m excesses 
provides better evidence for this rise. At both wavelengths, the excess 
has a broad peak for stars with ages of 10--30~Myr. At later times, the 
excess declines with time as $t^{-n}$ with $n \approx$ 0.5--1 
\citep[see also][]{dec03,gre03,rie05,rhe07a}.

To improve the statistics for 24 $\mu$m excesses around younger stars, 
\citet{cur08a} added {\it Spitzer} data for many young clusters to the
\citet{rie05} sample.  This expanded set of data provides unambiguous 
evidence for a rise in the typical 24~$\mu$m excess at stellar ages of 
5--10~Myr and a robust peak in the excess at stellar ages of 10--15~Myr. 
As in Fig. \ref{fig:su1}, the 24~$\mu$m excesses for this larger sample 
of A-type stars decline with age from $\sim$ 20~Myr to 1 Gyr. 

In addition to the long-term time evolution of mid-IR excess, the data 
also indicate a large range in the 24--70~$\mu$m excess at fixed stellar 
age \citep{rie05,car06,su06,cur08a}.  Although younger stars are more likely 
to have mid-IR excesses than older stars, there are many stars without 
excesses at every age. For ages $\lesssim$ 200~Myr, stars are equally 
likely to have any excess between zero and the maximum excess at that 
age. As stars age, they are less likely to have an excess close to the 
maximum excess at that age. Thus, the dispersion in the excess declines 
with time.

Our calculations provide a good match to the time evolution of the
amplitude of the 24--70~$\mu$m excesses. At both wavelengths, the 
models explain the rise in the amplitude at 5--10~Myr, the maximum 
at 10--20~Myr, and the slope of the power-law decline at late times. 
For models with $x_m$ = 1--3, the predicted excesses also agree with 
the maximum observed excesses.  Although there are a few stars with 
excesses larger than the model predictions, more than 99\% of the 
A stars in this sample have excesses within the range predicted in 
our calculations.

Our calculations also provide a natural explanation for a large range in 
the observed 24--70~$\mu$m excesses at fixed stellar age. At 70~$\mu$m, 
the maximum excess is roughly proportional to the initial disk mass (Eq.
(\ref{eq:f70})). Thus, a factor of ten range in initial disk masses yields
nearly a factor of ten range in the maximum excess at 70~$\mu$m. For
stars with ages 10--300~Myr, the {\it Spitzer} observations suggest a 
factor of $\sim$ 100 range in the 70~$\mu$m excess.  If this range is 
set by the initial disk mass, our models suggest initial disk masses
with $x_m$ = 0.03--3.

Variations in the initial disk radius can also produce a range in 
24--70~$\mu$m excesses at fixed stellar age (e.g., Fig. 
\ref{fig:f70x850-2}). For 2 \msun\ stars 
with ages $\sim$ 400--800~Myr, our results suggest that a factor of
three variation in the outer disk radius (e.g., 50--150 AU) yields
a factor of two (five) variation in the amplitude of the 24 $\mu$m 
(70 $\mu$m) excess.  Although the observed range in the amplitude of 
the 24 $\mu$m excess for older A stars agrees with this prediction, 
the range at 70 $\mu$m is much larger. Thus, variations in the initial 
disk radius can explain some of the observed range of excesses at 
24--70 $\mu$m.

Observations of the youngest stars support a large range in initial
disk masses and disk radii.  Submillimeter observations of dusty 
disks in the nearby Ophiuchus and Taurus-Auriga star-forming regions 
indicate a 2--3 ($\sim$ 1) order of magnitude range in the masses (radii) 
of disks surrounding young stars with typical ages of $\sim$ 1~Myr 
\citep[e.g.,][2007a,b]{ost95,mot01,and05}. Our models with $x_m$
= 3 have disk masses a little smaller than the maximum dust masses 
derived from the submillimeter surveys. Thus, the submillimeter data
imply disks with initial masses $0.01 \lesssim x_m \lesssim$ 5 and
initial disk radii 50 AU $\lesssim a_{out} \lesssim$ 1000 AU. Disks
with this range of initial masses and outer radii can produce the 
range of 24--70~$\mu$m excesses observed around nearby A-type stars.

Variations in the initial surface density distribution can also lead to a 
range in the 24--70~$\mu$m excess. In our calculations, we adopted a 
`standard' surface density relation with $\Sigma \propto a^{-3/2}$. 
Compared to this model, disks with shallower (steeper) surface density 
distributions have relatively more (less) mass at large semimajor axes.
The outer disk has cooler grains than the inner disk; thus, disks
with shallower (steeper) surface density distributions should produce
more (less) flux at longer wavelengths than our standard models. 
Because the relative fluxes at 24~$\mu$m and at 70~$\mu$m provide
a measure of the relative disk masses at different semimajor axes, 
color indices provide a natural measure of the gradient of the surface
density distribution.

\begin{figure}
\includegraphics[width=7.0in]{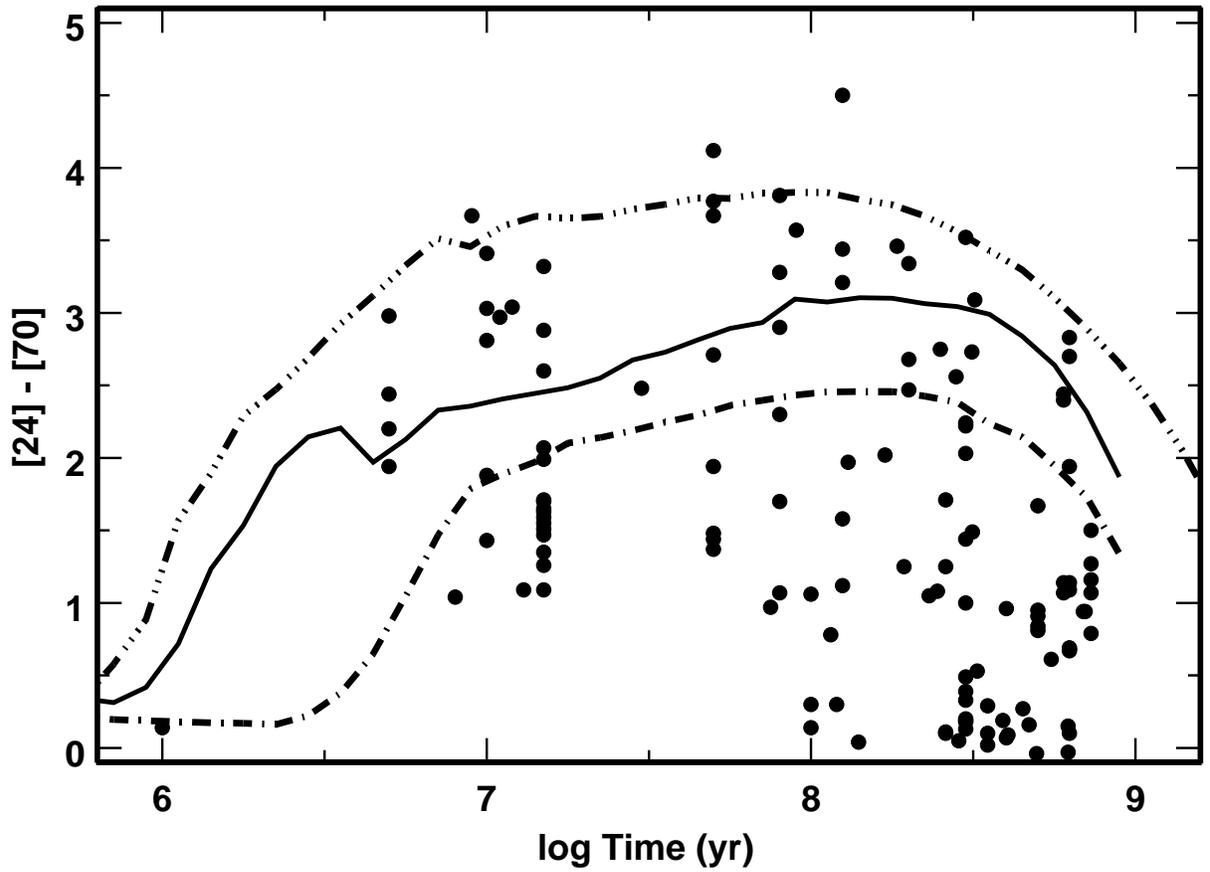}
\figcaption[f24.eps]
{Observations of the [24] - [70] color for nearby A-type 
stars with known ages \citep{su06}.
The lines show the predicted evolution of the excess for
debris disk models around 1.5 \msun\ stars
(triple dot-dashed line: $x_m$ = 2) and for
2 \msun\ stars
(dot-dashed line: $x_m$ = 1/3; solid line: $x_m$ = 1).
Observations for all but 2--3 stars fall within the
model predictions.
\label{fig:su3}}
\end{figure}

Fig. \ref{fig:su3} compares the predicted color evolution for disks
around 2~\msun\ stars with data from \citet{su06}. Although current 
observations do not probe the evolution well at 1--10~Myr, disks
have a large color range, [24]--[70] $\sim$ 1--3, for stars with 
ages $\sim$ 10~Myr. For older stars, the data suggest a slow rise 
in the maximum color from [24]--[70] $\sim$ 3 at 10~Myr to [24]--[70] 
$\sim$ 3.5 at $\sim$ 100~Myr. After $\sim$ 300~Myr, the maximum color
declines. For all stars older than $\sim$ 100~Myr, the range in color
is $\sim$ 3--4 mag.

Our models match the observed color evolution. At all ages, the
predicted colors for calculations with $x_m$ = 3 provide a clear
upper envelope to the observed colors. The predicted colors also
explain the slow rise in the maximum observed color for stars with
ages of 10--100~Myr. To explain the full range in observed colors
for 100~Myr to 1 Gyr old stars, we require disks with initial masses 
$x_m \approx$ 0.01--3. This range is similar to the range required 
for the time evolution of the 24~$\mu$m and 70~$\mu$m excesses.

The good match to the color observations suggests that the typical
initial surface density distribution is reasonably close to our 
adopted $\Sigma \propto a^{-3/2}$. For disks with shallower gradients,
we expect redder colors at later times. A few stars lie above our 
model predictions; however, most stars have bluer colors than models 
with $x_m$ = 3. Thus, few disks in these samples require shallower surface 
density distributions. Disks with steeper surface density distributions
can produce stars with blue colors, [24]--[70] $\sim$ 1, at late times. 
A large sample of A-type stars with [70]--[160] colors and spatially 
resolved observations of the radial dust distributions of these stars 
would provide a constraint on the initial surface density gradient.

Despite our success in matching these observations, other physical 
processes may be needed to explain the full diversity of debris disk 
properties for A stars with similar ages and luminosities. In their 
analysis of the large debris disk surrounding $\gamma$ Oph, 
\citet{su08} examine a dozen main sequence stars with A0--A3 spectral 
types, ages of 150--400~Myr, and fractional disk luminosities 
$L_d / L_{\star} \approx$ $10^{-5}$ to $10^{-4}$ \citep[see also][]{su06}. 
Although all of these stars have dust with $T \approx$ 50--100 K, 
Fomalhaut has a bright torus of dust with weak or negligible emission 
from a wind of small grains, Vega has a bright torus with a luminous 
wind of small grains, and $\gamma$ Oph has an extended disk 
($a_{out} \approx$ 500 AU) of dust apparently bound to the star. Some 
A0--A3 stars have warm inner disks with dust temperatures $\sim$ 100~K
to $\gtrsim$ 200~K; other A stars have no obvious warm dust emission. 
\citet{su08} conclude that collisional cascades in disks with a range 
of masses and other processes, such as the formation of giant planets 
or recent catastrophic collisions, combine to produce the wide range 
of observed properties in this sample.

In principle, our models can explain some of this diversity. The 
\citet{su08} sample contains A stars with a factor of five range in 
$L_{\star}$. Thus, these stars probably have a factor of five range 
in the blowout radius $r_2$ \citep[see also \S5.1;][]{art88,bac93}.  
If the protostellar disks around these stars had properties similar 
to those observed in Taurus-Auriga \citep[][2007a]{and05}, they 
probably had a factor of ten range in initial disk mass, a factor 
of three range in initial disk radius, and a 50\% range in the slope 
of the initial surface density distribution. Coupled with a similar 
dispersion in initial conditions for the terrestrial zones of these 
stars \citep{kb05}, our results suggest that this range in initial 
conditions can produce a broad diversity of debris disks. We have 
started a suite of calculations to address this issue.  Larger samples 
of A stars with resolved disks will provide crucial tests of these
calculations.

Other aspects of planet formation are also important.  If the cores 
of gas giant planets form before their parent stars reach the main 
sequence, we expect gas giants at 20--30 AU around 2--3 \msun\ stars 
\citep[e.g.,][]{kenn08}. Gas giants rapidly remove debris in the inner 
disk and impose structure in the debris beyond 30 AU 
\citep[e.g.,][]{wil02,mor05}. Because gas giants are common around 
evolved A stars \citep{john07}, gas giants probably play a significant
role in the evolution of debris disks around A stars.

Catastrophic collisions may also produce diversity among A star
debris disks \citep[e.g.,][]{wya02,su05}. Although debris from 
complete disruption of colliding planetesimals is unobservable in 
our simulations \citep[see also][]{kb05}, dynamical events similar
to those that produced the Late Heavy Bombardment in the Solar 
System probably are visible \citep[e.g.,][]{gom05}. Testing this 
idea requires numerical calculations that link the dynamics of 
massive planets with the collisional evolution of smaller objects 
\citep[e.g.,][]{cha03,kb06}.

We conclude that our debris disk models can explain the overall time
evolution of the IR excesses and IR colors of A-type main sequence 
stars at 24~$\mu$m and at 70~$\mu$m.  Our calculations for disks 
with $x_m$ = 1/3 to 3 around 2~\msun\ stars fit the overall level of 
the excesses and the trends with stellar age. Explaining the full range 
of observed IR excesses and IR colors requires a set of disks with 
$x_m$ = 0.01--3, as suggested from observations of disks around the 
youngest stars. Matching other properties of these stars -- including
the relative amount of emission from a warm inner disk, an outflowing 
wind of small grains, and a large outer disk -- requires calculations
that include a broader range of initial disk radii and gas giant and
terrestrial planet formation at $a_i <$ 30 AU 
\citep[e.g.,][]{kb05,naga07,kenn08,ida08,kre08}.

\subsection{Debris Disks Around Solar-type Stars}

Although most of the debris disks discovered with {\it IRAS} and {\it ISO} 
have A-type central stars, a few have F-type or G-type central stars 
with masses of 1--1.5 \msun\ \citep{bac93,lag00,dec03,son05,rhe07a}. More 
recent {\it Spitzer} observations reveal debris disks around many 
solar-type stars \citep{bry06,bei06,mey06,tri08,hil08}. Although 
several {\it Spitzer} programs concentrate on older solar-type stars 
as preparation for detailed planet searches, the range of ages is 
large enough to provide an initial test of our predictions.

To compare our model predictions with observations, we consider data for 
nearby solar-type stars from \citet{bei06} and \citet{hil08}. \citet{bei06}
observed $\sim$ 80 solar-type stars at 24~$\mu$m and at 70~$\mu$m using
MIPS on {\it Spitzer}. \citet{hil08} analyze $\sim$ 30 stars with 70 $\mu$m 
excesses out of a sample of 328 stars from the {\it Spitzer} Legacy Science 
Program, ``Formation and Evolution of Planetary Systems'' \citep{mey06}. 
After eliminating K-type and M-type stars from the \citet{bei06} study, 
the two programs contain $\sim$ 80 stars with ages of $\sim$ 10~Myr to 
$\sim$ 10 Gyr. Most of these stars have masses of 0.8--1.5 \msun. Thus, 
we compare these data with our results for debris disk evolution around 
1 \msun\ stars.

\begin{figure}
\includegraphics[width=7.0in]{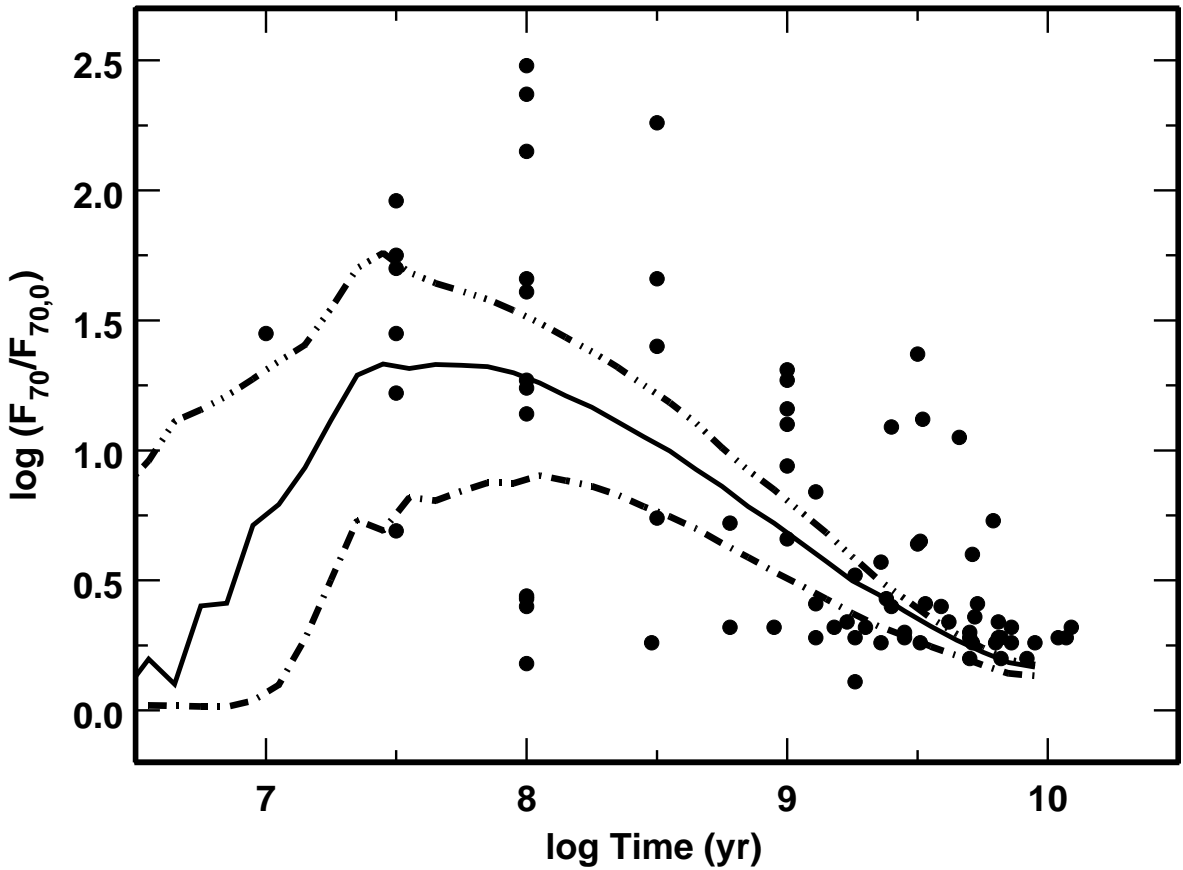}
\figcaption[f25.eps]
{Observations of the 70 $\mu$m excess for nearby solar-type 
stars with known ages \citep{bei06,hil08}. 
The lines show the predicted evolution of the excess for
debris disk models around 1 \msun\ stars
(dot-dashed line: $x_m$ = 1/3; solid line: $x_m$ = 1,
triple dot-dashed line: $x_m$ = 3).
Most stars fall within the loci defined by the calculations,
but many stars are 3--10 times brighter than model predictions.
\label{fig:g1}}
\end{figure}

\begin{figure}
\includegraphics[width=7.0in]{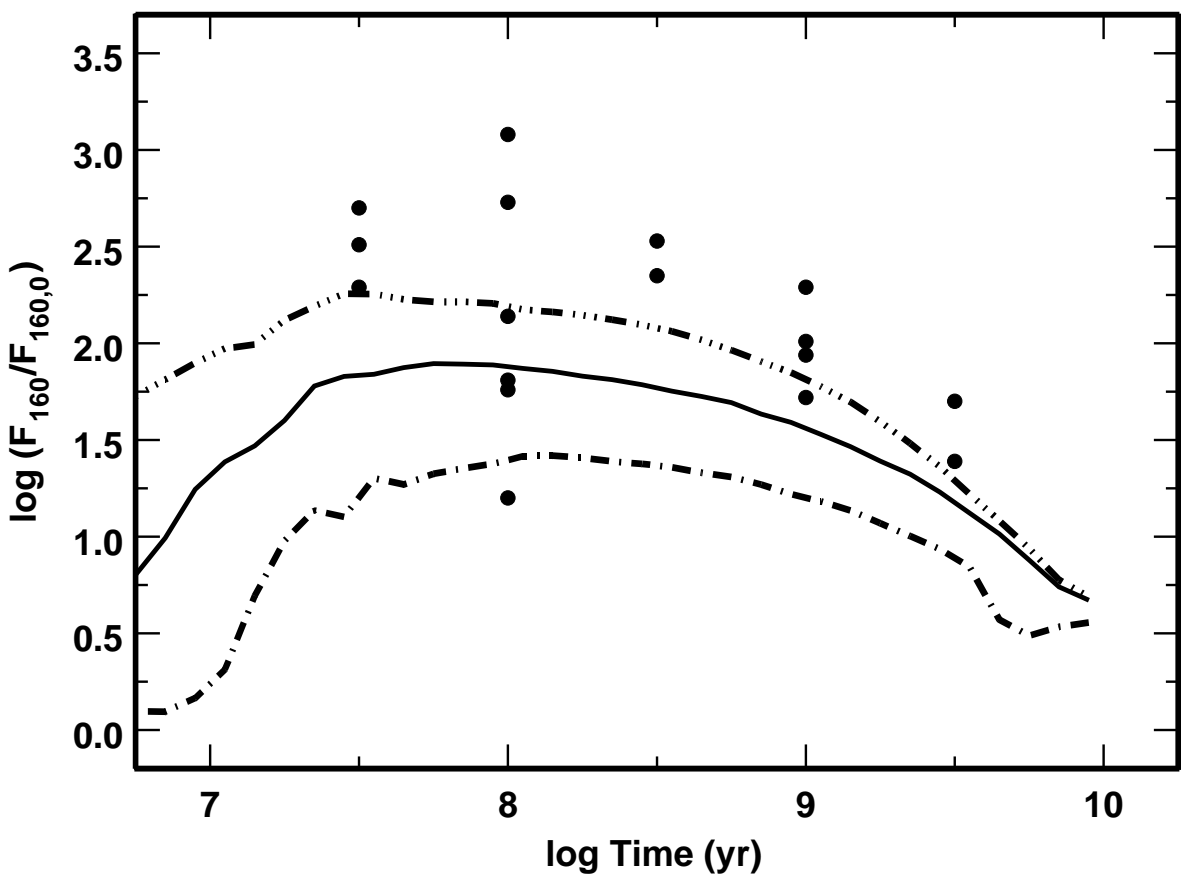}
\figcaption[f26.eps]
{As in Fig. \ref{fig:g1} for the 160~$\mu$m excess.
\label{fig:g2}}
\end{figure}

The observed 70--160~$\mu$m excesses of solar-type stars show trends similar 
to those observed in the evolution of A-type stars at 24--70~$\mu$m
(Fig. \ref{fig:g1}--\ref{fig:g2}). Although the statistics for solar-type
stars are poor for the youngest stars, the data suggest a rise in the 
70--160~$\mu$m excess at 10--100~Myr. The maximum in the 70 $\mu$m excess 
is comparable in magnitude but a factor of $\sim$ 10 later in time than 
the maximum 70~$\mu$m excess for A-type stars.  At 70~$\mu$m and at 
160~$\mu$m, the excess follows a roughly power-law decline with time
for older stars. Solar-type stars also have a large range in excess at 
all ages, with $F_{70}/F_{70,0} \approx$ 1--300 at 100~Myr and 
$F_{70}/F_{70,0} \approx$ 1--30 at 1--3~Gyr. For stars with similar 
ages, the range in the 70~$\mu$m excess is larger for solar-type stars 
than for A-type stars.

Our models match the observed trends for the IR excesses of solar-type 
stars. At 70~$\mu$m, 65\%--75\% of the observations lie within model
predictions; at 160~$\mu$m, more than half of the observations are
within model predictions.  For a range of initial disk masses
($x_m \approx$ 0.01--3) and outer radii ($a_{out} \approx$ 70--150 AU),
we predict a large range of excesses at all ages, as observed. These
model also explain the larger 70 $\mu$m excesses observed for solar-type
stars relative to A-type stars, the apparent maximum in the 70--160~$\mu$m 
excess at 30--100~Myr, and the general power-law decline in the excess 
flux for the oldest stars. 

Despite this general success, however, the models underpredict the 
largest observed excesses. At 70~$\mu$m, the brightest disks are a 
factor of 5--10 brighter than disks with $x_m$ = 3.  At 160~$\mu$m, 
the brightest systems are 3--5 times brighter than our most luminous 
disks. Although the sample of 160~$\mu$m sources is small, our models
underpredict the largest observed fluxes at all ages.

Changing two assumptions in our models yields a better match to the 
observed fluxes at 70--160~$\mu$m. For realistic grain properties, 
\citet{bur79} show that radiation pressure from the Sun cannot eject 
small grains from the Solar System. Reducing the minimum stable grain 
size from $r_2$ = 1~$\mu$m to $r_2$ = 0.1~$\mu$m increases our 
predicted 70~$\mu$m (160~$\mu$m) fluxes by a factor of 2--3 (1.5--2).
Submm observations of several debris disks imply $q \approx$ 0.6--1
for the slope of the radiative emissivity. If we adopt $q = 0.7$
instead of $q = 1$, our predicted 70--160~$\mu$m fluxes increase 
by factors of 2--4. Combining these two modifications increase our
predicted fluxes by a factor of $\sim$ 5--10 at both wavelengths.

Several observations could check whether these modifications of
our standard model are reasonable.
By analogy with the {\it Spitzer} Vega data, detection of an outflowing 
wind of small grains in a debris disk around a solar-type star provides
a clean measurement of $r_2$ and a better constraint on our predicted IR
excesses. Measurements of $q$ for larger samples of debris disks allows
a better assessment of our assumptions for the grain emissivity.

\section{CONCLUSIONS}

Our calculations provide a robust picture for the formation of planets 
and debris disks from a disk of icy planetesimals and set the context 
for the evolution of dusty debris in a dynamic system of planets. 
The results of this study provide a framework for interpreting 
existing observations of debris disks around 1--3 \msun\ stars and
suggest new observational tests of this picture.

We describe a suite of numerical calculations of planets growing 
from ensembles of icy planetesimals at 30--150~AU in disks around 
1--3~\msun\ stars. Using our hybrid multiannulus coagulation code, 
we solve for the evolution of sizes and orbits of objects with 
radii of $\sim$ 1~m to $\gtrsim$ 1000~km over the main sequence 
lifetime of the central star.  These results allow us to constrain 
the growth of planets as a function of disk mass, stellar mass, and 
semimajor axis. 

Debris disk formation is coincident with the formation of a planetary 
system. All calculations of icy planet formation at 30--150~AU lead 
to a collisional cascade which produces copious amounts of dust on 
timescales of 5--30 Myr. This dust is observable throughout the 
lifetime of the central star. Because we consider a broad range of 
input parameters, we derive the time evolution of (i) dust produced 
in the collisional cascade and (ii) the IR and submm emission from 
this dust as a function of disk mass, stellar mass, and time.

We divide the rest of this section into 
(i) theoretical considerations, 
(ii) observable consequences, and
(iii) observational tests. 
The theoretical considerations build on the highlights of icy planet 
formation in \S3.4. Observable consequences of the calculations 
follow from the discussion in \S4. The observational tests of
the models are described in \S5.

\subsection{Theoretical Considerations}

\begin{itemize}

\item Icy planet formation at 30--150 AU is self-limiting. Starting with 
a swarm of $\lesssim$ 1 km planetesimals, runaway growth produces a set 
of 100--500 km protoplanets. As the protoplanets grow, they stir up 
leftover planetesimals along their orbits. When the leftovers reach high 
$e$, collisions produce debris instead of mergers. Because protoplanets 
cannot accrete leftovers rapidly, a cascade of destructive collisions 
grinds the leftovers to dust. Poynting-Robertson drag and radiation 
pressure then remove the dust from the disk.

\item The maximum sizes of icy planets at 30--150 AU are remarkably 
independent of initial disk mass, stellar mass, and stellar age. 
For disks with $x_m$ = 1/3 to 3 around 1--3 \msun\ stars with ages 
$t = 0.1 - 1 t_{ms}$, the typical planet has $r_{max} \sim$ 1750~km 
and $m_{max} \sim 0.005$~\mearth. These objects contain $\lesssim$ 
3\%--4\% of the initial disk mass.  Although 
this result is also independent of the fragmentation parameters, the
finite main sequence lifetimes of 1--3 \msun\ stars limits the formation
of many large planets in the outer disk. Thus, the inner disk produces
many more Pluto-mass planets than the outer disk (Tables \ref{tab:rad1000.1}
and \ref{tab:rad1000.2}).

\item For stars close to the main sequence turnoff, stellar lifetimes and 
the collisional cascade limit the mass in solid objects at 30--150~AU. 
In the inner disk, the collisional cascade removes most of the leftover 
planetesimals before the central star evolves off the main sequence. 
Thus, the typical mass in small objects is $\sim$ 10\% of the initial 
mass at 30--40~AU.  In the outer disk, smaller collision rates produce 
a slower cascade. Thus, the central star evolves off the main sequence 
with $\sim$ 50\% of the initial mass remaining in 1--10~km planetesimals 
at 125--150~AU.

\item The collisional cascade produces copious amounts of dust. Dust 
begins to form during the transition from runaway to oligarchic growth
($t$ = 5--10~Myr), peaks when the first objects reach their maximum sizes 
($t$ = 10--30~Myr), and then slowly declines $t \gtrsim$ 30--50~Myr).  
The peak mass in 0.001--1~mm (0.001--1~m) particles is $\sim$ 1--2 lunar 
masses ($\sim$ 1~\mearth). Disks with initial masses $x_m$ = 1/3 to 3 
reach these peak masses when the age of the star is $\sim$ 10\% to 20\%
of its main sequence lifetime. Because the timescale to form dust
is short ($\sim$ 10--20~Myr), stars are surrounded by large disks
of debris at 30--150~AU throughout their main sequence lifetimes.

\item Radiative processes remove large amounts of mass from debris disks.
Radiation pressure produces a radial wind of small particles containing
$\sim$ 60\% to 70\% of the mass removed from the disk. Poynting-Robertson
drag pulls the rest of the lost mass into the inner disk. Because radiation
pressure is more important than Poynting-Robertson drag when collision rates 
are large, we expect more wind (inner disk) emission earlier (later) in 
the evolution.

\end{itemize}

\subsection{Observable Consequences}

We derive clear observational consequences of the collisional cascade.

\begin{itemize}

\item The dusty debris from the collisional cascade is directly 
observable.  For disks around 1--3~\msun\ stars, the maximum fractional 
dust luminosity of $L_d / L_{\star} \sim 2 \times 10^{-3}$ is comparable
to the maximum dust luminosities of known debris disks 
\citep{bac93,rie05,su06,rhe07a}. The dust temperature at the inner edge of 
a 30--150~AU disk scales with the temperature of the central star;
thus, the predicted 24~$\mu$m excess is very sensitive to the stellar
mass. At 70~$\mu$m, the predicted excesses scale roughly linearly 
with disk mass and stellar mass. The predicted 160--850~$\mu$m excesses 
depend on the disk mass but are nearly independent of the stellar mass.

\item For systems with little or no emission from terrestrial dust 
([5.8]--[8] $\lesssim$ 0.1), mid-IR color-color diagrams clearly 
distinguish debris disks around stars of different masses. In a 
[8]--[24] {\it vs} [24]--[70] diagram, 2--3 \msun\ (1--2 \msun) 
stars have red (blue) [8]--[24] and blue (red) [24]--[70] 
(Fig \ref{fig:cc-1}). In both cases, the color scales with the 
initial disk mass (Fig \ref{fig:cc-2}).  Optical colors and spectra 
generally provide good estimates for stellar mass; thus, these 
diagrams provide good tests of our model predictions. 

\end{itemize}

\subsection{Observational Tests}

We compare our predictions with observations of A-type stars and 
solar-type stars. 

\begin{itemize}

\item For A-type stars, our calculations are the first to explain the 
observed rise and fall of debris disk fluxes at 24~$\mu$m 
\citep[Fig. \ref{fig:su1};][]{cur08a,cur08b}. In our picture, the rise 
in debris disk emission corresponds to the transition from runaway 
growth -- when mergers of small planetesimals produce larger 
protoplanets -- to oligarchic growth -- when the collisional cascade 
begins to grind leftover planetesimals into dust. When oligarchs in 
the inner disk are close to their maximum sizes of $\sim$ 1750~km, 
the collisional cascade produces a maximum in debris disk emission.  
For a wide range of initial conditions, this maximum occurs at 10--20~Myr.  
As the collisional cascade moves out through the disk, smaller collision 
rates produce less dust which emits at lower temperatures. Thus, the 
24~$\mu$m excess falls with time. The predicted rate of decline,
$t^{-n}$ with $n \approx$ 0.6--0.8, is close to the observed rate
\citep[$n \approx$ 0.5--1;][]{gre03,rie05,rhe07a}.

\item At longer wavelengths, the maximum excess is larger and lasts 
longer than at 24~$\mu$m. Predicted mid-IR colors also increase
slowly with time. Although larger samples of A-type stars with 8~$\mu$m
photometry would provide a better test of our models, current data 
for the 70~$\mu$m excess and the evolution of the [24]--[70] color 
agree with our predictions (Fig. \ref{fig:su2}--\ref{fig:su3}).  For 
2--3 \msun\ stars near the main-sequence turnoff, our calculations 
also yield a clear maximum in the 850~$\mu$m flux. Large surveys, such 
as the proposed JCMT Legacy Survey \citep{mat07} and submm observations
with ALMA, Herschel and SOFIA, can test this prediction. 

\item For solar-type stars, our models match observations of most sources. 
The predicted evolution of the 70--160~$\mu$m excesses follows the 
observed rise at 10--100~Myr, the peak at $\sim$ 30--100~Myr, and 
the decline at $\gtrsim$ 300~Myr. Although $\sim$  70\% (55\%) of 
observed debris disks have fluxes that lie within model predictions,
our models underpredict fluxes for the brightest sources by a factor of 
5--10. Fluxes for models with $r_2 \approx$ 0.1~$\mu$m and $q \lesssim$ 
0.7 provide better matches to these observations. To guide our choices
for $r_2$ and $q$, we require spatially resolved images and submm
fluxes for these objects.

\item For 1--3 \msun\ stars with ages $\sim$ 0.1--1 Gyr, current data 
suggest that solar-type stars have a larger range of far-IR excesses
than A-type stars. In our models, faster debris disk evolution around 
A-type stars produces a smaller dispersion in far-IR excesses and 
colors for stars with ages of 100 Myr to 1 Gyr. Larger samples of 
debris disks can test this prediction in more detail.

\item We also consider observations of Vega, the prototypical debris disk.
If we adopt models with a blowout radius $r_2$ = 10~$\mu$m, we can 
match observations 
with a standard collisional cascade within a broad torus at 80--200~AU.
If the torus contains $\sim$ 1--5 \mearth\ in large objects with 
$r \gtrsim$ 1 cm, the cascade can generate (i) the observed ensemble 
of grains with $r \sim$ 200--300~$\mu$m within the torus and (ii) an
outflowing wind of small grains with $r \sim$ 1--50~$\mu$m. 

This conclusion differs from \citet{su06}, who postulate a recent
catastrophic collision between two large objects as the source of
the dusty Vega wind. Although the complete destruction of two large 
icy objects can produce a massive outflowing wind, our results 
suggest that the dusty wind is short-lived and cannot be rapidly 
replenished by the observed population of larger objects.  We show
that a steady-state collisional cascade can explain the {\it Spitzer} 
data \citep[see also][]{kb05}. If our interpretation is correct, 
sensitive observations at 1--10 mm should detect our proposed 
reservoir of larger objects.

\end{itemize}

Matching other observations of debris disks requires more realism in our 
planet formation calculations. Adding binary companions and giant planets 
provides ways to modify the evolution of the collisional cascade and 
to impose structure on rings and tori \citep[e.g.][]{wil02,mor05,qui06}. 
Extending the coagulation calculations to smaller sizes allows studies
of the formation of winds and other large structures. Although these
calculations have been prohibitively expensive in computing time, rapid 
advances in computing technology will make these additions possible in
the next few years.

Based on the results described here and in \citet{kb04b}, we conclude
that debris disks are the inevitable outcome of icy planet formation
in a disk of solid objects. The basic structures produced by this
model -- broad tori and narrow rings of dust that propagate out
through the disk \citep{kb04b} -- are consistent with observations 
\citep[e.g.][]{jay98,kal05,su06,fit07}. The model also explains the
time evolution of mid-IR colors and fluxes for debris disks around 
A-type and solar-type stars.

\acknowledgements
We acknowledge a generous allotment, $\sim$ 1000 cpu days, of computer 
time on the 1024 cpu Dell Xeon cluster `cosmos' at the Jet Propulsion 
Laboratory through funding from the NASA Offices of Mission to Planet 
Earth, Aeronautics, and Space Science.  We thank M. Werner for his strong 
support of this project.  We also acknowledge use of $\sim$ 250 cpu days 
on the CfA cluster `hydra.' Advice and comments from T. Currie, M. Geller, 
G. Kennedy, M. Meyer, G. Rieke, K. Su, and an anonymous referee greatly 
improved our presentation.  Portions of this project were supported by 
the {\it NASA } {\it Astrophysics Theory Program,} through grant 
NAG5-13278, the {\it NASA} {\it TPF Foundation Science Program,} through 
grant NNG06GH25G, and the {\it Spitzer Guest Observer Program,} through 
grant 20132.

\appendix

\section{APPENDIX}

\subsection{Growth rates}

In standard coagulation theory, protoplanets accrete material from 
a swarm of planetesimals at a rate \citep[e.g.,][]{saf69,lis87,wet93}
\begin{equation}
\dot{M} \propto \Sigma ~ \Omega ~ r^2 ~ \left [ 1 + (v_{esc}/v)^2 \right ] ~ ,
\end{equation}
where $r$ is the radius of a planetesimal,
$\Omega$ is the angular frequency of material in the disk,
$v$ is the random velocity of planetesimals, and $v_{esc}$ is
the escape velocity of the protoplanet. The $1 + (v_{esc}/v)^2$ 
term is the gravitational focusing factor. 

To derive the accretion time, we set $t = M / \dot{M}$ and substitute
the orbital period for the angular frequency,
\begin{equation}
t \propto ( \rho ~ r ~ P / \Sigma) ~ \left [ 1 + (v_{esc}/v)^2 \right ]^{-1} ~ ,
\end{equation}
where $\rho$ is the mass density of a planetesimal. Throughout runaway 
growth and the early stages of oligarchic growth $v_{esc}/v \gg 1$. 
Because we are interested in the time to produce planets with the
same $r$ and $\rho$ in disks with different $P$ and $\Sigma$, we 
eliminate $\rho$ and $r$.  Thus, the growth time is roughly
\begin{equation}
t \propto (P / \Sigma) (v/v_{esc})^2 ~ .
\end{equation}
This equation sets the typical timescale for planet growth in a disk of 
planetesimals. If $\Sigma \sim \Sigma_0 x_m a^{-3/2}$ (Eq. (\ref{eq:sigma}))
and $v/v_{esc}$ $\sim$ constant \citep[Fig. 1;][]{wet93,gol04},
\begin{equation}
t \propto a^3 x_m^{-1} \Sigma_0^{-1} ~ . 
\label{eq:tacc-app}
\end{equation}
This result is close to the $t \propto a^3 x_m^{-1.15} \Sigma_0^{-1}$
derived for the formation of the first 1000~km object in our 
calculations (e.g., Eq. (\ref{eq:t1000allm})).

To evaluate possible sources for the extra factor of $x_m^{-0.15}$ in our
derived accretion times, we consider the random velocity $v$ of accreted
planetesimals. Shorter growth times require smaller random velocities.
Thus, we consider processes that damp planetesimal velocities.  In our 
calculations, collisions and gas drag can reduce $v$; dynamical friction
and viscous stirring increase $v$.  At 30--150 AU, gas drag damps random 
velocities $\sim$ 10--20 times more rapidly than collisions \citep{gol04}. 
Thus, we ignore collisional damping and concentrate on gas drag.

\citet{raf04} investigated the dynamics of small planetesimals and growing 
protoplanets in a gaseous nebula. For the early stages of oligarchic growth,
the random velocity of planetesimals is
\begin{equation}
v/v_{esc} \propto \Sigma_{gas}^{-\gamma_1} ~ .
\end{equation}
Substituting this expression into Eq. (\ref{eq:tacc-app}) and adopting 
a constant gas-to-dust ratio, $\Sigma_{gas} \sim \Sigma$, we derive
\begin{equation}
t \propto a^3 x_m^{-\gamma2} \Sigma_0^{-1} ~ ,
\label{eq:tacc-gd}
\end{equation}
with $\gamma_2 = 2\gamma_1 + 1$.  For typical conditions in planetesimal 
disks, \citet{raf04} derived $\gamma_1 \approx$ 1/6 to 1/5. Thus, 
$\gamma_2 \approx$ $1.3$ to $1.4$, close to the exponent of $1.15$ 
derived in our calculations.

Our treatment of gas drag probably reduces the exponent of $x_m$ in Eq.
(\ref{eq:tacc-gd}) from the predicted $1.3$--$1.4$ to $1.15$. In our
simulations, we
assume the gas density declines exponentially on a timescale $t_{gas}$ 
= 10~Myr. With typical growth times of 20--40 Myr, the gas density is 
$\sim$ 1\% to 10\% of its initial value when the first 1000 km objects
form in the inner disk. Thus, gas drag cannot reduce planetesimal 
random velocities as efficiently as predicted in Eq. \ref{eq:tacc-gd}.
Reducing drag lowers the exponent. With gas depletion timescales
$\sim$ 25\% to 50\% of the growth time, we expect an exponent of
$\gamma_2 \approx$ $1.1$ to $1.2$, similar to the $\gamma_2 = 1.15$
in our calculations.  

\subsection{Radiation from dust}

In the Appendix of \citet{kb04a}, we briefly described our simple algorithm 
for the evolution of particles with sizes smaller than the smallest object 
-- $r \sim$ 1 m -- followed in the multiannulus coagulation code.  This 
algorithm yields the optical depth in very small grains ejected from the 
system and the optical depth in larger grains evolving under the influence 
of collisions and Poynting-Robertson drag. The optical depth in both grain 
populations allows us to derive the time evolution of the disk luminosity 
and surface brightness in bolometric units. Here, we describe the derivation 
of grain temperature for these populations that yields the predicted time 
evolution of the broadband spectral energy distributions of debris disks.

As in \citet{kb04a}, we divide objects with sizes smaller than $\sim$ 1 m
into very small grains, small grains, and large grains. In each annulus 
$k$ of our calculation, radiation pressure ejects very small grains with 
radii between $r_1$ and $r_2$.  If $\rho_g$ is the mass density of these 
grains and $\dot{M}_k$ is the production rate of very small grains in each 
annulus, the very small grains have an integrated optical depth
\begin{equation}
\tau_{s} = \frac{3 ( \sqrt{r_2/r_1} - 1 )}{8 \pi \rho_g r_2 ( 1 - \sqrt{r_1/r_2} ) } \sum_{i=1}^{N} ~ \left [ \sum_{k=1}^{i} \left ( \frac{\dot{M}_k}{v_{Kk} h_k } \right ) \left ( \frac{1}{a_{b,k}} - \frac{1}{a_{b,k+1}} \right ) \right ]~ ,
\end{equation}
where 
$a_{b,k}$ is the inner boundary of an annulus centered at $a_k$,
$h_k$ is the vertical scale height in units of the semimajor axis,
and $v_{K,k}$ is the orbital velocity in annulus $k$.

For small ($r$ = $r_2$ to 1 mm) and large grains ($r$ = 1 mm to 1 m), 
we derive the optical depth $\tau_k$ in each annulus.  To derive the 
radial surface brightness and total disk luminosity, we follow 
\citet{kh87} and derive the amount of stellar radiation absorbed
by each annulus. We assume a spherical, limb-darkened star with radius 
$R_{\star}$, luminosity $L_{\star}$, and limb-darkening parameter 
$\epsilon_0$ = 0.6. For a point $P$ at the outer boundary of annulus 
$k$ with height $h_P$ above the disk midplane, rays from the star enter 
the annulus at a scale height $h_{in}$ above (below) the midplane.  We 
compute the length $l$ of the path through the disk and derive the optical 
depth along this path as $\tau_p$ = ($l/\Delta a_k$)$\tau_k$, where 
$\Delta a_k$ is the width of the annulus.  The radiation absorbed along 
this path is $e^{-\tau_p} I_0$, where $I_0$ is the flux incident on the 
boundary of the annulus. Numerical integrations over the stellar surface 
and the vertical extent of an annulus yield the amount of flux absorbed 
by each annulus, which we convert to relative surface brightness. A final
numerical integration over the radial extent of the disk yields the ratio 
of the disk luminosity to the stellar luminosity, $L_d/L_\star$.

To derive the spectral energy distribution of the disk, we make several
assumptions. Consistent with observations of scattered light from resolved 
debris disks \citep{bac93,lag00},
we adopt a single albedo $\omega$ = 0.25 for all grains. For all $\lambda$,
the luminosity in scattered light is then $\omega L_d / L_{\star}$; the 
thermal luminosity emitted by all grains is $(1 - \omega) L_d / L_{\star}$. 
In each annulus $k$, we assume grains emit at a temperature $T_{i,k}$, where 
the index $i$ refers to discrete bins in grain size. To derive equilibrium
temperatures for these grains, we assume the grains have an absorption
efficiency $\epsilon_a \propto (\lambda / \lambda_0)^p$ and radiative
efficiency $\epsilon_r \propto (\lambda / \lambda_0)^q$. For most grains
in our calculations, the grain size is larger than the peak wavelength 
of radiation emitted by the central star. Thus, the grains efficiently
absorb stellar photons and $p$ = 0. Large grains with $r \gg \lambda$ emit 
as blackbodies and have $q$ = 0. Smaller grains radiate less efficiently
and have $q \approx$ 1.

\clearpage

\begin{deluxetable}{lcccccc}
\tablecolumns{7}
\tablewidth{0pc}
\tabletypesize{\footnotesize}
\tablenum{1}
\tablecaption{Grid of Debris Disk Calculations\tablenotemark{a}}
\tablehead{
  \colhead{} &
  \multicolumn{6}{c}{Stellar Mass in $M_{\odot}$} 
\\
  \colhead{$x_m$} &
  \colhead{~~1.0\tablenotemark{b}~~} &
  \colhead{~~1.0\tablenotemark{c}~~} &
  \colhead{~~1.5~~} &
  \colhead{~~2.0~~} &
  \colhead{~~2.5~~} &
  \colhead{~~3.0~~}
}
\startdata
0.33 & 41 & 19 & 15 & 20 & 18 & 18 \\
0.50 & 49 & 18 & 18 & 17 & 17 & 17 \\
1.00 & 49 & 15 & 22 & 17 & 15 & 15 \\
2.00 & 41 & 15 & 30 & 17 & 16 & 15 \\
3.00 & 45 & 18 & 12 & 22 & 15 & 21 \\
\\
$t_{ms}$\tablenotemark{d} & 10.00 & 10.00 & 2.90 & 1.22 & 0.65 & 0.39 \\ 
\enddata
\tablenotetext{a}{Number of independent calculations for each 
combination of $M_{\star},x_m$}
\tablenotetext{b}{32 annulus models at 30--70~AU and at 70--150~AU}
\tablenotetext{c}{64 annulus models at 30--150~AU}
\tablenotetext{d}{Main sequence lifetime in Gyr \cite{dem04}}
\label{tab:modgrid}
\end{deluxetable}
\clearpage

\begin{deluxetable}{lcccccccc}
\tablecolumns{9}
\tablewidth{0pc}
\tabletypesize{\footnotesize}
\tablenum{2}
\tablecaption{Median number of Plutos at t = $t_{ms}$/3 for disks around 1 \msun\ stars}
\tablehead{
  \colhead{$x_m$} &
  \colhead{30--37~AU} &
  \colhead{37--45~AU} &
  \colhead{45--55~AU} &
  \colhead{55--67~AU} &
  \colhead{67--82~AU} &
  \colhead{82--100~AU} &
  \colhead{100--123~AU} &
  \colhead{123--146~AU}
}
\startdata
0.33 &  40 &  44 &  33 &  32 &  19 &  12 &   3 &   1 \\
0.50 &  62 &  65 &  39 &  49 &  33 &  25 &  10 &   1 \\
1.00 & 111 & 110 &  73 &  73 &  47 &  55 &  26 &   5 \\
2.00 & 172 & 194 & 134 & 155 & 116 &  84 &  58 &  33 \\
3.00 & 165 & 260 & 172 & 251 & 137 & 109 &  85 &  44 \\
\enddata
\label{tab:rad1000.1}
\end{deluxetable}
\clearpage

\begin{deluxetable}{lcccccccc}
\tablecolumns{9}
\tablewidth{0pc}
\tabletypesize{\footnotesize}
\tablenum{3}
\tablecaption{Median number of Plutos at t = $t_{ms}$/3 for disks around 1.5--3 \msun\ stars}
\tablehead{
  \colhead{$x_m$} &
  \colhead{30--37~AU} &
  \colhead{37--45~AU} &
  \colhead{45--55~AU} &
  \colhead{55--67~AU} &
  \colhead{67--82~AU} &
  \colhead{82--100~AU} &
  \colhead{100--123~AU} &
  \colhead{123--146~AU}
}
\startdata
\cutinhead{1.5 \msun}
0.33 & 46  & 46  & 45  & 28  & 16  &  3  &  1  &  1 \\
0.50 & 69  & 85  & 68  & 48  & 53  & 25  &  2  &  1 \\
1.00 & 102 & 136 & 128 & 101 &  98 &  67 &  28 &   2 \\
2.00 & 158 & 243 & 240 & 261 & 211 & 123 & 169 &  47 \\
3.00 & 201 & 239 & 301 & 295 & 381 & 165 & 198 &  97 \\
\cutinhead{2.0 \msun}
0.33 &  55 & 51 & 48  &30 &  2 &  1 &  1 &  3 \\
0.50 & 115 & 87 & 85  &51 & 20 & 10 &  1 &  2 \\
1.00 & 172 &187 &231 &123 & 82 & 50 &  1 &  3 \\
2.00 & 261 &230 &366 &236 &324 &204 & 77 &  5 \\
3.00 & 259 &262 &398 &264 &446 &295 &171 & 48 \\
\cutinhead{2.5 \msun}
0.33 & 105 & 92 & 63 & 55 & 19 &  1 &  5 &  2 \\
0.50 & 152 &139 &123 &100 & 53 & 10 &  5 &  5 \\
1.00 & 164 &174 &234 &198 &125 & 90 &  7 &  6 \\
2.00 & 223 &230 &278 &243 &158 &108 & 10 &  8 \\
3.00 & 353 &434 &490 &495 &477 &615 &313 &157 \\
\cutinhead{3.0 \msun}
0.33 & 133 &121 & 62 & 36 &  6 &  5 &  1 &  1 \\
0.50 & 127 &153 &129 &103 & 50 &  2 &  2 &  1 \\
1.00 & 199 &230 &239 &258 &171 & 66 &  6 &  5 \\
2.00 & 224 &353 &376 &392 &342 &265 &173 &  5 \\
3.00 & 428 &598 &479 &657 &570 &756 &578 &272 \\
\enddata
\label{tab:rad1000.2}
\end{deluxetable}
\clearpage

\begin{deluxetable}{lccccc}
\tablecolumns{6}
\tablewidth{0pc}
\tabletypesize{\footnotesize}
\tablenum{4}
\tablecaption{Predicted Excesses for Disks Around 1 \msun\ Stars\tablenotemark{a}}
\tablehead{
  \colhead{log $t$ (yr)} &
  \colhead{log $L_d/L_{\star}$} &
  \colhead{log $F_{24}/F_{24,0}$} &
  \colhead{log $F_{70}/F_{70,0}$} &
  \colhead{log $F_{160}/F_{160,0}$} &
  \colhead{log $F_{850}/F_{850,0}$}
}
\startdata
\cutinhead{$x_m$ = 0.33}
 5.05 & -4.57 & 0.000 & 0.053 & 0.145 & 0.090 \\
 5.15 & -4.57 & 0.000 & 0.053 & 0.145 & 0.090 \\
 5.25 & -4.63 & 0.000 & 0.047 & 0.131 & 0.082 \\
 5.35 & -4.65 & 0.000 & 0.045 & 0.127 & 0.080 \\
 5.45 & -4.67 & 0.000 & 0.043 & 0.123 & 0.077 \\
\cutinhead{$x_m$ = 1.0}
 5.05 & -4.18 & 0.000 & 0.096 & 0.255 & 0.172 \\
 5.15 & -4.18 & 0.000 & 0.096 & 0.255 & 0.172 \\
 5.25 & -4.29 & 0.000 & 0.090 & 0.243 & 0.164 \\
 5.35 & -4.31 & 0.000 & 0.088 & 0.238 & 0.161 \\
 5.45 & -4.33 & 0.000 & 0.086 & 0.233 & 0.158 \\
\cutinhead{$x_m$ = 3.0}
 5.05 & -4.02 & 0.000 & 0.162 & 0.464 & 0.357 \\
 5.15 & -4.06 & 0.000 & 0.150 & 0.410 & 0.307 \\
 5.25 & -4.10 & 0.000 & 0.138 & 0.355 & 0.257 \\
 5.35 & -4.12 & 0.000 & 0.133 & 0.354 & 0.248 \\
 5.45 & -4.14 & 0.000 & 0.127 & 0.332 & 0.238 \\
\enddata
\label{tab:mod-1p0}
\tablenotetext{a}{The electronic version of this paper contains the
complete version of this Table.}
\end{deluxetable}
\clearpage

\begin{deluxetable}{lccccc}
\tablecolumns{6}
\tablewidth{0pc}
\tabletypesize{\footnotesize}
\tablenum{5}
\tablecaption{Predicted Excesses for Disks Around 1.5 \msun\ Stars\tablenotemark{a}}
\tablehead{
  \colhead{log $t$ (yr)} &
  \colhead{log $L_d/L_{\star}$} &
  \colhead{log $F_{24}/F_{24,0}$} &
  \colhead{log $F_{70}/F_{70,0}$} &
  \colhead{log $F_{160}/F_{160,0}$} &
  \colhead{log $F_{850}/F_{850,0}$}
}
\startdata
\cutinhead{$x_m$ = 0.33}
 5.05 & -4.40 & 0.001 & 0.100 & 0.202 & 0.110 \\
 5.15 &	-4.45 & 0.001 & 0.090 & 0.184 & 0.099 \\
 5.25 & -4.51 & 0.001 & 0.078 & 0.163 & 0.088 \\
 5.35 &	-4.52 & 0.001 & 0.074 & 0.157 & 0.085 \\
 5.45 &	-4.53 & 0.001 & 0.070 & 0.151 & 0.082 \\
\cutinhead{$x_m$ = 1.0}
 5.05 & -4.01 & 0.001 & 0.162 & 0.321 & 0.188 \\
 5.15 & -4.19 & 0.001 & 0.148 & 0.299 & 0.178 \\
 5.25 & -4.22 & 0.001 & 0.140 & 0.287 & 0.170 \\
 5.35 & -4.24 & 0.001 & 0.133 & 0.275 & 0.163 \\
 5.45 & -4.24 & 0.001 & 0.128 & 0.266 & 0.157 \\
\cutinhead{$x_m$ = 3.0}
 5.05 & -3.73 & 0.002 & 0.206 & 0.406 & 0.261 \\
 5.15 & -4.01 & 0.002 & 0.199 & 0.394 & 0.251 \\
 5.25 & -4.04 & 0.002 & 0.192 & 0.382 & 0.242 \\
 5.35 & -4.05 & 0.002 & 0.181 & 0.361 & 0.227 \\
 5.45 & -4.08 & 0.002 & 0.167 & 0.339 & 0.211 \\
\enddata
\label{tab:mod-1p5}
\tablenotetext{a}{The electronic version of this paper contains the
complete version of this Table.}
\end{deluxetable}

\begin{deluxetable}{lccccc}
\tablecolumns{6}
\tablewidth{0pc}
\tabletypesize{\footnotesize}
\tablenum{6}
\tablecaption{Predicted Excesses for Disks Around 2.0 \msun\ Stars\tablenotemark{a}}
\tablehead{
  \colhead{log $t$ (yr)} &
  \colhead{log $L_d/L_{\star}$} &
  \colhead{log $F_{24}/F_{24,0}$} &
  \colhead{log $F_{70}/F_{70,0}$} &
  \colhead{log $F_{160}/F_{160,0}$} &
  \colhead{log $F_{850}/F_{850,0}$}
}
\startdata
\cutinhead{$x_m$ = 0.33}
 5.05 & -4.28 & 0.002 & 0.128 & 0.215 & 0.109 \\
 5.15 & -4.34 & 0.002 & 0.118 & 0.201 & 0.102 \\
 5.25 & -4.41 & 0.002 & 0.108 & 0.191 & 0.096 \\
 5.35 & -4.43 & 0.002 & 0.103 & 0.183 & 0.092 \\
 5.45 & -4.46 & 0.002 & 0.098 & 0.175 & 0.087 \\
\cutinhead{$x_m$ = 1.0}
 5.05 & -3.89 & 0.003 & 0.195 & 0.334 & 0.185 \\
 5.15 & -4.00 & 0.003 & 0.186 & 0.320 & 0.177 \\
 5.25 &  4.11 & 0.003 & 0.177 & 0.309 & 0.170 \\
 5.35 & -4.16 & 0.003 & 0.175 & 0.307 & 0.168 \\
 5.45 & -4.18 & 0.003 & 0.161 & 0.286 & 0.156 \\
\cutinhead{$x_m$ = 3.0}
 5.05 & -3.71 & 0.005 & 0.248 & 0.421 & 0.250 \\
 5.15 & -3.98 & 0.005 & 0.239 & 0.409 & 0.240 \\
 5.25 & -4.02 & 0.004 & 0.227 & 0.389 & 0.226 \\
 5.35 & -4.02 & 0.004 & 0.207 & 0.358 & 0.205 \\
 5.45 & -4.07 & 0.004 & 0.192 & 0.333 & 0.188 \\
\enddata
\label{tab:mod-2p0}
\tablenotetext{a}{The electronic version of this paper contains the
complete version of this Table.}
\end{deluxetable}

\begin{deluxetable}{lccccc}
\tablecolumns{6}
\tablewidth{0pc}
\tabletypesize{\footnotesize}
\tablenum{7}
\tablecaption{Predicted Excesses for Disks Around 2.5 \msun\ Stars\tablenotemark{a}}
\tablehead{
  \colhead{log $t$ (yr)} &
  \colhead{log $L_d/L_{\star}$} &
  \colhead{log $F_{24}/F_{24,0}$} &
  \colhead{log $F_{70}/F_{70,0}$} &
  \colhead{log $F_{160}/F_{160,0}$} &
  \colhead{log $F_{850}/F_{850,0}$}
}
\startdata
\cutinhead{$x_m$ = 0.33}
 5.05 & -4.18 & 0.004 & 0.170 & 0.259 & 0.120 \\
 5.15 & -4.22 & 0.004 & 0.153 & 0.233 & 0.108 \\
 5.25 & -4.34 & 0.004 & 0.137 & 0.207 & 0.096 \\
 5.35 & -4.36 & 0.004 & 0.130 & 0.197 & 0.093 \\
 5.45 & -4.38 & 0.004 & 0.123 & 0.188 & 0.086 \\
\cutinhead{$x_m$ = 1.0}
 5.05 & -3.97 & 0.008 & 0.237 & 0.352 & 0.181 \\
 5.15 & -4.08 & 0.007 & 0.219 & 0.329 & 0.168 \\
 5.25 & -4.09 & 0.000 & 0.000 & 0.000 & 0.000 \\
 5.35 & -4.14 & 0.006 & 0.201 & 0.306 & 0.154 \\
 5.45 & -4.14 & 0.005 & 0.187 & 0.287 & 0.144 \\
\cutinhead{$x_m$ = 3.0}
 5.05 & -3.83 & 0.010 & 0.285 & 0.424 & 0.230 \\
 5.15 & -3.95 & 0.009 & 0.270 & 0.403 & 0.217 \\
 5.25 & -3.98 & 0.009 & 0.258 & 0.387 & 0.206 \\
 5.35 & -4.03 & 0.009 & 0.248 & 0.372 & 0.197 \\
 5.45 & -3.48 & 0.087 & 0.749 & 0.887 & 0.558 \\
\enddata
\label{tab:mod-2p5}
\tablenotetext{a}{The electronic version of this paper contains the
complete version of this Table.}
\end{deluxetable}
\clearpage

\begin{deluxetable}{lccccc}
\tablecolumns{6}
\tablewidth{0pc}
\tabletypesize{\footnotesize}
\tablenum{8}
\tablecaption{Predicted Excesses for Disks Around 3.0 \msun\ Stars\tablenotemark{a}}
\tablehead{
  \colhead{log $t$ (yr)} &
  \colhead{log $L_d/L_{\star}$} &
  \colhead{log $F_{24}/F_{24,0}$} &
  \colhead{log $F_{70}/F_{70,0}$} &
  \colhead{log $F_{160}/F_{160,0}$} &
  \colhead{log $F_{850}/F_{850,0}$}
}
\startdata
\cutinhead{$x_m$ = 0.33}
 5.05 & -4.08 & 0.011 & 0.173 & 0.229 & 0.099 \\
 5.15 & -4.11 & 0.010 & 0.164 & 0.213 & 0.091 \\
 5.25 & -4.14 & 0.009 & 0.152 & 0.198 & 0.083 \\
 5.35 & -4.16 & 0.008 & 0.140 & 0.183 & 0.076 \\
 5.45 & -4.19 & 0.007 & 0.134 & 0.174 & 0.072 \\
\cutinhead{$x_m$ = 1.0}
 5.05 & -3.96 & 0.015 & 0.254 & 0.326 & 0.150 \\
 5.15 & -4.01 & 0.014 & 0.239 & 0.310 & 0.142 \\
 5.25 & -4.04 & 0.013 & 0.227 & 0.297 & 0.135 \\
 5.35 & -4.07 & 0.012 & 0.216 & 0.282 & 0.127 \\
 5.45 & -4.10 & 0.011 & 0.204 & 0.269 & 0.120 \\
\cutinhead{$x_m$ = 3.0}
 5.05 & -3.83 & 0.020 & 0.318 & 0.410 & 0.199 \\
 5.15 & -3.88 & 0.019 & 0.301 & 0.390 & 0.189 \\
 5.25 & -3.92 & 0.030 & 0.321 & 0.389 & 0.187 \\
 5.35 & -3.38 & 0.113 & 0.661 & 0.740 & 0.416 \\
 5.45 & -3.01 & 0.226 & 0.934 & 0.987 & 0.592 \\
\enddata
\label{tab:mod-3p0}
\tablenotetext{a}{The electronic version of this paper contains the
complete version of this Table.}
\end{deluxetable}
\clearpage

\begin{deluxetable}{lccc}
\tablecolumns{4}
\tablewidth{0pc}
\tabletypesize{\normalsize}
\tablenum{9}
\tablecaption{Debris disk loci in color-color space}
\tablehead{
  \colhead{$M_{\star}$ (\msun)} &
  \colhead{~~~~x$_0$,y$_0$~~~~} &
  \colhead{~~~~x$_u$,y$_u$~~~~} &
  \colhead{~~~~x$_l$,y$_l$~~~~}
}
\startdata
1.0 & 0.00,0.0 & 4.0,0.1 & 5.00,0.00 \\
1.5 & 1.50,0.0 & 4.0,1.5 & 4.50,0.25 \\ 
2.0 & 1.25,0.0 & 2.5,2.5 & 3.50,0.50 \\
2.5 & 1.00,0.0 & 2.0,4.0 & 3.00,1.00 \\
3.0 & 1.00,0.0 & 1.5,5.0 & 2.25,1.75 \\
\enddata
\label{tab:cc-loci}
\end{deluxetable}
\clearpage

\begin{deluxetable}{lcccccccccc}
\tablecolumns{11}
\tablewidth{0pc}
\tabletypesize{\footnotesize}
\tablenum{10}
\tablecaption{Vega Debris Disk Model\tablenotemark{a}}
\tablehead{
  \colhead{Source} &
  \colhead{$F_{24,\star}$} &
  \colhead{$F_{70,\star}$} &
  \colhead{$F_{160,\star}$} &
  \colhead{$F_{850,\star}$} &
  \colhead{$F_{24,disk}$} &
  \colhead{$F_{70,disk}$} &
  \colhead{$F_{160,disk}$} &
  \colhead{$F_{850,disk}$} &
  \colhead{$M_{d,1-50}$} &
  \colhead{$\dot{M}$}
}
\startdata
Vega & 7.2 & 0.8 & 0.16 & 0.006 & 1.5 & 7.0 & 4.0 & 0.091 & 3.0 & 30\tablenotemark{b} \\
\\
Model 1\tablenotemark{c} & 7.2 & 0.8 & 0.16 & 0.006 & 4.2 & 10.0 & 3.0 & 0.05 & 0.9 & 0.2 \\
Model 2\tablenotemark{d} & 7.2 & 0.8 & 0.16 & 0.006 & 7.2 & 23.0 & 8.0 & 0.11 & 3.8 & 0.6 \\
Model 3\tablenotemark{e} & 7.2 & 0.8 & 0.16 & 0.006 & 10.9 & 15.0 & 4.0 & 0.05 & 2.4 & 0.3 \\
Model 4\tablenotemark{f} & 7.2 & 0.8 & 0.16 & 0.006 & 15.5 & 23.0 & 5.0 & 0.07 & 4.8 & 0.8 \\
\enddata
\tablenotetext{a}{Fluxes ($F$) are in units of Jy; 
dust mass in 1--50 $\mu$m particles ($M_{d,1-50}$) 
is in units of $10^{-3}$ \mearth; dust production 
rate ($\dot{M}$) is in units of $10^{21}$ g yr$^{-1}$.
}
\tablenotetext{b}{Dust production rate from \citet{su05}. Our analysis
suggests a smaller dust production rate, $\dot{M} \gtrsim 0.3 \times 10^{21}$ 
g yr$^{-1}$.}
\tablenotetext{c}{Debris disk model with $M_{\star}$ = 2 \msun, $x_m$ = 1/3, $t$ = 200~Myr}
\tablenotetext{d}{As in note (c) for $M_{\star}$ = 2 \msun, $x_m$ = 1, $t$ = 200~Myr}
\tablenotetext{e}{As in note (c) for $M_{\star}$ = 2.5 \msun, $x_m$ = 1/2, $t$ = 200~Myr}
\tablenotetext{f}{As in note (c) for $M_{\star}$ = 2.5 \msun, $x_m$ = 1, $t$ = 200~Myr}
\label{tab:Vega}
\end{deluxetable}
\clearpage

%
\clearpage

\clearpage

\clearpage

\clearpage

\clearpage
\clearpage

\clearpage

\clearpage

\clearpage

\begin{thebibliography}{}

\bibitem[Absil et al. (2006)]{abs06} Absil, O., et al.\ 2006, \aap, 452, 237 

\bibitem[Adachi et al. (1976)]{ada76} Adachi, I., Hayashi, C., \& Nakazawa, K. 
1976, Progress of Theoretical Physics 56, 1756

\bibitem[Allen (1976)]{all76} Allen, C. W. 1976, {\it Astrophysical Quantities,}
Athlone, London, p. 197

\bibitem[Andrews \& Williams (2005)]{and05} Andrews, S.~M., \& Williams, 
J.~P.\ 2005, \apj, 631, 1134

\bibitem[Andrews \& Williams (2007a)]{and07a} Andrews, S.~M., \& Williams, J.~P.\ 2007a, \apj, 659, 705 

\bibitem[Andrews \& Williams (2007b)]{and07b} Andrews, S.~M., \& Williams, 
J.~P.\ 2007b, \apj, 671, 1800

\bibitem[Artymowicz (1988)]{art88} Artymowicz, P. 1988, \apj, 335, L79

\bibitem[Artymowicz (1997)]{art97}Artymowicz, P. 1997, ARE\&PS, 25, 175

\bibitem[Asphaug \& Benz (1996)]{asph96} Asphaug, E., \& Benz, 
W.\ 1996, Icarus, 121, 225 

\bibitem[Aufdenberg et al. (2006)]{auf06} Aufdenberg, J.~P., et al.\ 2006, 
\apj, 645, 664 

\bibitem[Augereau \& Beust (2006)]{aug06} Augereau, J.-C., \& Beust, H.\ 2006, \aap, 455, 987 

\bibitem[Augereau et al. (1999)]{au99} Augereau, J. C., Lagrange, A.-M.,
Mouillet, D., Papaloizou, J. C. B., \& Grorod, P. A. 1999, A\&A, 348, 557

\bibitem[Aumann et al. (1984)]{aum84} Aumann, H.~H., et al.\ 1984, \apjl, 
278, L23 

\bibitem [Backman \& Paresce (1993)]{bac93} Backman, D. E., \& Paresce, F. 
1993, in {\it Protostars and Planets III}, eds. E. H. Levy \& J. I. Lunine,
Tucson, Univ of Arizona, p. 1253

\bibitem[Barge \& Pellat (1991)]{bar91} Barge, P., \& Pellat, R. 1991,
Icarus, 93, 270

\bibitem[Beichman et al. (2005)]{bei05} Beichman, C.~A., et al.\ 2005, 
\apj, 626, 1061 

\bibitem[Beichman et al. (2006)]{bei06} Beichman, C.~A., et al.\ 2006, 
\apj, 652, 1674 

\bibitem[Benz \& Asphaug (1999)]{ben99} Benz, W., \& Asphaug, E. 1999,
Icarus, 142, 5

\bibitem[Brandeker et al. (2004)]{bra04} Brandeker, A., Liseau, R., 
Olofsson, G., \& Fridlund, M.\ 2004, \aap, 413, 681 

\bibitem[Bromley \& Kenyon (2006)]{bk06} Bromley, B., \& Kenyon, S. J.
2006, \aj, 131, 2737 

\bibitem[Brownlee et al. (1997)]{bro97} Brownlee, D.~E., et al.\ 1997, 
Meteoritics \& Planetary Science, vol.~32, page A22, 32, 22 

\bibitem[Bryden et al. (2006)]{bry06} Bryden, G., et al.\ 2006, \apj, 636, 1098 

\bibitem[Burns,~Lamy, \& Soter (1979)]{bur79} Burns, J. A., Lamy, P. L., 
\& Soter, S. 1979, Icarus, 40, 1

\bibitem[Campo Bagatin et al. (1994)]{cam94} Campo Bagatin, A., 
Cellino, A., Davis, D.~R., Farinella, P., \& Paolicchi, P.\ 1994

\bibitem[Carpenter et al. (2006)]{car06} Carpenter, J.~M., Mamajek, E.~E., 
Hillenbrand, L.~A., \& Meyer, M.~R.\ 2006, \apjl, 651, L49 

\bibitem[Chambers (2001)]{cha01} Chambers, J. E. 2001, Icarus, 152, 205

\bibitem[Chambers (2006)]{cha06} Chambers, J.\ 2006, Icarus, 180, 496

\bibitem[Charnoz \& Morbidelli (2003)]{cha03} Charnoz, S., \& Morbidelli, A.\ 2003, Icarus, 166, 141 

\bibitem[Chen et al. (2005)]{che05} Chen, C.~H., et al.\ 2005, \apj, 634, 1372 

\bibitem[Chen et al. (2006)]{che06} Chen, C.~H., et al.\ 2006, 
\apjs, 166, 351 

\bibitem[Ciesla (2007)]{cie07} Ciesla, F.~J.\ 2007, \apjl, 654, L159 

\bibitem[Currie et al. (2007a)]{cur07a} Currie, T., et al.\ 2007, \apj, 659, 599 

\bibitem[Currie et al. (2007b)]{cur07b} Currie, T., Kenyon, S.~J., Rieke, G., 
Balog, Z., \& Bromley, B.~C.\ 2007, \apjl, 663, L105 

\bibitem[Currie et al. (2007c)]{cur07c} Currie, T., Kenyon, S.~J., Balog, Z., 
Bragg, A., \& Tokarz, S.\ 2007, \apjl, 669, L33 

\bibitem[Currie et al. (2008a)]{cur08a} Currie, T., Kenyon, S.~J., 
Balog, Z., Rieke, G., Bragg, A., \& Bromley, B.\ 2008, \apj, 672, 558

\bibitem[Currie et al. (2008b)]{cur08b} Currie, T., Plavchan, P., \&
Kenyon, S. J. 2008b, \apj, submitted

\bibitem[Davis et al. (1985)]{dav85} Davis, D. R., Chapman, C. R., 
Weidenschilling, S. J., \& Greenberg, R. 1985, Icarus, 62, 30

\bibitem[Decin et al. (2003)]{dec03} Decin, G., Dominik, C., 
Waters, L. B. F. M., \& Waelkens, C. 2003, ApJ, 598, 636

\bibitem[Demarque et al. (2004)]{dem04} Demarque, P., Woo, J.-H., 
Kim, Y.-C., \& Yi, S.~K.\ 2004, \apjs, 155, 667 

\bibitem[Dent et al. (2000)]{den00} Dent, W.~R.~F., Walker, H.~J., 
Holland, W.~S., \& Greaves, J.~S.\ 2000, \mnras, 314, 702 

\bibitem[Dohnanyi (1969)]{doh69} Dohnanyi, J. W. 1969, 
J. Geophys. Res., 74, 2531

\bibitem[Dominik \& Decin (2003)]{dom03} Dominik, C., \& Decin, G. 2003,
ApJ, 598, 626

\bibitem[Dullemond \& Dominik (2005)]{dul05} Dullemond, C.~P., 
\& Dominik, C.\ 2005, \aap, 434, 971 

\bibitem[Durda \& Dermott (1997)]{dur97} Durda, D.~D., \& Dermott, S.~F.\ 1997, Icarus, 130, 140 

\bibitem[Elliot et al. (2003)]{ell03} Elliot, J.~L., Person, 
M.~J., \& Qu, S.\ 2003, \aj, 126, 1041 

\bibitem[Elliot et al. (2007)]{ell07} Elliot, J.~L., et al.\ 2007, \aj, 134, 1 

\bibitem[Fitzgerald et al. (2007)]{fit07} Fitzgerald, M.~P., 
Kalas, P.~G., \& Graham, J.~R.\ 2007, \apj, 670, 557 

\bibitem[Garaud (2007)]{gar07} Garaud, P.\ 2007, \apj, 671, 2091 

\bibitem[Goldreich,~Lithwick, \& Sari (2004)]{gol04} Goldreich, P., 
Lithwick, Y., \& Sari, R. 2004, ARA\&A, 42, 549

\bibitem[Golimowski et al. (1993)]{gol93} Golimowski, D.~A., 
Durrance, S.~T., \& Clampin, M.\ 1993, \apjl, 411, L41 

\bibitem[Gomes et al. (2005)]{gom05} Gomes, R., Levison, H.~F., 
Tsiganis, K., \& Morbidelli, A.\ 2005, \nat, 435, 466 

\bibitem[Gorlova et al. (2007)]{gor07} Gorlova, N., Balog, Z., Rieke, G.~H., 
Muzerolle, J., Su, K.~Y.~L., Ivanov, V.~D., \& Young, E.~T.\ 2007, \apj, 670, 516 

\bibitem[Gorlova et al. (2006)]{gor06} Gorlova, N., Rieke, G.~H., 
Muzerolle, J., Stauffer, J.~R., Siegler, N., Young, E.~T., \& 
Stansberry, J.~H.\ 2006, \apj, 649, 1028

\bibitem[Greaves et al. (1998)]{gre98} Greaves, J. S. et al. 1998, 
ApJ, 506, L133

\bibitem[Greaves et al. (2000)]{gre00} Greaves J. S., 
Mannings V. \& Holland, W. S. 2000b, Icarus, 143, 155

\bibitem[Greaves \& Wyatt (2003)]{gre03} Greaves, J. S., \& Wyatt, M. C. 
2003, MNRAS, 345, 1212

\bibitem[Greenberg et al. (1984)]{gre84} Greenberg, R., Weidenschilling, S. J.,
Chapman, C. R., \& Davis, D. R. 1984, Icarus, 59, 87

\bibitem[Greenberg et al. (1991)]{gre91} Greenberg, R., Bottke, W.,
Carusi, A., Valsecchi, G. B. 1991, Icarus, 94, 98

\bibitem[Greenzweig \& Lissauer (1990)]{grz90}
Greenzweig, Y., \& Lissauer, J. J. 1990, Icarus, 87, 40

\bibitem[Greenzweig \& Lissauer (1992)]{grz92}
Greenzweig, Y., \& Lissauer, J. J. 1992, Icarus, 100, 440

\bibitem[Grigorieva et al. (2007)]{gri07} Grigorieva, A., Artymowicz, P., 
\& Th{\'e}bault, P.\ 2007, \aap, 461, 537 

\bibitem[Gr{\"u}n et al. (1995)]{gru95} Gr{\"u}n, E., et al. 1995, \planss, 43, 971 

\bibitem[Habing et al. (2001)]{hab01} Habing, H. J., et al. 2001, A\&A, 365, 545

\bibitem[Hahn et al. (2002)]{hah02} Hahn, J.~M., Zook, H.~A., Cooper, B., 
\& Sunkara, B.\ 2002, Icarus, 158, 360 

\bibitem[Haisch,~Lada, \& Lada (2001)]{hai01} Haisch, K., Lada, E. A.,
\& Lada, C. J. 2001, ApJ, 553, 153

\bibitem[Hayashi (1981)]{hay81} Hayashi, C. 1981, Prog Theor Phys Suppl, 70, 35

\bibitem[Hern{\'a}ndez et al. (2006)]{her06} Hern{\'a}ndez, J., Brice{\~n}o, 
C., Calvet, N., Hartmann, L., Muzerolle, J., \& Quintero, A.\ 2006, \apj, 652, 472 

\bibitem[Hillenbrand et al. (2008)]{hil08} Hillenbrand, L.~A., et al.\ 2008, 
ArXiv e-prints, 801, arXiv:0801.0163 

\bibitem[Holland et al. (1998)]{hol98} Holland, W.~S., et al.\ 1998, \nat, 392, 788 

\bibitem[Holland et al. (2003)]{hol03} Holland, W. S., et al. 2003, ApJ, 582, 1141

\bibitem[Hornung et al. (1985)]{hor85} Hornung, P., Pellat, R., \& Barge, P.  1985, Icarus, 64, 295

\bibitem[Iben (1967)]{ibe67} Iben, I. Jr, 1967, ARA\&A, 5, 571

\bibitem[Ida (1990)]{ida90} Ida, S.\ 1990, Icarus, 88, 129 

\bibitem[Ida \& Lin (2008)]{ida08} Ida, S., \& Lin, D.~N.~C.\ 2008, 
ArXiv e-prints, 802, arXiv:0802.1114 

\bibitem[Ida \& Makino (1992)]{ida92} Ida, S., \& Makino, J.\ 1992, Icarus, 96, 107 

\bibitem[Ida \& Makino (1993)]{ida93} Ida, S., \& Makino, J. 1993, Icarus, 106, 210

\bibitem[Inaba et al. (2001)]{ina01} Inaba, S. H., Tanaka, H., Nakazawa, K.,
Wetherill, G. W., \& Kokubo, E. 2001, Icarus, 149, 235

\bibitem[Jayawardhana et al. (1998)]{jay98} Jayawardhana, R. et al. 1998,
ApJ, 503, L79

\bibitem[Johnson et al. (2007)]{john07} Johnson, J.~A., et al.\ 2007, \apj, 665, 785 

\bibitem[Kalas (1998)]{kal98} Kalas, P. 1998, Earth, Moon, \& Planets, 81, 27 

\bibitem[Kalas (2005)]{kal05} Kalas, P.\ 2005, \apjl, 635, L169

\bibitem[Kalas et al. (2005)]{kal05b} Kalas, P., Graham, J.~R., \& Clampin, M.\ 
2005, \nat, 435, 1067 

\bibitem[Kalas et al. (2006)]{kal06} Kalas, P., Graham, J.~R., Clampin, M.~C., 
\& Fitzgerald, M.~P.\ 2006, \apjl, 637, L57 

\bibitem[Kalas et al. (2004)]{kal04} Kalas, P., Liu, M.~C., \& 
Matthews, B.~C.\ 2004, Science, 303, 1990 

\bibitem[Kennedy \& Kenyon (2008)]{kenn08} Kennedy, G.~M., \& Kenyon, S.~J.\ 2008, \apj, 673, 502 

\bibitem[Kenyon \& Bromley (2001)]{kb01} Kenyon, S. J., \& Bromley, B. C.
2001, AJ, 121, 538

\bibitem[Kenyon \& Bromley (2002a)]{kb02a} Kenyon, S. J., \& Bromley, B. C.
2002a, AJ, 123, 1757

\bibitem[Kenyon \& Bromley (2002b)]{kb02b} Kenyon, S. J., \& Bromley, B. C.
2002b, ApJ, 577, L35

\bibitem[Kenyon \& Bromley (2004a)]{kb04a} Kenyon, S. J., \& Bromley, B. C.,
2004a, AJ, 127, 513

\bibitem[Kenyon \& Bromley (2004b)]{kb04b} Kenyon, S. J., \& Bromley, B. C.,
2004b, ApJ, 602, L133

\bibitem[Kenyon \& Bromley (2004c)]{kb04c} Kenyon, S. J., \& Bromley, B. C.,
2004c, AJ, 128, 1916

\bibitem[Kenyon \& Bromley (2005)]{kb05} Kenyon, S.~J., \& Bromley, B.~C.\ 2005, \aj, 130, 269 

\bibitem[Kenyon \& Bromley (2006)]{kb06} Kenyon, S.~J., \& 
Bromley, B.~C.\ 2006, \aj, 131, 1837 

\bibitem[Kenyon et al. (2008)]{kbod08} Kenyon, S. J., Bromley, B. C., O'Brien, 
D. C., \& Davis, D. R. 2008, to appear in {\it The Solar System Beyond Neptune,}
edited by A. Barucci, H. Boehnhardt, D. Cruikshank, \& A. Morbidelli, Tucson, 
Univ. of Arizona Press, in press

\bibitem[Kenyon \& Hartmann (1987)]{kh87} Kenyon, S. J., \& 
Hartmann, L. 1987, ApJ, 323, 714

\bibitem[Kenyon \& Hartmann (1995)]{kh95}  Kenyon, S.~J., \& Hartmann, L. W.,
1995, ApJS, 101, 117

\bibitem[Kenyon \& Luu (1998)]{kl98} Kenyon, S. J., \& Luu, J. X.
1998, AJ, 115, 2136 

\bibitem[Kenyon \& Luu (1999)]{kl99} Kenyon, S. J., \& Luu, J. X.
1999, AJ, 118, 1101

\bibitem[Kenyon et al. (1999)]{ken99} Kenyon, S. J., Wood, K., 
Whitney, B. A., \& Wolff, M. 1999, ApJ, 524, L119

\bibitem[Kim et al. (2005)]{kim05} Kim, J.~S., et al.\ 2005, \apj, 632, 659 

\bibitem[Kimura et al. (2002)]{kim02} Kimura, H., Okamoto, H., \& Mukai, 
T.\ 2002, Icarus, 157, 349 

\bibitem[Knapp \& Morris (1985)]{kna85} Knapp, G.~R., \& Morris, M.\ 1985, \apj, 292, 640 

\bibitem[Kobayashi \& Ida (2001)]{kob01} Kobayashi, H., \& Ida, S. 2001,
Icarus, 153, 416

\bibitem[Koeberl (2003)]{koe03} Koeberl, C. 2003, EM\&P, 92, 79

\bibitem[Koerner et al. (1998)]{koe98} Koerner, D. W., Ressler, M. E., 
Werner, M. W., \& Backman, D. E. 1998, ApJ, 503, L83

\bibitem[Koerner,~Sargent, \& Ostroff  (2001)]{koe01} Koerner, D. W., 
Sargent, A. I., \& Ostroff, N. A. 2001, ApJ, 560, L181

\bibitem[Kokubo \& Ida (1998)]{kok98} Kokubo, E., \& Ida, S. 1998,
Icarus, 131, 171

\bibitem[Kretke et al. (2008)]{kre08} Kretke, K.~A., Lin, D.~N.~C., 
Garaud, P., \& Turner, N.~J.\ 2008, ArXiv e-prints, 806, arXiv:0806.1521 

\bibitem[Krivov et al. (2000)]{kri00} Krivov, A. V., Mann, I., \& Krivova, 
N. A. 2000, A\&A, 362, 1127

\bibitem[Krivov et al. (2006)]{kri06} Krivov, A.~V., 
L{\"o}hne, T., \& Srem{\v c}evi{\'c}, M.\ 2006, \aap, 455, 509 

\bibitem[Kuchner \& Holman (2003)]{kuc03} Kuchner, M. J., \& Holman, M. J.
2003, ApJ, 588, 1110

\bibitem[Lada (1999)]{lad99} Lada, C. J. 1999, in
{\it The Physics of Star Formation and Early Stellar Evolution,}
edited by C. J. Lada and N. Kylafis, Dordrecht, Kluwer, p. 143

\bibitem[Lagrange et al. (2000)]{lag00} Lagrange, A.-M., Backman, D.,
\& Artymowicz, P. 2000, in Protostars \& Planets IV, eds.  V. Mannings, 
A. P. Boss, \& S. S. Russell, Tucson, Univ. of Arizona, in press

\bibitem[Landgraf et~al. (2002)]{lan02} Landgraf, M., Liou, J.-C., Zook, H.~A., 
\& Gr{\"u}n, E. 2002, \aj, 123, 2857

\bibitem[Larwood (1997)]{lar97} Larwood, J. D. 1997, MNRAS, 290, 490

\bibitem[Larwood \& Kalas (2001)]{lar01} Larwood, J. D., \& Kalas, P. G.
2001, MNRAS, 323, 402

\bibitem[Leinhardt et al. (2008)]{lein08} Leinhardt, Z.~M., Stewart, S.~T., 
\& Schultz, P.~H.\ 2008, to appear in {\it The Solar System Beyond Neptune,}
edited by A. Barucci, H. Boehnhardt, D. Cruikshank, \& A. Morbidelli, Tucson,
Univ. of Arizona Press, in press (ArXiv e-prints, 705, arXiv:0705.3943)

\bibitem[Levison \& Stewart (2001)]{lev01} Levison, H.~F., \& 
Stewart, G.~R.\ 2001, Icarus, 153, 224

\bibitem[Lin \& Papaloizou (1979)]{lin79}Lin, D. N. C., \&
Papaloizou, J. C. B. 1979, MNRAS, 186, 799

\bibitem[Lissauer (1987)]{lis87} Lissauer, J. J. 1987, Icarus, 69, 249

\bibitem[Lissauer \& Stewart (1993)]{lis93} Lissauer, J. J., \&
Stewart, G. R. 1993,  In {\it Protostars and Planets III,}
edited by E. H. Levy and J. I. Lunine, U. of Arizona Press, Tucson, 1061

\bibitem[Lisse et al. (2007a)]{lis07a} Lisse, C.~M., Beichman, C.~A., 
Bryden, G., \& Wyatt, M.~C.\ 2007, \apj, 658, 584 

\bibitem[Lisse et al. (2007c)]{lis07c} Lisse, C.~M., Chen, C.~H., 
Wyatt, M.~C., \& Morlok, A.\ 2007, ArXiv e-prints, 710, arXiv:0710.0839 

\bibitem[Lisse et al. (2007b)]{lis07b} Lisse, C.~M., Kraemer, K.~E., 
Nuth, J.~A., Li, A., \& Joswiak, D.\ 2007, Icarus, 191, 223 

\bibitem[Liu (2004)]{liu04a} Liu, M.~C.\ 2004, Science, 305, 1442 

\bibitem[Liu et al. (2004)]{liu04b} Liu, M.~C., Matthews, 
B.~C., Williams, J.~P., \& Kalas, P.~G.\ 2004, \apj, 608, 526 

\bibitem[Liu et al. (2004)]{wliu04} Liu, W.~M., et al.\ 2004, 
\apjl, 610, L125 

\bibitem[L{\"o}hne et al. (2008)]{loh08} L{\"o}hne, T., Krivov, A.~V., 
\& Rodmann, J.\ 2008, \apj, 673, 1123

\bibitem[Marsh et al. (2006)]{mar06} Marsh, K.~A., Dowell, C.~D., 
Velusamy, T., Grogan, K., \& Beichman, C.~A.\ 2006, \apjl, 646, L77 

\bibitem[Matthews et al. (2007)]{mat07} Matthews, B.~C., et al.\ 2007, \pasp, 119, 842 

\bibitem[Melosh,~Vickery, \& Tonks (1993)]{mel93} Melosh, H. J.,
Vockery, A. M., \& Tonks, W. B. 1993, in
{\it Protostars and Planets III}, eds. E. H. Levy \& J. I. Lunine,
Tucson, Univ of Arizona, p. 1339

\bibitem[Meyer et al. (2007)]{mey07} Meyer, M.~R., Backman, 
D.~E., Weinberger, A.~J., \& Wyatt, M.~C.\ 2007, Protostars and Planets V, 573 

\bibitem[Meyer et al. (2006)]{mey06} Meyer, M.~R., et al.\ 2006, \pasp, 118, 1690 
\bibitem[Meyer et al. (2008)]{mey08} Meyer, M.~R., et al.\ 2008, 
\apjl, 673, L181 

\bibitem[Mo{\'o}r et al. (2006)]{moor06} Mo{\'o}r, A., {\'A}brah{\'a}m, P., 
Derekas, A., Kiss, C., Kiss, L.~L., Apai, D., Grady, C., \& Henning, T.\ 
2006, \apj, 644, 525 

\bibitem[Moro-Mart{\'{\i}}n \& Malhotra (2005)]{mor05} 
Moro-Mart{\'{\i}}n, A., \& Malhotra, R.\ 2005, \apj, 633, 1150 

\bibitem[Moro-Martin et al. (2007)]{mor08} Moro-Martin, A., Wyatt, M.~C., 
Malhotra, R., \& Trilling, D.~E.\ 2007, ArXiv Astrophysics e-prints, 
arXiv:astro-ph/0703383

\bibitem[Motte \& Andr{\'e}(2001)]{mot01} Motte, F., \& Andr{\'e}, P.\ 
2001, \aap, 365, 440 

\bibitem[Mouillet et al. (1997)]{mou97} Mouillet, D., Larwood, J. D.,
Papaloizou, J. C. B., \& Lagrange, A.-M. 1997, MNRAS, 292, 896

\bibitem[Nagasawa et al. (2007)]{naga07} Nagasawa, M., Thommes, 
E.~W., Kenyon, S.~J., Bromley, B.~C., \& Lin, D.~N.~C.\ 2007, Protostars 
and Planets V, 639 

\bibitem[Najita \& Williams (2005)]{naj05} Najita, J., \& Williams, J.~P.\ 2005, \apj, 635, 625 

\bibitem[Natta et al. (2000)]{nat2000} Natta, A., Grinin, V., \& 
Mannings, V.\ 2000, Protostars and Planets IV, 559

\bibitem[Nesvorn{\'y} et al. (2006)]{nes06} Nesvorn{\'y}, D., 
Vokrouhlick{\'y}, D., Bottke, W.~F., \& Sykes, M.\ 2006, Icarus, 181, 107 

\bibitem[Nomura \& Nakagawa (2006)]{nom06} Nomura, H., \& Nakagawa, Y.\ 2006, 
\apj, 640, 1099 

\bibitem[Ohtsuki (1992)]{oht92} Ohtsuki, K. 1992, Icarus, 98, 20

\bibitem[Ohtsuki,~Stewart, \& Ida (2002)]{oht02} Ohtsuki, K.,
Stewart, G. R., \& Ida, S. 2002, Icarus, 155, 436

\bibitem[Osterloh \& Beckwith (1995)]{ost95} Osterloh, M., \&
Beckwith, S. V. W. 1995, ApJ, 439, 288

\bibitem[Pan \& Sari (2005)]{pan05} Pan, M., \& Sari, R. 2005,
Icarus, 173, 342

\bibitem[Press et al. (1992)]{pre92} Press, W. H., Flannery, B. P., 
Teukolsky, S. A., \& Vetterling, W. T. 1992, {\it Numerical Recipes, 
The Art of Scientific Computing,} Cambridge, Cambridge

\bibitem[Quillen (2006)]{qui06} Quillen, A.~C.\ 2006, \mnras, 372, L14 

\bibitem[Quillen et al. (2007)]{qui07} Quillen, A.~C., Morbidelli, A., 
\& Moore, A.\ 2007, \mnras, 380, 1642 

\bibitem[Rafikov (2003)]{raf03} Rafikov, R.~R.\ 2003, \aj, 125, 942 

\bibitem[Rafikov (2004)]{raf04} Rafikov, R.~R.\ 2004, \aj, 128, 1348 

\bibitem[Rhee et al. (2007a)]{rhe07a} Rhee, J.~H., Song, I., Zuckerman, B., 
\& McElwain, M.\ 2007, \apj, 660, 1556 

\bibitem[Rhee et al. (2007b)]{rhe07b} Rhee, J.~H., Song, I., \& 
Zuckerman, B.\ 2007, ArXiv e-prints, 711, arXiv:0711.2111 

\bibitem[Rieke et al. (2005)]{rie05} Rieke, G. H., Su, K. Y. L.,
Stansberry, J. A., Trilling, D., Bryden, G., Muzerolle, J., White, B.,
Gorlova, N., Young, E. T., Beichman, C. A., Stapelfeldt, K. R., \&
hines, D. C. 2005, ApJ, 620, 1010

\bibitem[Safronov (1969)]{saf69} Safronov, V. S. 1969, Evolution of
the Protoplanetary Cloud and Formation of the Earth and Planets,
Nauka, Moscow [Translation 1972, NASA TT F-677]

\bibitem[Schneider et al. (1999)]{sch99} Schneider, G., et al. 
1999, ApJ, 513, L127

\bibitem[Scholz et al. (2006)]{sch2006} Scholz, A., 
Jayawardhana, R., \& Wood, K.\ 2006, \apj, 645, 1498 

\bibitem[Sicilia-Aguilar et al. (2006)]{sic06} 
Sicilia-Aguilar, A., et al.\ 2006, \apj, 638, 897 

\bibitem[Siegler et al. (2007)]{sie07} Siegler, N., Muzerolle, J., 
Young, E.~T., Rieke, G.~H., Mamajek, E.~E., Trilling, D.~E., Gorlova, N., 
\& Su, K.~Y.~L.\ 2007, \apj, 654, 580 

\bibitem[Smith \& Terrile (1984)]{smi84}
Smith, B. A., \& Terrile, R. J. 1984, Science, 226, 1421

\bibitem[Song et al. (2005)]{son05} Song, I., Zuckerman, B., 
Weinberger, A.~J., \& Becklin, E.~E.\ 2005, \nat, 436, 363 

\bibitem[Spaute et al. (1991)]{spa91} Spaute, D., Weidenschilling, S. J.,
Davis, D. R., \& Marzari, F. 1991, Icarus, 92, 147 

\bibitem[Stapelfeldt et al. (2004)]{sta04} Stapelfeldt, K.~R., et al.\ 
2004, \apjs, 154, 458

\bibitem[Stauffer et al. (2005)]{stau05} Stauffer, J.~R., et al.\ 2005, \aj, 130, 1834 

\bibitem[Stern \& Colwell (1997)]{st97} Stern, S. A., \&
Colwell, J. E. 1997a,  AJ, 114, 841

\bibitem[Stewart \& Ida (2000)]{ste00} Stewart, G. R., \& Ida, S. 2000, 
Icarus, 143, 28

\bibitem[Su et al. (2005)]{su05} Su, K.~Y.~L., et al.\ 2005, 
\apj, 628, 487 

\bibitem[Su et al. (2006)]{su06} Su, K.~Y.~L., et al.\ 2006, 
\apj, 653, 675 

\bibitem[Su et al. (2008)]{su08} Su, K.~Y.~L., Rieke, G.~H., 
Stapelfeldt, K.~R., Smith, P.~S., Bryden, G., Chen, C.~H., 
\& Trilling, D.~E.\ 2008, \apjl, 679, L125 

\bibitem[Swindle (1993)]{swi93} Swindle, T. D. 1993, in
{\it Protostars and Planets III}, eds. E. H. Levy \& J. I. Lunine,
Tucson, Univ of Arizona, p. 867

\bibitem[Takeuchi \& Artymowicz (2001)]{tak01} Takeuchi, T., \&
Artymowicz, P. 2001, ApJ, 557, 990

\bibitem[Telesco et al. (1988)]{tel88} Telesco, C.~M., Decher, R., Becklin, 
E.~E., \& Wolstencroft, R.~D.\ 1988, \nat, 335, 51 

\bibitem[Telesco et al. (2000)]{tel00} Telesco, C.~M., et al.\ 
2000, \apj, 530, 329

\bibitem[Th{\'e}bault \& Augereau (2007)]{the07} Th{\'e}bault, 
P., \& Augereau, J.-C.\ 2007, \aap, 472, 169

\bibitem[Th\'ebault,~Augereau, \& Beust (2003)]{the03}
Th\'ebault, P., Augereau, J. C., \& Beust, H. 2003, A\&A, 408, 775

\bibitem[Trilling et al. (2007)]{tri07} Trilling, D.~E., et 
al.\ 2007, \apj, 658, 1289 

\bibitem[Trilling et al. (2008)]{tri08} Trilling, D.~E., et 
al.\ 2008, \apj, 674, 1086

\bibitem[Wadhwa \& Russell (2000)]{wad00} Wadhwa, M., \& Russell, S. S.
2000, in {\it Protostars abd Planets IV}, eds. V. Mannings, A. P. Boss,
\& S. S. Russell, Tucson, Univ. of Arizona, p. 995

\bibitem[Weidenschilling (1977a)]{wei77a} Weidenschilling, S. J. 1977a,
Astrophys Sp Sci, 51, 153

\bibitem[Weidenschilling (1977b)]{wei77b} Weidenschilling, S. J. 1977b,
MNRAS, 180, 57

\bibitem[Weidenschilling (1989)]{wei89} Weidenschilling, S.~J.\ 1989, Icarus, 80, 179 

\bibitem[Weidenschilling et al. (1997)]{wei97} Weidenschilling, S. J., 
Spaute, D., Davis, D. R., Marzari, F., \& Ohtsuki, K. 1997, Icarus, 128, 429

\bibitem[Wetherill (1980)]{wet80} Wetherill, G. W. 1980, ARA\&A, 18, 77

\bibitem[Wetherill \& Stewart (1993)]{wet93} Wetherill, G. W., \& Stewart, 
G. R. 1993,  Icarus, 106, 190

\bibitem[Williams \& Wetherill (1994)]{wil94} Williams, D. R., \&
Wetherill, G. W. 1994, Icarus, 107, 117

\bibitem[Williams \& Andrews (2006)]{wil06} Williams, J.~P., \& 
Andrews, S.~M.\ 2006, \apj, 653, 1480 

\bibitem[Wilner et al. (2002)]{wil02} Wilner, D. J., Holman, M. J.,
Kuchner, M. J., \& Ho, P. T. P. 2002, ApJ, 569, 115

\bibitem[Wolf \& Hillenbrand (2003)]{wol03} Wolf, S., \& Hillenbrand, L.~A.\ 2003, \apj, 596, 603 

\bibitem[Wood et al. (2002)]{woo02} Wood, K., Lada, C. J., Bjorkman, J. E.,
Kenyon, S. J., Whitney, B., Wolff, M. J. 2002, ApJ, 567, 1183

\bibitem[Wyatt (2003)]{wya03b} Wyatt, M. C. 2003, ApJ, 598, 1321

\bibitem[Wyatt (2005)]{wya05} Wyatt, M.~C.\ 2005, \aap, 433, 1007 

\bibitem[Wyatt \& Dent (2002)]{wya02} Wyatt, M. C., \& Dent, W. R. F. 
2002, MNRAS, 334, 589

\bibitem[Wyatt,~Dent, \& Greaves (2003)]{wya03a} Wyatt, M. C., Dent, W. R. F.,
\& Greaves, J. S. 2003, MNRAS, 342, 867

\bibitem[Wyatt et al. (1999)]{wya99} Wyatt, M. C., Dermott, S. F.,
Telesco, C. M., Fisher, R. S., Grogan, K., Holmes, E. K., \& Pi\~na, R. K.
1999, ApJ, 527, 918

\bibitem[Wyatt et al. (2007a)]{wya07a} Wyatt, M.~C., Smith, R., Greaves, J.~S., 
Beichman, C.~A., Bryden, G., \& Lisse, C.~M.\ 2007, \apj, 658, 569 

\bibitem[Wyatt et al. (2007b)]{wya07b} Wyatt, M.~C., Smith, R., Su, K.~Y.~L., 
Rieke, G.~H., Greaves, J.~S., Beichman, C.~A., \& Bryden, G.\ 2007, \apj, 663, 365 

\bibitem[Yin et al. (2002)]{yin02} Yin, Q., Jacobsen, S. B., Yamashita, K.,
Blichert-Toft, J., T\'elouk, P.; Albar\`ede, F. 2002, Nature, 418, 949

\bibitem[Young \& Binzel (1994)]{you94} Young, E.~F., \& 
Binzel, R.~P.\ 1994, Icarus, 108, 219 

\bibitem[Zuckerman (2001)]{zuc01} Zuckerman, B. 2001, ARA\&A, 39, 549

\end{thebibliography}
\end{document}